\documentclass[10pt]{article}

\usepackage{amsfonts,amssymb,amsmath,mathrsfs,latexsym,bbm,dsfont,amsthm}
\usepackage{color}
\usepackage{graphicx}
\usepackage{pstricks,pstricks-add}
\usepackage{pst-plot}
\usepackage[scriptsize,nooneline,hang]{caption}
\usepackage[hang,nooneline,scriptsize]{subfigure}
\usepackage{rotating}
\usepackage[all]{xy}

\usepackage[abs]{overpic}

\begin{document}

\title{Geodesic Motion in the (Charged) Doubly Spinning Black Ring Spacetime}

\author{Saskia Grunau, Valeria Kagramanova, Jutta Kunz\\
Institut f\"ur Physik, Universit\"at Oldenburg,
D--26111 Oldenburg, Germany
}

\maketitle

\begin{abstract}
In this article we analyze the geodesics of test particles and light in the five dimensional (charged) doubly spinning black ring spacetime. Apparently it is not possible to separate the Hamilton-Jacobi-equation for (charged) doubly spinning black rings in general, so we concentrate on special cases: null geodesics in the ergosphere and geodesics on the two rotational axes of the (charged) doubly spinning black ring. We present analytical solutions to the geodesic equations for these special cases. Using effective potential techniques we study the motion of test particles and light and discuss the corresponding orbits.
\end{abstract}

\section{Introduction}

A promising candidate to solve the problem of quantum gravity is the string theory. For its internal consistency more than four dimensions are required. This led to growing interest in higher dimensional solutions and especially higher dimensional black holes (see e.g. \cite{Emparan:2008eg}). The higher dimensional analogon of the stationary axisymmetric Kerr black hole was found by Myers and Perry \cite{Myers:1986un}. They also conjectured the existence of higher dimensional black holes with a non-spherical event horizon. Emparan and Reall \cite{Emparan:2001wn} found such a solution with a topology of $S^1 \times S^2$, the rotating black ring. Since there is a certain range of values for the angular momentum where five dimensional black holes as well as two branches of black rings exist, the uniqueness theorem is no longer valid in higher dimensions.

The first black ring solution found possesses a single angular momentum. A black ring with two angular momenta rotating in two independent planes was constructed by Pomeransky and Sen'kov \cite{Pomeransky:2006bd}.

In 2003 Elvang presented a charged singly spinning black ring solution \cite{Elvang:2003yy}. The doubly spinning version of the charged black ring was found by Hoskisson \cite{Hoskisson:2008qq}. Gal'tsov and Scherbluk \cite{Gal'tsov:2009da} constructed a three-charge doubly spinning black ring, but unless the Dirac-Misner string is removed by setting two of the charges to zero, this solution is unbalanced. Various further black ring solutions are known (see e.g.  \cite{Elvang:2004rt,Elvang:2004ds,Chen:2012kd,Rocha:2012vs}).\\

The geodesic equations for test particles and light in Myers-Perry black hole spacetimes are separable \cite{Kubiznak:2006kt,Page:2006ka,Frolov:2006pe}. Apparently this is not possible in black ring spacetimes, in general. However one can separate the equations of motion in special cases: geodesics on the two rotational axes (which are actually planes) of the charged doubly spinning black ring and zero energy null geodesics (which can only exist in the ergosphere) \cite{Hoskisson:2007zk,Durkee:2008an}.

The geodesic motion in the singly spinning black ring spacetime was studied by Hoskisson \cite{Hoskisson:2007zk}. He discussed the separability of the Hamilton-Jacobi equation and studied numerically the motion in the equatorial plane and the rotational axis of the black ring. Elvang, Emparan and Virmani \cite{Elvang:2006dd} analyzed some aspects of null geodesics in the equatorial plane. Also zero energy null geodesics in the singly spinning dipole black ring spacetime were studied by Armas \cite{Armas:2010pw}. Igata, Ishihara and Takamori \cite{Igata:2010ye} found numerically stable bound orbits on and near the rotational axis.

Analytical solutions of the equations of motion for zero energy null geodesics, geodesics in the equatorial plane and on the rotational axis of the singly spinning black ring were given in \cite{Grunau:2012ai}, where also the possible orbits were discussed.

Geodesics of the doubly spinning black ring were studied by Durkee \cite{Durkee:2008an}, who separated the equations of motion for zero energy null geodesics and on the two rotational axes and analyzed the effective potential.

But so far neither numerical nor analytical solutions of the equations of motion have been given in the (charged) doubly spinning black ring spacetime.\\

Hagihara \cite{Hagihara:1931} was the first to solve the equations of motion in a black hole spacetime analytically. He found the solution of the equations of motion in the Schwarzschild spacetime in terms of the elliptic Weierstra{\ss} $\wp$ function.

When the cosmological constant is added to the Schwarzschild metric, the geodesic equations are of hyperelliptic type. The analytical solution of the geodesic equation in Schwarzschild-(anti) de Sitter spacetimes was found in \cite{Hackmann:2008zz}. Also in higher dimensional Myers-Perry spacetime \cite{Enolski:2010if} and in higher dimensional Schwarzschild, Schwarzschild-(anti)de Sitter, Reissner-Nordstr\"om and Reissner-Nordstr\"om -(anti) de Sitter spactime \cite{Hackmann:2008tu} the equations of motion were solved analytically.

Mathematically speaking the integration of the equations of motion can be reduced to the solution of the Jacobi inversion problem. The solution can be found by restricting the problem to the theta-divisor, the set of zeros of the theta function. This method was developed in 2003 by Enolski, Pronine and Richter \cite{Enolski:2003} to solve the problem of the double pendulum.

In this article analytic solutions of the geodesic equations in the charged doubly spinning black ring spacetime are presented (for special cases). In the case $E=m=0$ the equations of motion are of elliptic type, however on the two rotational axes of the doubly spinning black ring, the equations are of hyperelliptic type.

\section{(Charged) Doubly Spinning Black Ring Spacetime}
\label{sec:spacetime}

\begin{figure}[h]
 \centering
 \includegraphics[width=12cm]{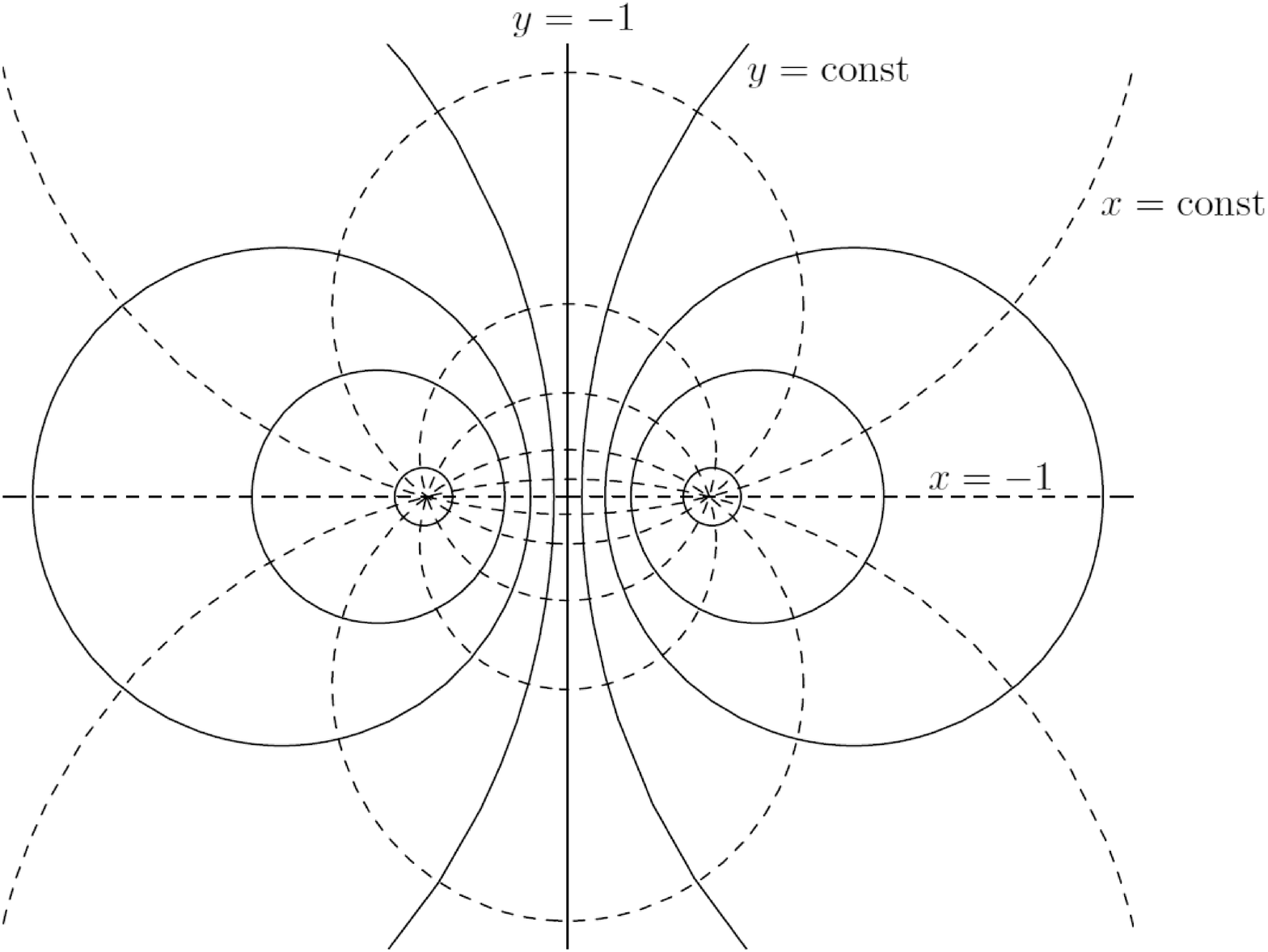}
 \caption{Toroidal coordinates (or ring coordinates) on a cross section at constant angles $\phi$ and $\psi$. Solid circles correspond to $y=\mathrm{const.}$ and dashed circles corredspond to  $x=\mathrm{const.}$. \cite{Emparan:2006mm}}
 \label{pic:ringcoord}
\end{figure}

The metric of an uncharged doubly spinning black ring can be written in the form \cite{Durkee:2008an}
\begin{eqnarray}
 \mathrm{d} s^2 &=& -\frac{H(y,x)}{H(x,y)} (\mathrm{d}t + \Omega)^2 + \frac{R^2H(x,y)}{(x-y)^2(1-\nu)^2} \left[ \frac{\mathrm{d}x^2}{G(x)} - \frac{\mathrm{d}y^2}{G(y)} \right. \nonumber \\ 
 && \left. + \frac{A(y,x)\mathrm{d}\phi^2 - 2L(x,y)\mathrm{d}\phi\mathrm{d}\psi - A(x,y)\mathrm{d}\psi^2}{H(x,y)H(y,x)} \right] \,.
\end{eqnarray}
The metric is given in toroidal coordinates (see \ref{pic:ringcoord}) where $-1 \leq x \leq 1$, $-\infty < y \leq -1$ and $-\infty < t < \infty$.  $\phi$ and $\psi$ are $2\pi$-periodic. The metric functions are
\begin{eqnarray}
 G(x) &=& (1-x^2)(1+\lambda x+\nu x^2) \nonumber \\
 H(x,y) &=& 1+ \lambda^2 -\nu^2 + 2\lambda\nu (1-x^2)y + 2x\lambda (1-y^2\nu^2) + x^2y^2\nu (1-\lambda^2-\nu^2) \nonumber \\
 L(x,y) &=& \lambda \sqrt{\nu} (x-y)(1-x^2)(1-y^2)[1+\lambda^2-\nu^2+2(x+y)\lambda\nu -xy\nu (1-\lambda^2-\nu^2)] \nonumber \\
 A(x,y) &=& G(x)(1-y^2) [((1-\nu^2)-\lambda^2)(1+\nu)+y\lambda (1-\lambda^2 + 2\nu -3\nu^2)] \nonumber \\
        && + G(y)[2\lambda^2 + x\lambda((1-\nu)^2+\lambda^2) + x^2((1-\nu)^2-\lambda^2)(1+\nu) \nonumber \\
        && + x^3\lambda (1-\lambda^2-3\nu^2+2\nu^3) + x^4\nu (1-\nu)(1-\lambda^2-\nu^2)] \, .
\end{eqnarray}
The parameters $R$, $\lambda$ and $\nu$ describe the shape, mass and angular momenta of the ring and lie in the range $0\leq \nu <1$ and respectively $2\sqrt{\nu} \leq \lambda <1+\nu$. If $\nu=0$ the metric reduces to the singly spinning black ring.

A ring-like curvature singularity is located at $y=-\infty$. The metric has a coordinate singularity at $G(y)=0$, so there are two horizons
\begin{eqnarray}
 y_{h+} &=& \frac{-\lambda + \sqrt{\lambda ^2 -4\nu}}{2\nu} \\
 y_{h-} &=& \frac{-\lambda - \sqrt{\lambda ^2 -4\nu}}{2\nu} \, .
\label{eqn:horizonte}
\end{eqnarray}

\begin{figure}[h]
 \centering
 \subfigure[$\nu=0.1$ and $\lambda=0.7$: ring-like ergosphere in $a$-$b$ coordinates]{
   \includegraphics[width=7cm]{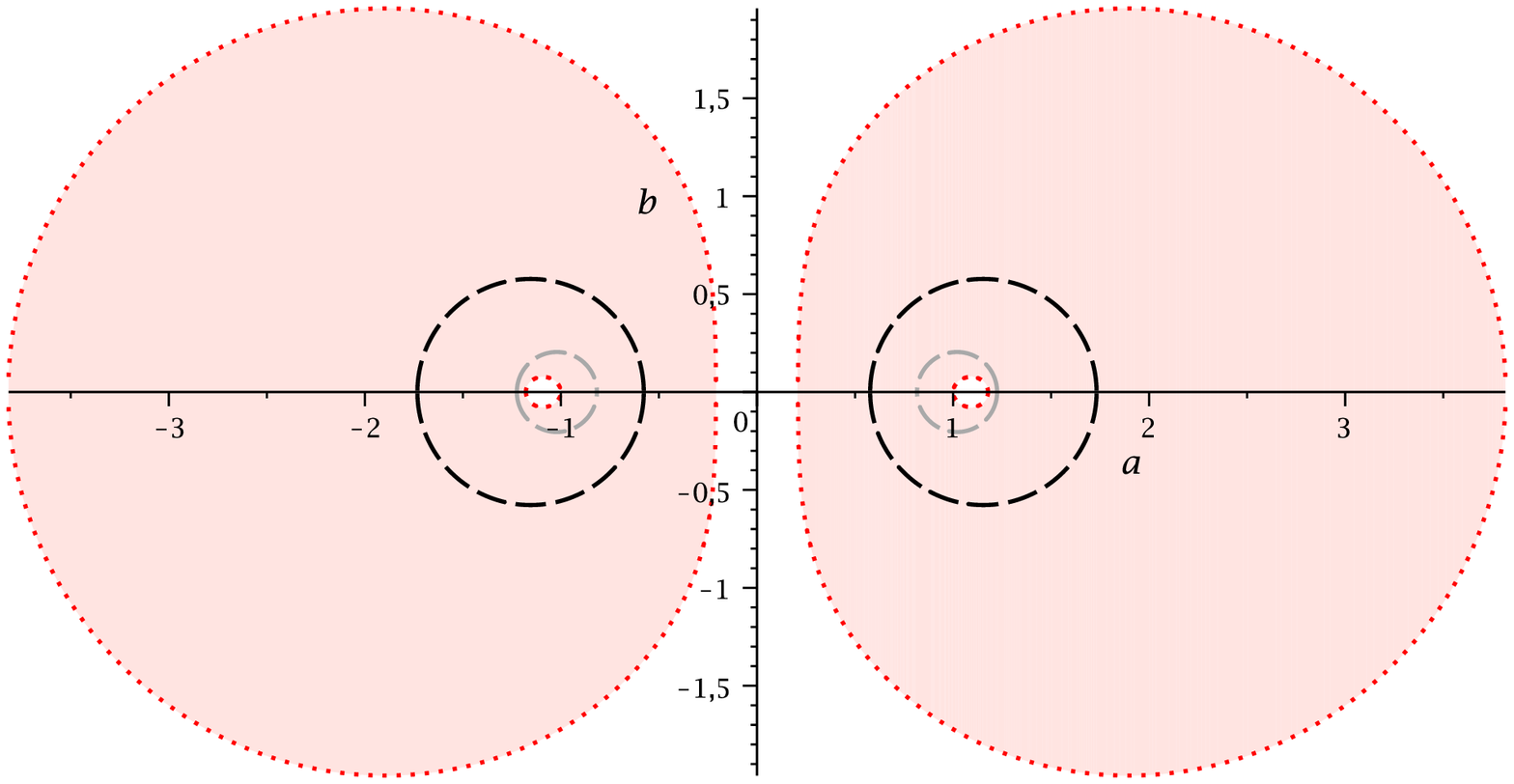}
   \label{pic:ergosphere-a}
 }
 \qquad\qquad
 \subfigure[$\nu=0.1$ and $\lambda=0.7$: ring-like ergosphere in $x_3$-$x_4$ coordinates]{
   \includegraphics[width=5cm]{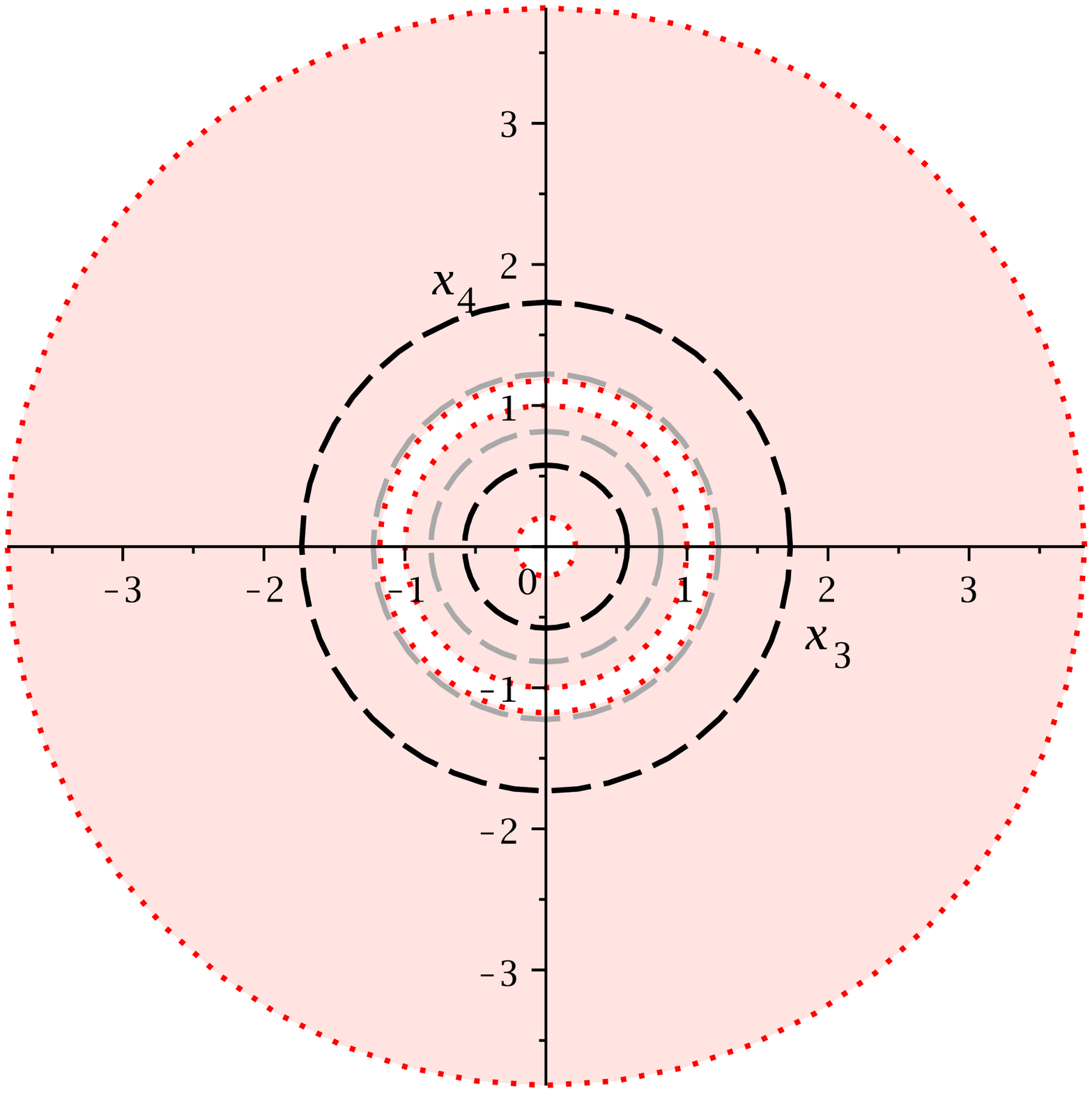}
   \label{pic:ergosphere-b}
 }
 \subfigure[$\nu=0.2$ and $\lambda=0.9$: spherical ergosphere in $a$-$b$ coordinates]{
   \includegraphics[width=7cm]{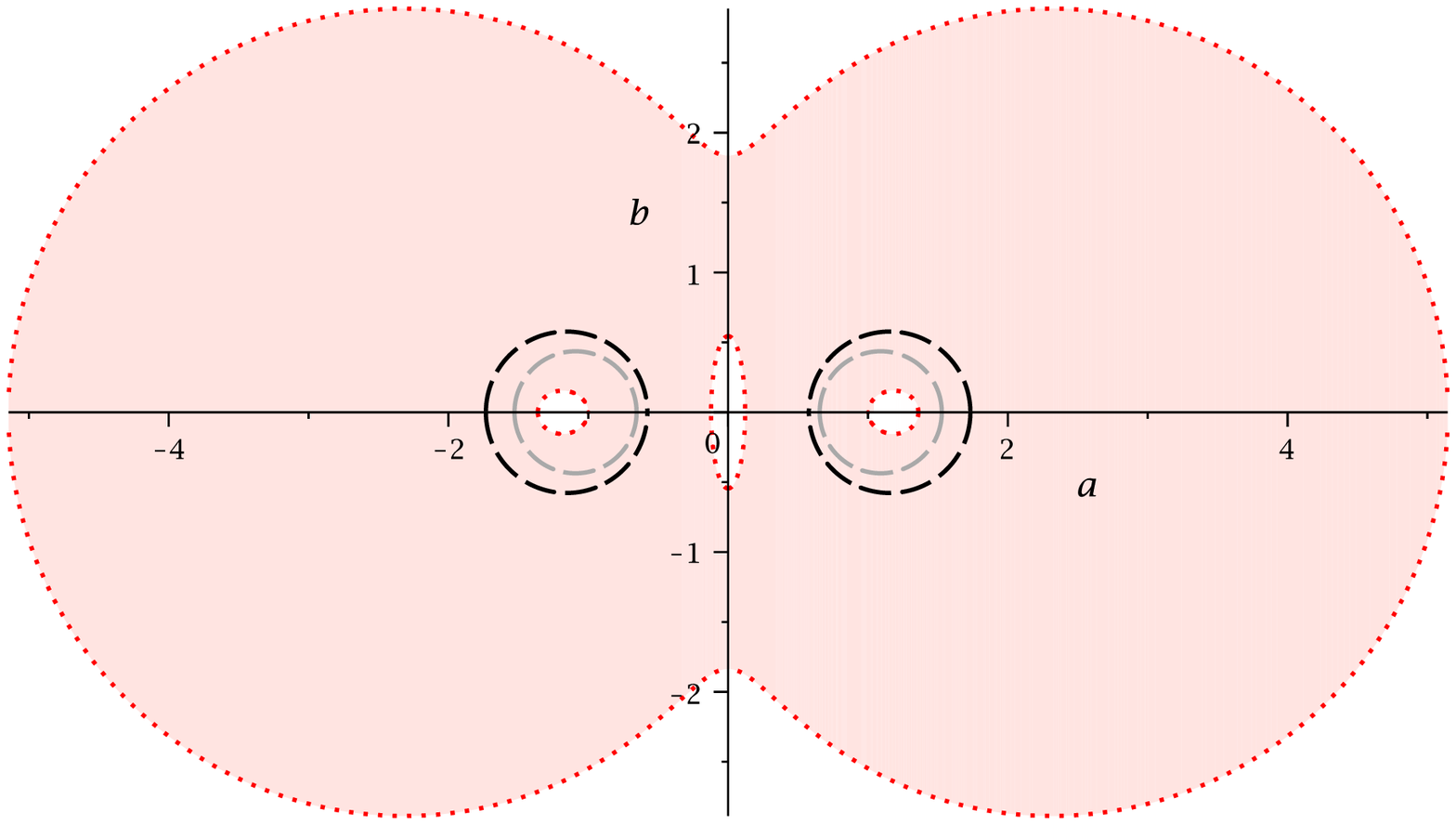}
    \label{pic:ergosphere-c}
 }
 \qquad\qquad
 \subfigure[$\nu=0.2$ and $\lambda=0.9$: spherical ergosphere in $x_3$-$x_4$ coordinates]{
   \includegraphics[width=5cm]{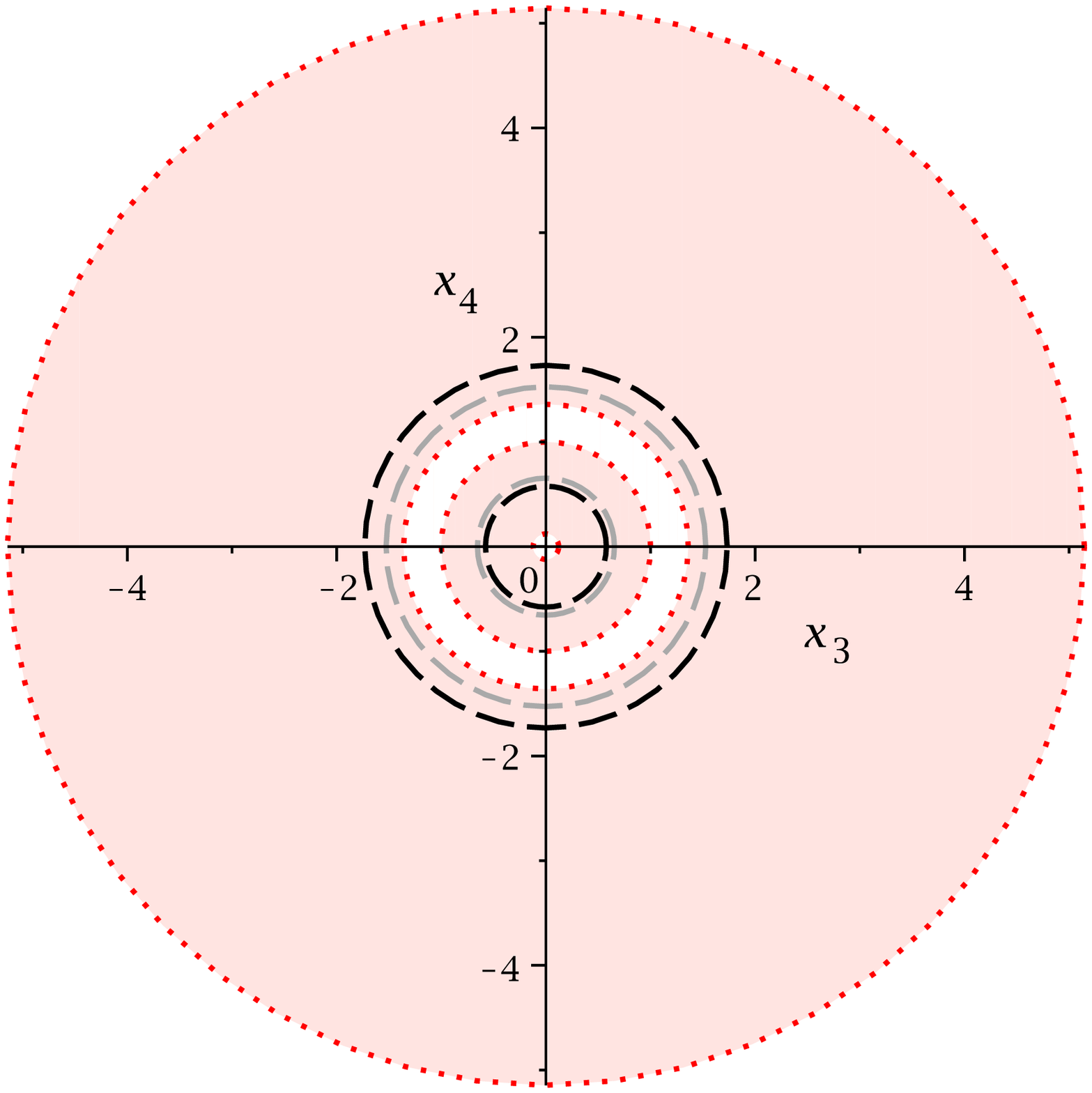}
   \label{pic:ergosphere-d}
 }
\caption{Different shapes of the ergosphere plotted in $a$-$b$ coordinates and $x_3$-$x_4$ coordinates. The $a$-$b$ coordinates describe a plane of constant angles $\phi$ and $\psi$. The $x_3$-$x_4$ coordinates describe the equatorial plane of the ring where $x=\pm1$. A more detailed explanation of the different coordinates and planes will be given in section \ref{sec:orbits}. The ergosphere is depicted as a light red area with a red dotted border. There is no ergosphere in the white region. The black and grey dashed circles are the inner and outer horizons of the black ring.}
 \label{pic:ergosphere}
\end{figure}

The ergosphere is determined by $H(y,x)=0$. In the singly spinning case $\nu=0$ the ergosphere has $S^1\times S^2$ topology like the horizon, but in the doubly spinning case $\nu \neq 0$ the shape of the ergosphere becomes more complicated. Depending on the parameters $\nu$ and $\lambda$ the shape of the ergosphere varies from ring-like $S^1\times S^2$ topology to spherical $S^3\cup S^3$ topology (see \cite{Durkee:2008an} for a detailed description). There also exists an ergosphere-free region behind the inner horizon ($y<y_{h-}$), which as not mentioned in \cite{Durkee:2008an}.
An example for a ring-like ergosphere is shown in figure \ref{pic:ergosphere-a} and \ref{pic:ergosphere-b} and an example for a spherical ergosphere is depicted in figure \ref{pic:ergosphere-c} and \ref{pic:ergosphere-d} .

The doubly spinning black ring has two angular momenta and thus the rotation is described by
\begin{equation}
 \Omega =\Omega _\psi \mathrm{d}\psi + \Omega _\phi \mathrm{d}\phi \, ,
\end{equation}
where
\begin{equation}
 \Omega _\psi = - \frac{R\lambda\sqrt{2((1+\nu)^2-\lambda^2)}}{H(y,x)}\frac{1+y}{1-\lambda +\nu} (1+\lambda -\nu + x^2y\nu (1-\lambda-\nu) + 2\nu x(1-y))
\end{equation}
and
\begin{equation}
 \Omega _\phi = - \frac{R\lambda\sqrt{2((1+\nu)^2-\lambda^2)}}{H(y,x)}(1-x^2)y\sqrt{\nu} \, .
\end{equation}

One can split the polynomials $A(x,y)$ and $L(x,y)$ into $x$- and $y$-parts which will be useful later:
\begin{eqnarray}
 A(x,y) &=& G(x)\alpha (y) + G(y)\beta (x) \nonumber \\
 L(x,y) &=& G(x)\delta(y) - G(y)\delta (x) \, ,
 \label{eqn:A-L-kurz}
\end{eqnarray}
where
\begin{eqnarray}
 \alpha (\xi) &=& \nu (1-\xi^2)[-(1+\lambda^2)-\nu (1-\nu)+\lambda\xi (2-3\nu) - (1-\lambda^2)\xi^2] \nonumber \\
 \beta (\xi) &=& (1+\lambda^2) + \lambda\xi (1+(1-\nu)^2) - \nu\xi^2(2\lambda^2+\nu (1-\nu)) - \lambda\nu^2\xi^3(3-2\nu) \nonumber \\
             && -\nu^2\xi^4(1-\lambda^2+\nu(1-\nu)) \nonumber \\
 \delta (\xi) &=& \lambda\sqrt{\nu}(1-\xi^2)(\lambda-(1-\nu^2)\xi -\lambda\nu\xi^2) \, .
\end{eqnarray}

In \cite{Gal'tsov:2009da} a three-charged version of the doubly spinning black ring was constructed. Here we use the balanced version where two of the charges are set to zero to remove the Dirac-Misner string. Compared to \cite{Gal'tsov:2009da} the metric of the charged doubly spinning black ring is written in a slightly different way so that the metric functions $G(x)$, $H(x,y)$, $A(x,y)$, $L(x,y)$ stay the same as in the uncharged case from \cite{Durkee:2008an}.

The metric is
\begin{eqnarray}
 \mathrm{d} s^2 &=& - D(x,y)^{-2/3}\frac{H(y,x)}{H(x,y)} (\mathrm{d}t + c\Omega)^2 + D(x,y)^{1/3}\frac{R^2H(x,y)}{(x-y)^2(1-\nu)^2} \left[ \frac{\mathrm{d}x^2}{G(x)} - \frac{\mathrm{d}y^2}{G(y)} \right. \nonumber \\ 
 && \left. + \frac{A(y,x)\mathrm{d}\phi^2 - 2L(x,y)\mathrm{d}\phi\mathrm{d}\psi - A(x,y)\mathrm{d}\psi^2}{H(x,y)H(y,x)} \right] \, .
\end{eqnarray} 
The function $D(x,y)$ is defined as
\begin{equation}
 D(x,y)=c^2-s^2\frac{H(y,x)}{H(x,y)} = 1+s^2\frac{2\lambda(1-\nu)(x-y)(1-\nu xy)}{H(x,y)} \, .
\end{equation}
The charge is represented by the parameters
\begin{equation}
 c=\cosh (\alpha) \quad \text{and} \quad s=\sinh (\alpha) \quad \text{with} \quad \alpha \in \mathbb{R} \, .
\end{equation}
The horizons have the same coordinates as in the uncharged case (see equation \eqref{eqn:horizonte}) and the ergosphere is still determined by $H(y,x)=0$. If $c=1$ and $s=0$ then $D(x,y)=1$ and one obtains the metric of the uncharged black ring.\\

The non-vanishing components of the inverse metric of the charged doubly spinnig black ring are
\begin{equation}
 \begin{split}
   g^{tt} &= -D(x,y)^{2/3}\frac{H(x,y)}{H(y,x)}+c^2D(x,y)^{-1/3}\frac{(x-y)^2}{R^2H(x,y)} \left[ \frac{\Omega_\phi^2A(y,x) - 2\Omega_\phi\Omega_\psi L(x,y) - \Omega_\psi^2 A(x,y)}{G(x)G(y)} \right]  \\
   g^{t\phi} &= cD(x,y)^{-1/3}\frac{(x-y)^2}{R^2H(x,y)} \frac{\Omega_\psi L(x,y) - \Omega_\psi A(x,y)}{G(x)G(y)} \\
   g^{t\psi} &= cD(x,y)^{-1/3}\frac{(x-y)^2}{R^2H(x,y)} \frac{\Omega_\phi L(x,y) + \Omega_\psi A(y,x)}{G(x)G(y)} \\
   g^{\phi\phi} &= D(x,y)^{-1/3}\frac{(x-y)^2}{R^2H(x,y)} \frac{A(x,y)}{G(x)G(y)}\\
   g^{\psi\psi} &= -D(x,y)^{-1/3}\frac{(x-y)^2}{R^2H(x,y)} \frac{A(y,x)}{G(x)G(y)}\\
   g^{\phi\psi} &= -D(x,y)^{-1/3}\frac{(x-y)^2}{R^2H(x,y)} \frac{L(x,y)}{G(x)G(y)}\\
   g^{xx} &= D(x,y)^{-1/3}\frac{(x-y)^2 (1-\nu)^2}{R^2H(x,y)} G(x) \\
   g^{yy} &= -D(x,y)^{-1/3}\frac{(x-y)^2 (1-\nu)^2}{R^2H(x,y)} G(y) \, .
 \end{split}
\end{equation}

The metric of the charged doubly spinning black ring and its Hamiltonian $\mathscr{H} = \frac{1}{2}g^{ab}p_ap_b$ do not depend on the coordinates $t$, $\phi$ and $\psi$, so we have three conserved momenta $p_a=g_{ab}\dot{x}^b$ with the associated killing vector fields $\partial / \partial t$, $\partial / \partial \phi$ and $\partial / \partial \psi$. A dot denotes the derivative with respect to an affine parameter $\tau$.
\begin{eqnarray}
 p_t &=& -D(x,y)^{-2/3}\frac{H(y,x)}{H(x,y)} (\dot{t}+c\Omega _\phi \dot{\phi} +c\Omega _\psi \dot{\psi}) \equiv -E \label{eqn:charge-t-impuls}\\
 p_\phi &=& -c\Omega _\phi E -D(x,y)^{1/3}\frac{R^2}{H(y,x)(x-y)^2(1-\nu)^2}(-A(y,x)\dot{\phi}+L(x,y)\dot{\psi}) \equiv \Phi \label{eqn:charge-phi-impuls}\\
 p_\psi &=& -c\Omega _\psi E -D(x,y)^{1/3}\frac{R^2}{H(y,x)(x-y)^2(1-\nu)^2}(A(x,y)\dot{\psi}+L(x,y)\dot{\phi}) \equiv \Psi \label{eqn:charge-psi-impuls}
\end{eqnarray}
$E$ is the energy, $\Phi$ and $\Psi$ are the angular momenta in $\phi$- an $\psi$-direction. The conjugate momenta in $x$- and $y$-direction are:
\begin{eqnarray}
 p_x &=& D^{1/3}\frac{R^2H(x,y)\dot{x}}{(x-y)^2(1-\nu)^2G(x)} \label{eqn:charge-x-impuls}\\
 p_y &=& -D^{1/3}\frac{R^2H(x,y)\dot{y}}{(x-y)^2(1-\nu)^2G(y)} \label{eqn:charge-y-impuls}
\end{eqnarray}

To obtain the equations of motion for an uncharged particle in the charged doubly spinning black ring spacetime we need the Hamilton-Jacobi equation:
\begin{equation}
 \frac{\partial S}{\partial \tau} + \mathscr{H} \left( x^a, \frac{\partial S}{\partial x^b}\right) = 0 \, .
 \label{eqn:hamilton-jacobi}
\end{equation}
We already have three constants of motion ($E$, $\Phi$ and $\Psi$) and the mass shell condition $g^{ab}p_a p_b=-m^2$ gives us a fourth , so we can make the ansatz
\begin{equation}
 S(\tau, t, x, y, \phi, \psi) = \frac{1}{2}m^2\tau -Et +\Phi\phi +\Psi\psi + S_x(x)+S_y(y) .
 \label{eqn:s-ansatz}
\end{equation}
Inserting this ansatz into (\ref{eqn:hamilton-jacobi}) gives
\begin{equation}
 \begin{split}
 0 &= m^2 -D(x,y)^{2/3}\frac{H(x,y)}{H(y,x)} E^2 + D(x,y)^{-1/3}\frac{(x-y)^2(1-\nu)^2}{R^2H(x,y)} \left[ G(x) \left( \frac{\partial S}{\partial x} \right)^2 
   - G(y) \left( \frac{\partial S }{\partial y} \right)^2  \right. \\
   & \left. + \frac{ A(x,y) \left(\Phi +c\Omega _\phi E\right)^2 -2L(x,y) \left(\Phi +c\Omega _\phi E\right) \left(\Psi +c\Omega _\psi E\right) 
   -A(y,x) \left(\Psi +c\Omega _\psi E\right)^2 }{(1-\nu)^2G(x)G(y)} \right] \, .
 \label{eqn:hj-c-d-ring}
 \end{split}
\end{equation}
The Hamilton-Jacobi equation does not seem to be separable in general. However, it is possible to separate the equation in the special case $E=m=0$. These zero energy null geodesics are only realisable in the ergoregion. 

We can also obtain equations of motion for geodesics on the $\phi$- and $\psi$-axis by setting $y=-1$ ($\psi$-axis) or $x=\pm 1$ ($\phi$-axis). The plane $y=-1$ which is called the $\psi$-axis corresponds to one of the two rotational axes of the doubly spinning black ring. The plane  $x=\pm 1$ which is called the $\phi$-axis, is the equatorial plane of the black ring and also the second rotational axis. \\

In the next chapters we will study these special cases and solve the corresponding equations of motion analytically.

\section{Null geodesics in the Ergosphere}

For $E=m=0$ it is possible to separate the Hamilton-Jacobi equation:
\begin{equation}
 G(x) \left( \frac{\partial S}{\partial x} \right)^2  -\frac{-\beta (x)\Phi ^2 -2\delta (x)\Phi\Psi + \alpha (x)\Psi ^2}
 {(1-\nu)^2G(x)}= 
 G(y) \left( \frac{\partial S }{\partial y} \right)^2 -\frac{\alpha (y)\Phi ^2 -2\delta (y)\Phi\Psi -\beta (y)\Psi ^2}
 {(1-\nu)^2G(y)}
\label{eqn:ham-jac-cdoub}
\end{equation}
With a separation constant $k$, the equation (\ref{eqn:ham-jac-cdoub}) splits into two:
\begin{eqnarray}
 \left( \frac{\partial S}{\partial x} \right)^2 &=& - \frac{U(x)}{(1-\nu)^2G(x)^2} \\
 \left( \frac{\partial S}{\partial y} \right)^2 &=& - \frac{V(y)}{(1-\nu)^2G(y)^2} \, ,
\end{eqnarray}
where
\begin{eqnarray}
 U(x) &=& -[-\beta (x)\Phi ^2 -2\delta (x)\Phi\Psi + \alpha (x)\Psi ^2 + kG(x)] \\
 V(y) &=& -[\alpha (y)\Phi ^2 -2\delta (y)\Phi\Psi -\beta (y)\Psi ^2 + kG(y)] .
\end{eqnarray}
Using $p_a = \frac{\partial S}{\partial x^a}$ and (\ref{eqn:charge-t-impuls})-(\ref{eqn:charge-y-impuls}) the separated Hamilton-Jacobi equation gives the equations of motion:
\begin{eqnarray}
 \frac{\text{d}x}{\text{d}\gamma} &=& \sqrt{X(x)} 	\label{eqn:charge-x-Gleichung}\\
 \frac{\text{d}y}{\text{d}\gamma} &=& - \sqrt{Y(y)} 	\label{eqn:charge-y-Gleichung} \\ 
 \frac{\text{d}\phi}{\text{d}\gamma} &=& \frac{[A(x,y)\Phi - L(x,y)\Psi]}{G(x)G(y)} 	\label{eqn:charge-phi-Gleichung}\\
 \frac{\text{d}\psi}{\text{d}\gamma} &=&\frac{[-L(x,y)\Phi - A(y,x)\Psi]}{G(x)G(y)} 	\label{eqn:charge-psi-Gleichung}\\
 \frac{\text{d}t}{\text{d}\gamma} &=& -c\cdot\Omega_\phi(x,y)\frac{\text{d}\phi}{\text{d}\gamma}-c\cdot\Omega_\psi(x,y)\frac{\text{d}\psi}{\text{d}\gamma} \, , \label{eqn:charge-t-Gleichung}
\end{eqnarray}
where 
\begin{eqnarray}
 X(x) &=& -(1-\nu)^2 U(x) \\
 Y(y) &=& -(1-\nu)^2 V(y)
\end{eqnarray}
are polynomials of fourth order. We also introduced the Mino-time \cite{Mino:2003yg} $d\gamma = D(x,y)^{-1/3}\frac{(x-y)^2}{R^2 H(x,y)} d\tau$.\\

From (\ref{eqn:charge-x-Gleichung})-(\ref{eqn:charge-t-Gleichung}) we see that the $x$-, $y$-, $\phi$- and $\psi$-equation of motion do not depend on the charge. Therefore the charge of the black ring has no effect on the possible orbits for null geodesics in the ergosphere.

In the sections below we will solve the equations (\ref{eqn:charge-x-Gleichung})-(\ref{eqn:charge-y-Gleichung}) analytically. The $t$-equation (\ref{eqn:charge-t-Gleichung}) cannot be solved in general because it cannot be separated into $x$- and $y$-parts. Only in the special case $\nu=0$ which corresponds to the singly spinning black ring, it is possible to solve the $t$-equation (see \cite{Grunau:2012ai})

\subsection{Classification of geodesics}

Equation (\ref{eqn:charge-x-Gleichung}) and (\ref{eqn:charge-y-Gleichung}) can be written as
\begin{eqnarray}
 \frac{1}{(1-\nu)^2} \left( \frac{\text{d}x}{\text{d}\gamma} \right) ^2 + U(x) &=& 0 \\
 \frac{1}{(1-\nu)^2} \left( \frac{\text{d}y}{\text{d}\gamma} \right) ^2 + V(y) &=& 0 \, .
\end{eqnarray}
$U(x)$ and $V(y)$ can be regarded as effective potentials (see \cite{Durkee:2008an}). To get real solutions for the $x$- and $y$-equation the effective potentials have to be negative. The zeros of the effective potentials (because $E=0$) and hence $X$ and $Y$ mark the turning points of the motion of light or a test particle (in this case we only have light since $m=0$). Thus the number of zeros determines the type of the orbit.

Since $X$ and $Y$ are polynomials of fourth order it is difficult to calculate the zeros analytically. A good way to determine the number of zeros are parametric diagrams.

For the $x$-motion we use the equations $U(x)=0$ and $\frac{\mathrm{d}U(x)}{\mathrm{d}x}=0$ to construct a parametric $\Phi$-$\Psi$-diagram. The diagram for the $y$-motion can be found analogously. If both $\Phi$-$\Psi$-diagrams are shown in the same plot they form regions for the different numbers of zeros. Here one has to keep in mind that $-1 \leq x \leq 1$ and $-\infty < y \leq -1$ to find the allowed regions.

\begin{figure}
 \centering
 \subfigure[$k=1$, $\nu =0.1$ and $\lambda =0.7$ \newline For small $\nu$ and $\lambda$ there are regions (black) where $V(y)$ has only one zero.]{
   \label{pic:nu0.1-lambda0.7-c1}
   \begin{overpic}[width=7cm]{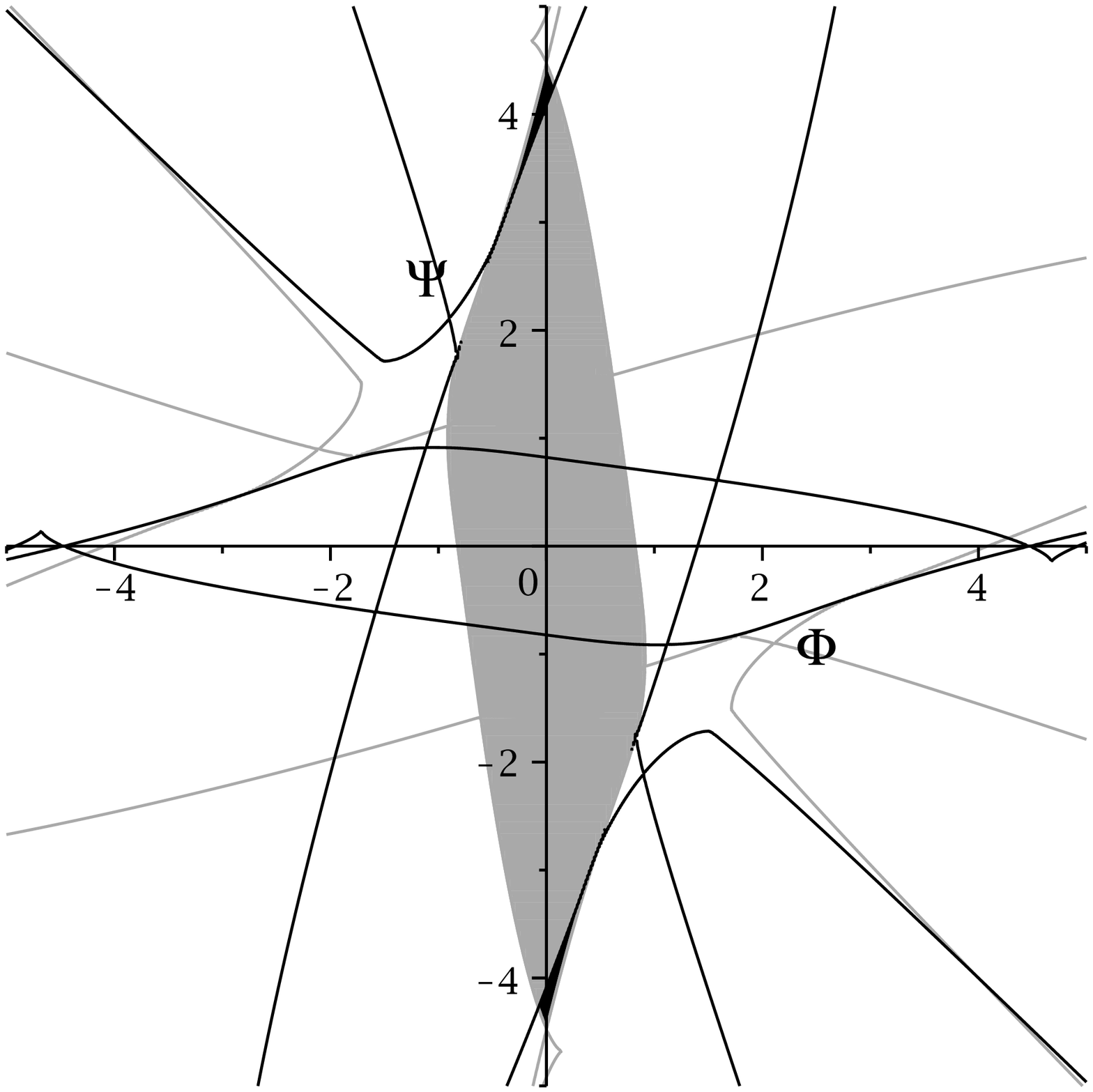}
     \put(135,135){\includegraphics[width=2.5cm]{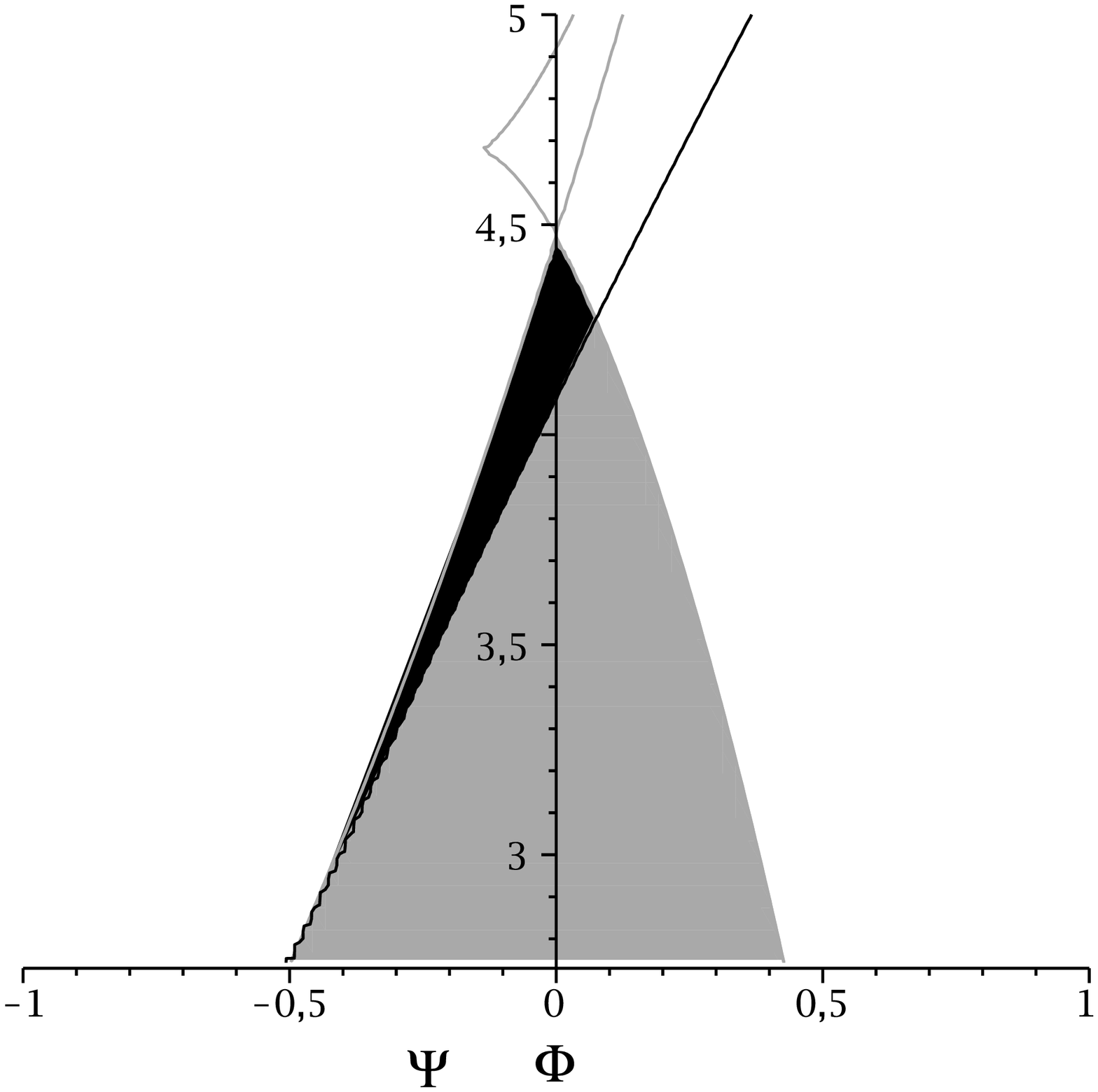}}
     \put(0,0){\includegraphics[width=2.5cm]{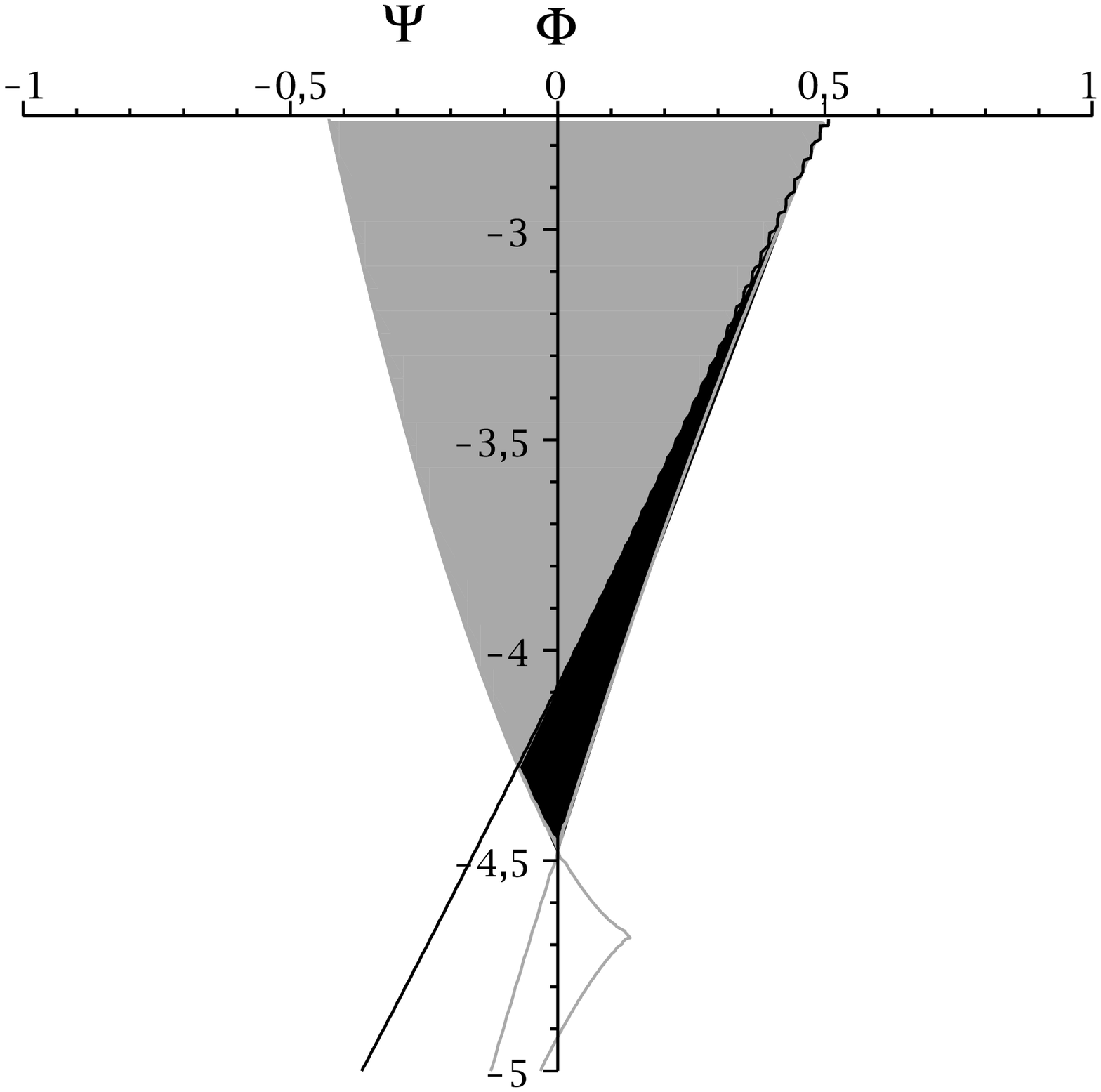}}
     \thicklines
     \put(100,175){\circle{23}}
     \put(112,175){\vector(1,0){40}}
     \put(100,20){\circle{23}}
     \put(88,20){\vector(-1,0){40}}
   \end{overpic}
 } \subfigure[$k=1$, $\nu =0.1$ and $\lambda =0.9$ \newline For larger $\lambda$, $V(y)$ has always two zeros.]{
   \label{pic:nu0.1-lambda0.9-c1}
   \includegraphics[width=7cm]{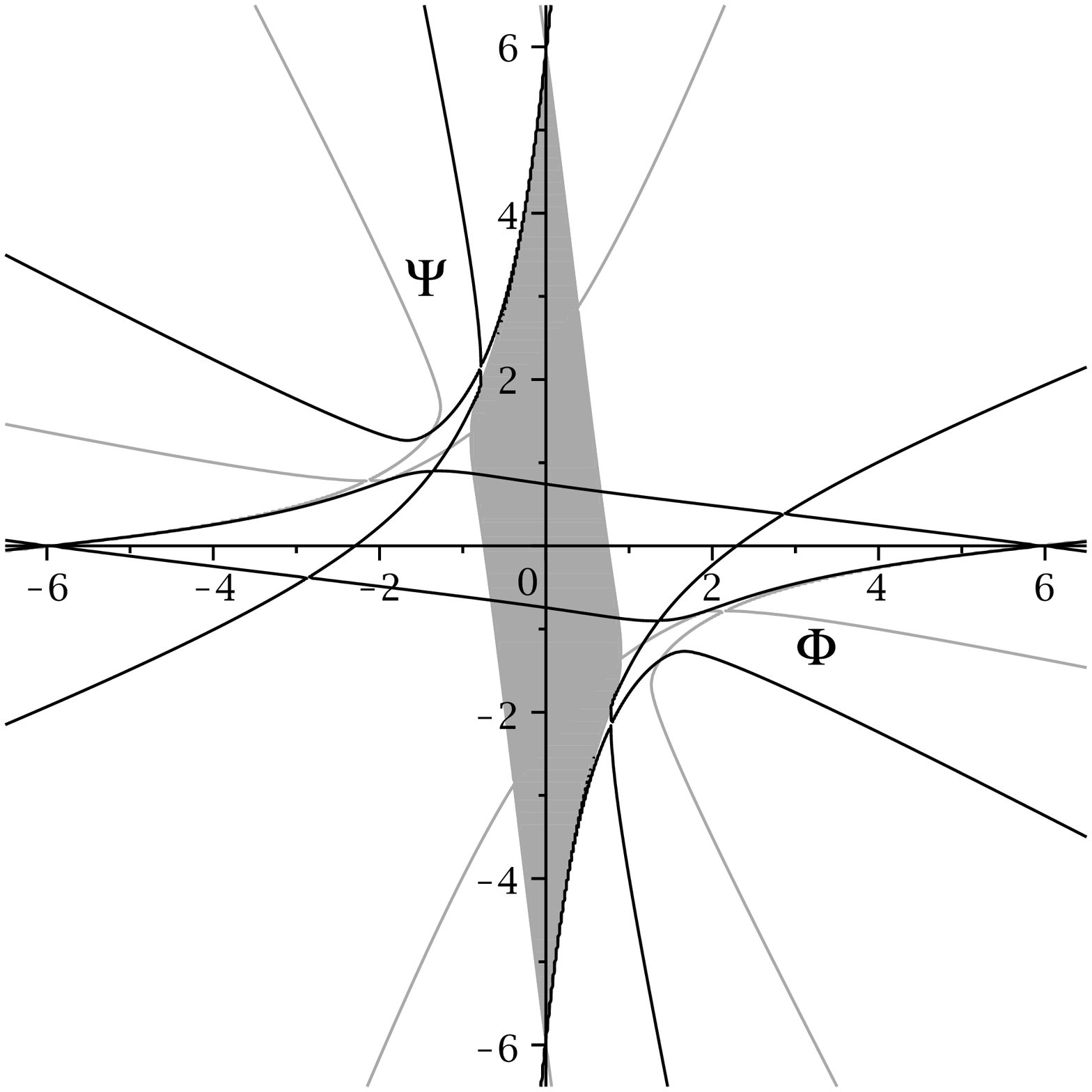}
 }
 \subfigure[$k=0.1$, $\nu =0.1$ and $\lambda =0.9$ \newline If the separation constant $k$ becomes smaller, the allowed region also becomes smaller.]{
   \label{pic:nu0.1-lambda0.9-c0.1}
   \includegraphics[width=7cm]{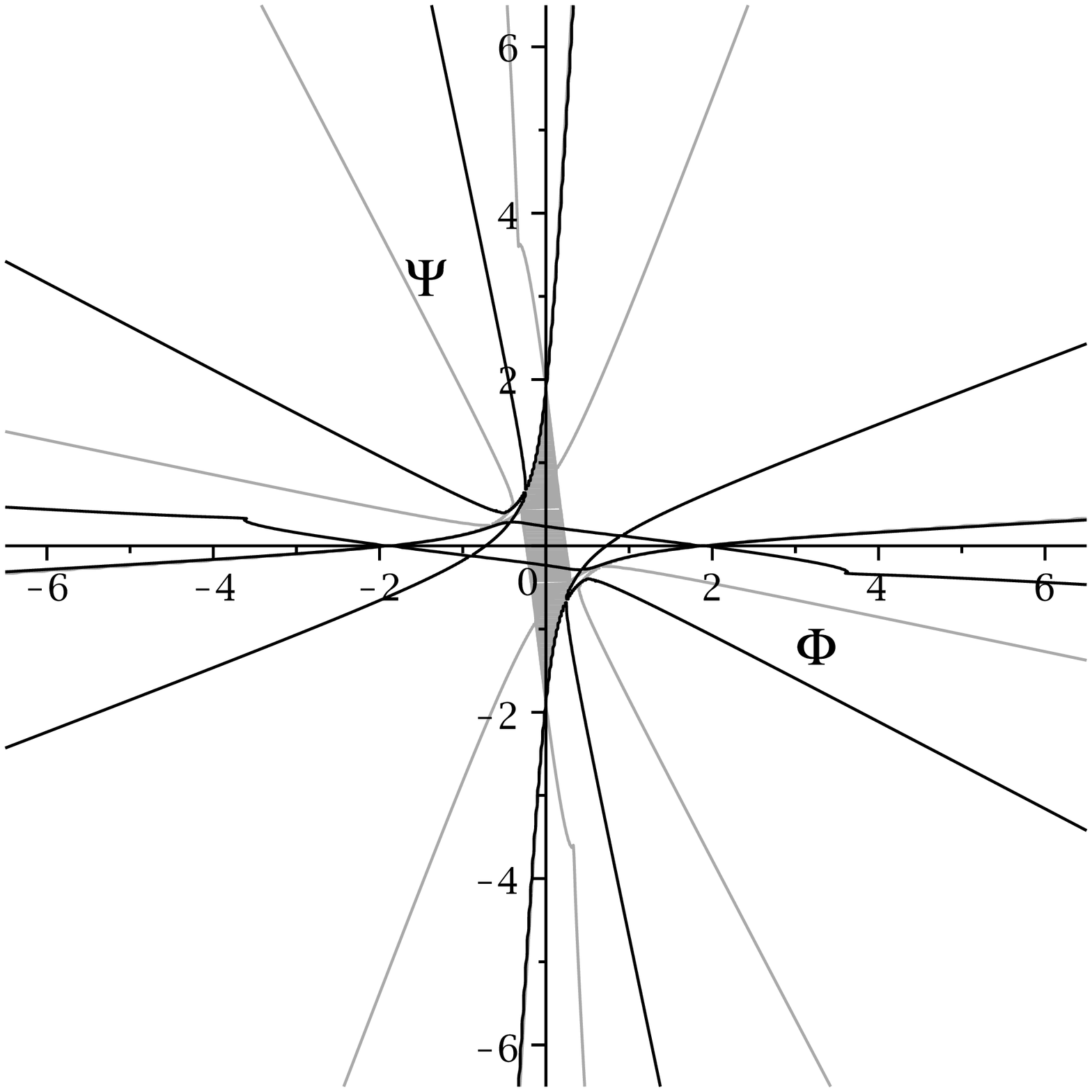}
 }
 \subfigure[$k=0$, $\nu =0.1$ and $\lambda =0.9$ \newline If $k=0$ the allowed region vanishes.]{
   \label{pic:nu0.1-lambda0.9-c0}
   \includegraphics[width=7cm]{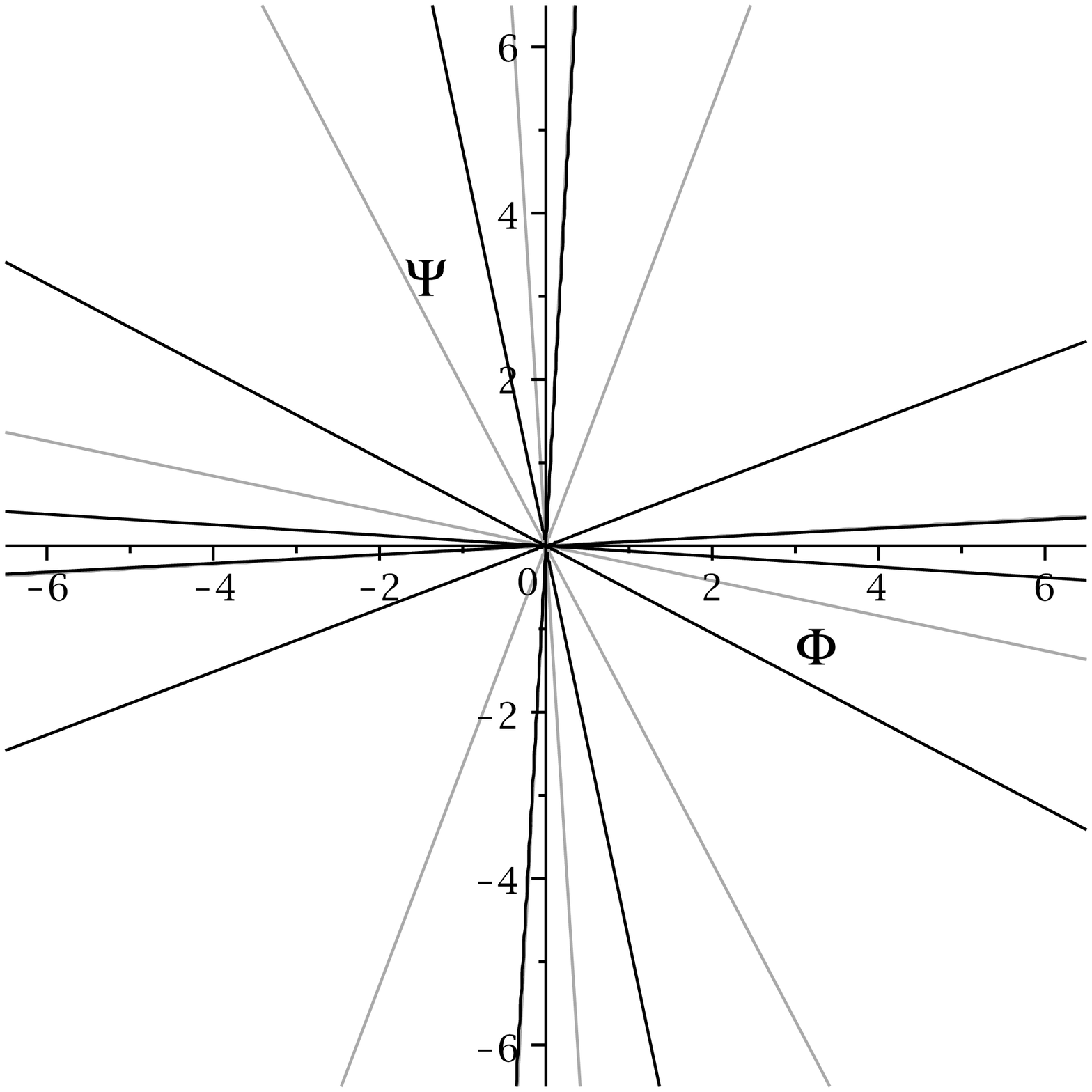}
 }
 \caption{Parametric $\Psi$-$\Phi$-diagrams of the charged doubly spinning black ring for $E=m=0$. In the allowed (coloured in grey and black) region $U(x)$ has two zeros. $V(y)$ has two zeros in the grey and one in the blue region. In the special cases $\Phi =0$ and/or $\Psi =0$, three zeros of $U(x)$ and $V(y)$ are possible (see text). }
 \label{pic:parameterplot1}
\end{figure}

Figure \ref{pic:parameterplot1} shows the parametric $\Phi$-$\Psi$-diagram for the $x$-motion (grey lines) and $y$-motion (black lines). Inside the grey and black coloured region different types of orbits are possible. Here $U(x)$ has zeros in the range $-1 \leq x \leq 1$ and $V(y)$ has zeros in the range $-\infty < y \leq -1$. Outside the coloured regions $U(x)$ has no zeros or is positive for $-1 \leq x \leq 1$ and therefore no motion possible.

If $|\Phi|>0$ and $\Psi$ lies in the coloured region the effective potential $U(x)$ has always two zeros between $-1$ and $1$. If $\Phi=0$ the effective potential has the zeros $-1$ and $1$ because
\begin{equation}
 U(\pm 1)=(1-\nu)^2 (1+\nu \pm \lambda)^2 \Phi^2 \, .
 \label{eqn:charge-Upm1}
\end{equation}
Also a third zero between $-1$ and $1$ is possible in the case $\Phi=0$.

From (\ref{eqn:charge-Upm1}) it is clear that $U(\pm 1)$ is always positive for $|\Phi| >0$, which means that there is a potential barrier preventing the photons from reaching the ``equatorial plane'' at $x=\pm 1$. One can also conclude that $U(x)$ is negative between its two zeros and the $x$-motion takes place between these two values.

Inside the grey region of the parametric diagram $V(y)$ has two zeros. For small $\nu$ and $\lambda$ a new region appears (the black region in figure \ref{pic:nu0.1-lambda0.7-c1}), here $V(y)$ has only one zero.

At $y=-1$ we have
\begin{equation}
  V(- 1)=(1-\nu)^2 (1+\nu - \lambda)^2 \Psi^2 \, .
\end{equation}
For $\Psi=0$ one of the zeros lies at $y=-1$, in this case $V(y)$ has up to three zeros. If $|\Psi| >0$ a potential barrier prevents photons from reaching $y=-1$ since $V(- 1)>0$ for $|\Psi| >0$.

The influence of the separation constant $k$ can be seen in figure \ref{pic:nu0.1-lambda0.9-c0.1} and \ref{pic:nu0.1-lambda0.9-c0}. If $k$ becomes smaller  the coloured region, where orbits are possible, of the $\Psi$-$\Phi$-diagram shrinks. For $k=0$ the coloured region vanishes, now only solutions with $\Phi=\Psi=0$ are possible, and for $k<0$ no orbits are possible.\\

Possible orbits for null geodesics in the ergosphere of a (charged) doubly spinning black ring are:
\begin{enumerate}
 \item \textit{Terminating Orbits} (TO):\\
  The photon approaches the black ring, crosses both horizons and falls into the singularity.
 \item \textit{Bound Orbits} (BO):\\
  The photon circles the black ring on a periodic bound orbit. In the ergosphere these orbits can only exist on the axis of $\psi$-direction ($y=-1$), moreover a special condition for the separation constant has to be fulfilled.
 \item \textit{Many-World Bound Orbits} (MBO):\\
  The photon circles the black ring on a periodic bound orbit, but crosses both horizons several times on its flight. Everytime both horizons are traversed twice, the photon emerges into another universe.
\end{enumerate}

\begin{figure}
 \centering
 \subfigure[Potential $U(x)$]{
   \includegraphics[width=6cm]{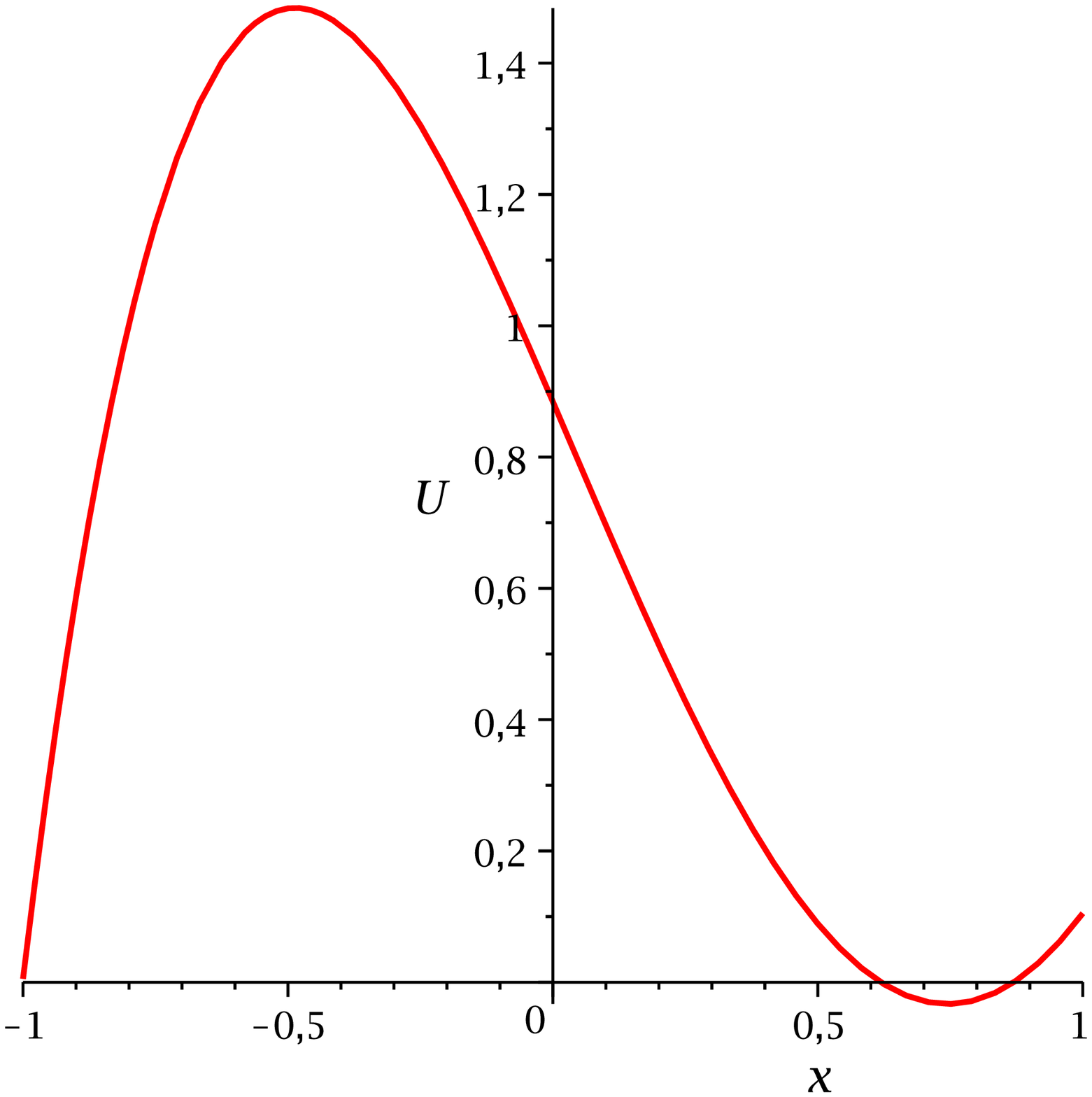}
 }
 \subfigure[Potential $V(y)$]{
   \includegraphics[width=6cm]{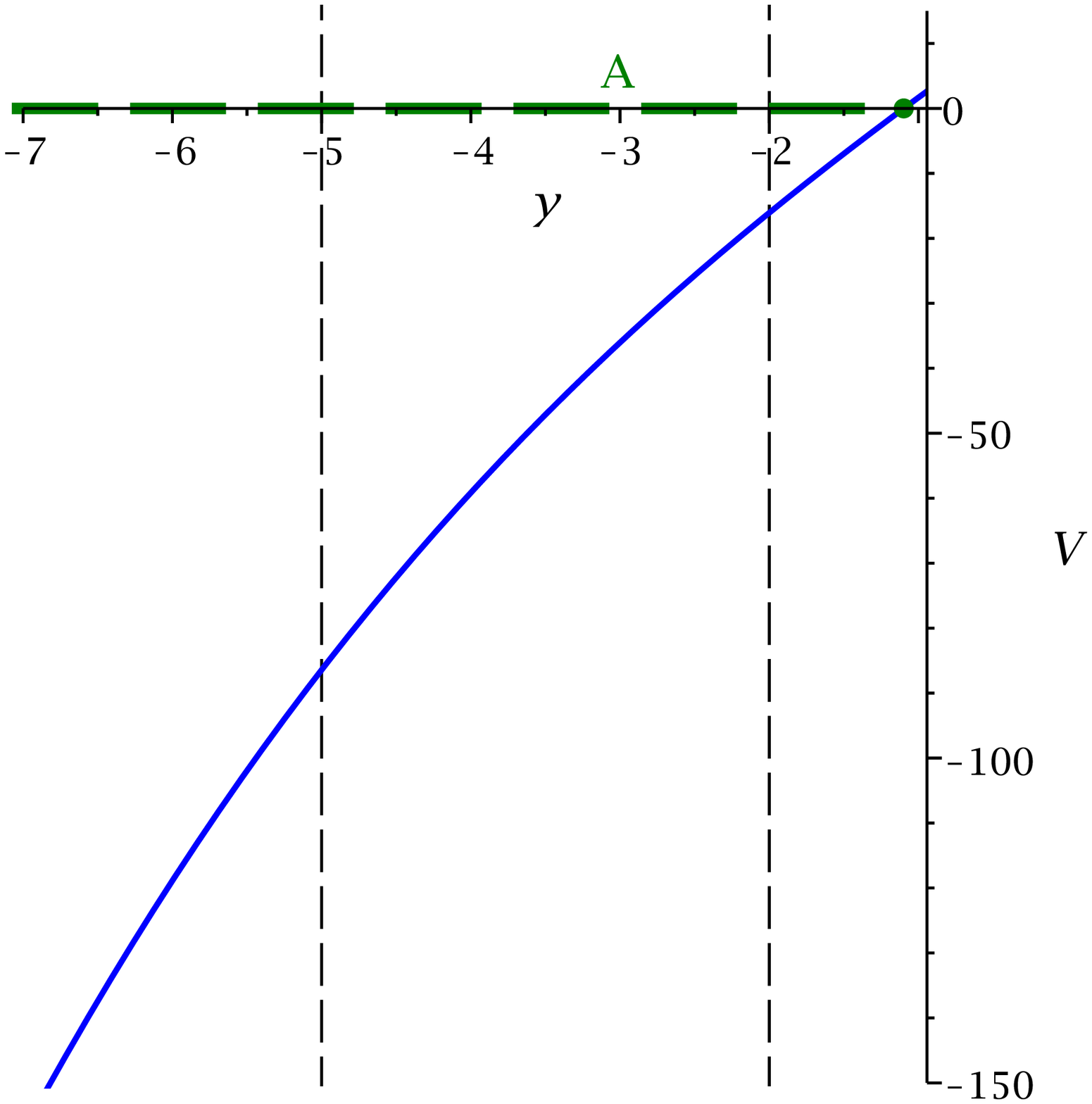}
 }
 \caption{$k=1$, $\nu =0.1$, $\lambda =0.7$, $\Phi =-0.2$ and $\Psi =3.6$ \newline
 Effective potentials $U(x)$ (red) and $V(y)$ (blue). The vertical black dashed lines indicate the position of the horizons. The horizontal green dashed line represents the energy and the green dot shows the position of the turning point. In the right picture an example of an orbit of type A can be seen.}
 \label{pic:pot_a}
\end{figure}

\begin{figure}
 \centering
 \subfigure[Potential $U(x)$]{
   \includegraphics[width=6cm]{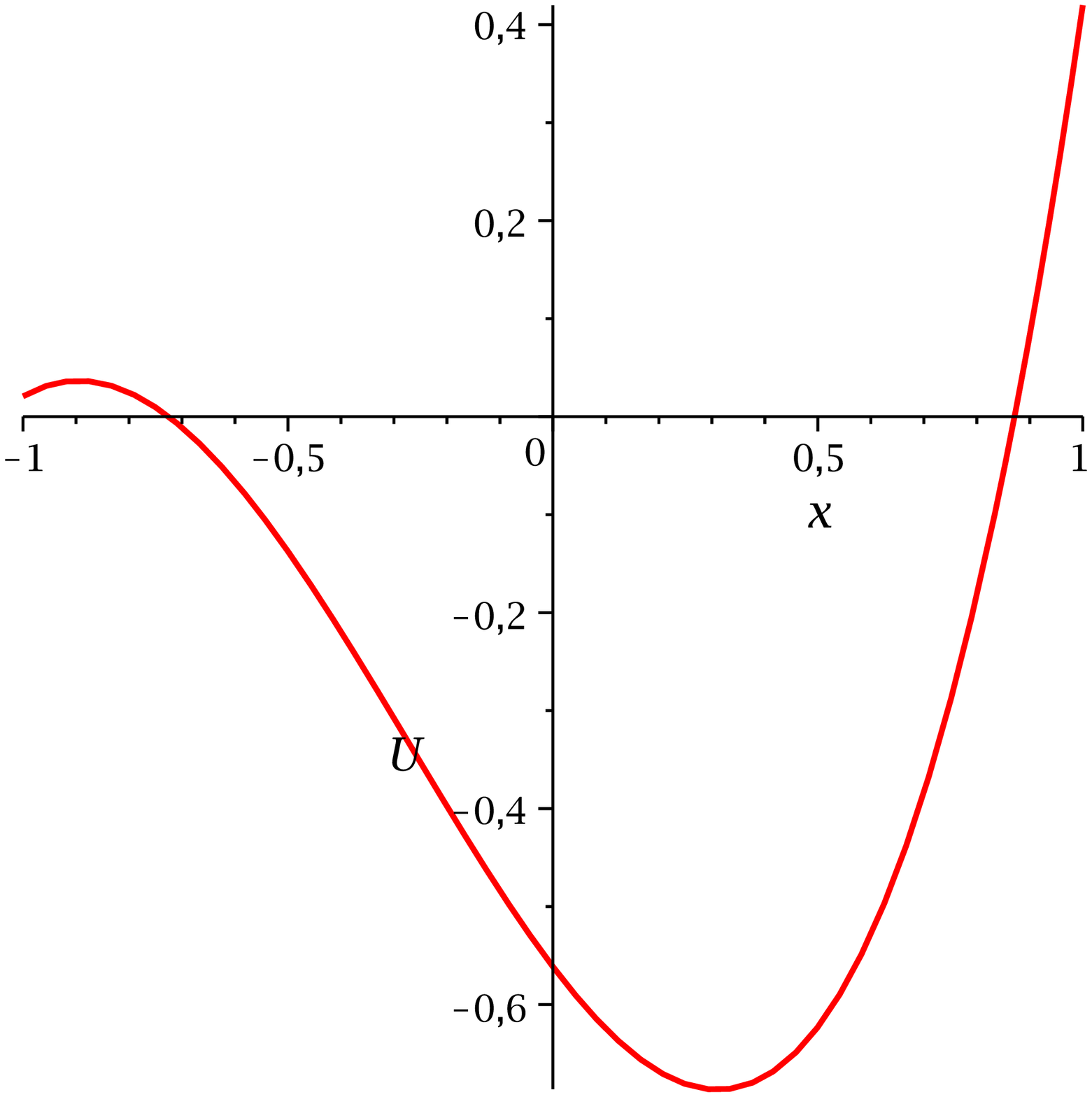}
 }
 \subfigure[Potential $V(y)$]{
   \includegraphics[width=6cm]{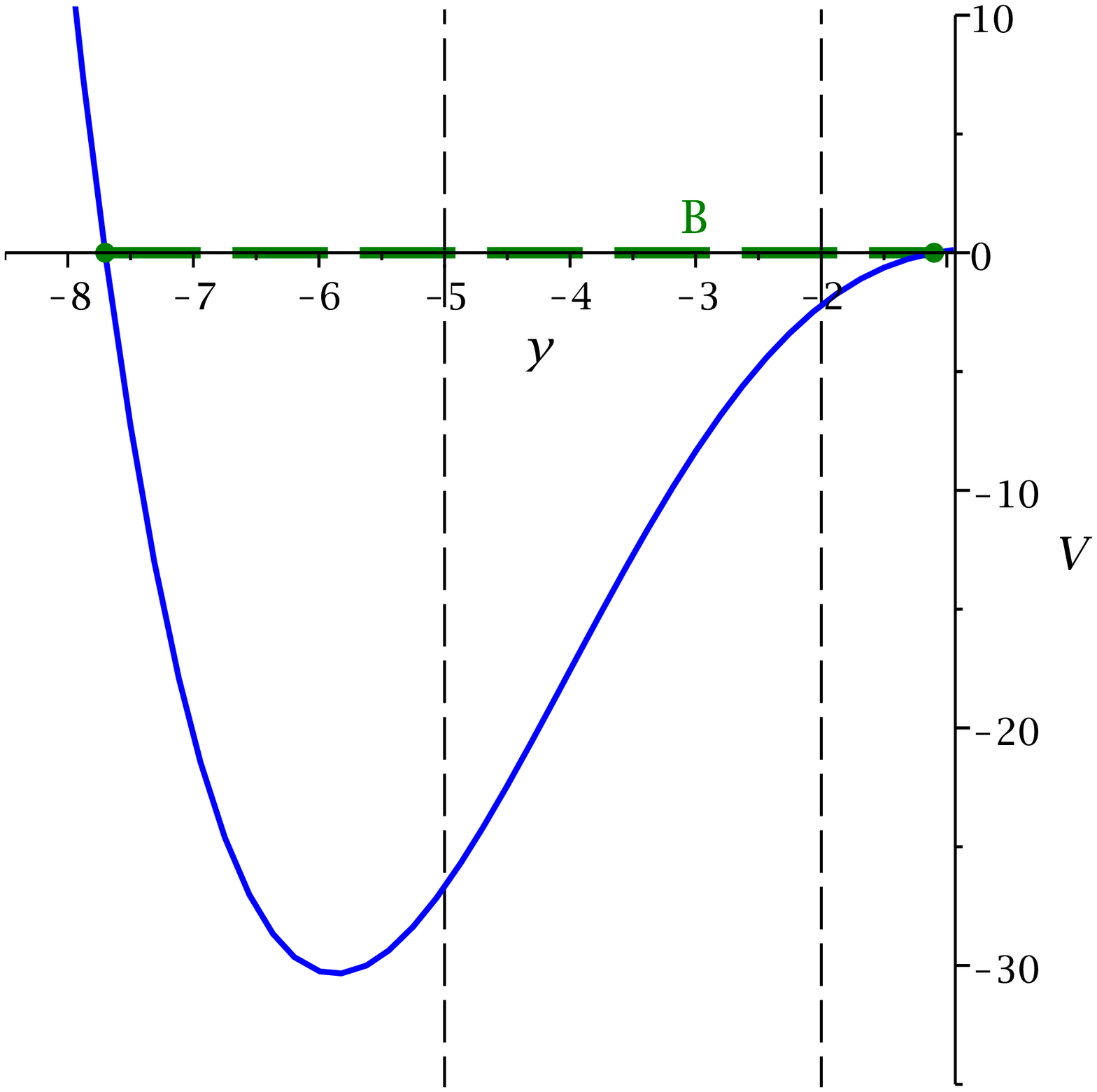}
 }
 \caption{$k=1$, $\nu =0.1$, $\lambda =0.7$, $\Phi =0.4$ and $\Psi =0.8$ \newline
 Effective potentials $U(x)$ (red) and $V(y)$ (blue). The vertical black dashed lines indicate the position of the horizons. The horizontal green dashed line represents the energy and the green dots show the position of the turning points. In the right picture an example of an orbit of type B can be seen.}
 \label{pic:pot_b}
\end{figure}

\begin{figure}
 \centering
 \subfigure[Potential $U(x)$]{
   \includegraphics[width=6cm]{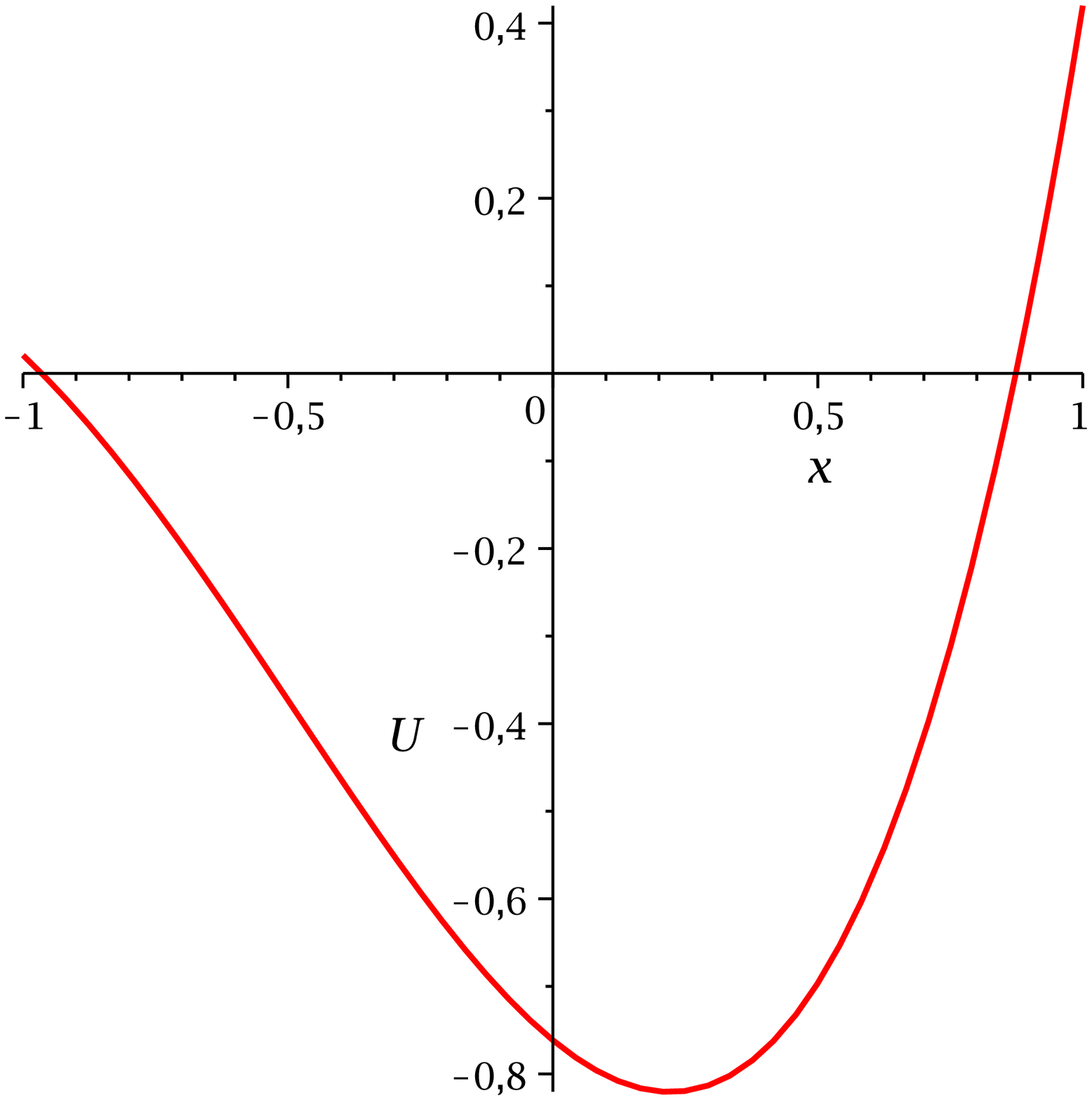}
 }
 \subfigure[Potential $V(y)$]{
   \includegraphics[width=6cm]{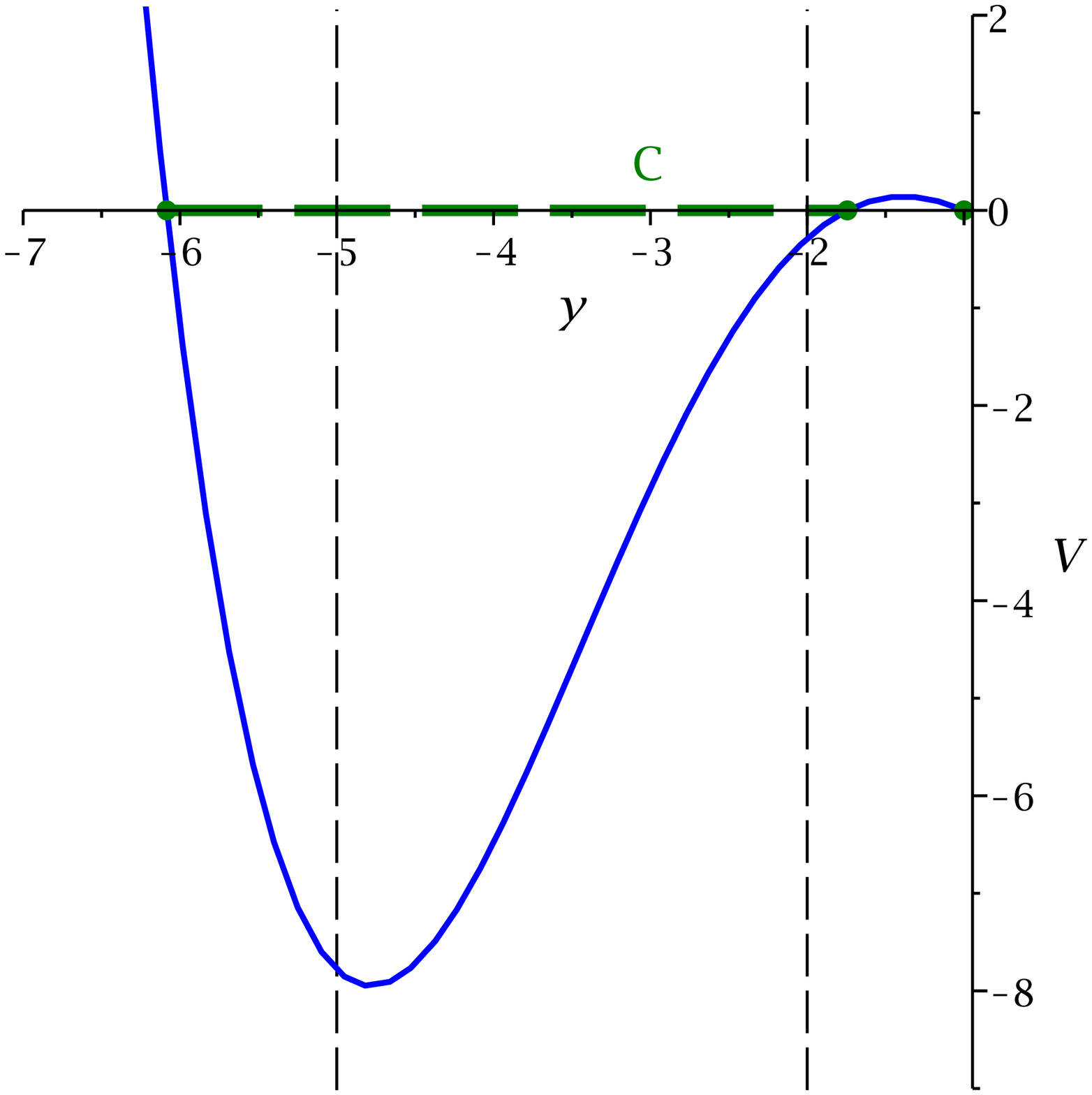}
 }
 \caption{$k=1$, $\nu =0.1$, $\lambda =0.7$, $\Phi =0.4$ and $\Psi =0$ \newline
 Effective potentials $U(x)$ (red) and $V(y)$ (blue). The vertical black dashed lines indicate the position of the horizons. The horizontal green dashed line represents the energy and the green dots show the position of the turning points. In the right picture an example of an orbit of type C can be seen. The bound orbit at $y=-1$ is not allowed for the chosen parameters $\nu$ and $\lambda$, since $U(x)$ has no zeros if one chooses the right value for the separation constant (equation (\ref{eqn:carter})).}
 \label{pic:pot_c}
\end{figure}

From the effective potential $U(x)$ we see that the $x$-motion is always bound. Either the motion takes place between two zeros or $x$ is constant at $-1$ (this is possible if $U(x)$ has three zeros in the special case $\Phi=0$). To determine the type of the orbit we have to consider $V(y)$, too. Depending on the number of zeros of $V(y)$ and accordingly $Y(y)$ three different orbit types are possible. Table \ref{tab:typen-orbits} shows an overview of all possible orbit types, some examples of the effective potential can be seen in figure \ref{pic:pot_a}, \ref{pic:pot_b} and \ref{pic:pot_c} .
\begin{itemize}
 \item Type A:\\
  $Y(y)$ has one zero in the range $-\infty < y \leq -1$. $X(x)$ has two positive zeros between $x=-1$ and $x=1$. Only TOs are possible.
 \item Type B:\\
  $Y(y)$ has two zeros. Since the orbit crosses both horizons, MBOs are possible.
 \item Type C:\\
  This orbit type is only possible for $\Psi=0$. $Y(y)$ has three zeros. Between two of the zeros a MBO can be found, the third zero is located at $y=-1$ and belongs to a BO. $X(x)$ has a positive and a negative zero. If additionally $\Phi=0$ the turning points of the MBO lie directly on the horizons since then $V(y)=-cG(y)$.

  The BO lies on the axis of $\psi$-rotation ($y=-1$). Since the shape of the ergosphere varies from ring-like to spherical, sometimes the axis $y=-1$ is not in the ergosphere and sometimes parts of the axis are inside the ergosphere. If the axis $y=-1$ is not in the ergosphere, no such bound orbit should be possible. For this reason the separation constant $k$ has to fulfill a specific condition if $y=-1$:
  \begin{equation}
   k_{\rm axis}=\Phi^2\,\frac{\nu [2+\nu(1-\nu)+\lambda(2-3\nu)]}{1-\lambda+\nu} \, .
   \label{eqn:carter}
  \end{equation}
 If $k=k_{\rm axis}$ then every BO at $y=-1$ which would be lying outside the ergosphere is forbidden by $X(x)$ ($X$ will be positive and will have no zeros). A BO at $y=-1$ in the ergosphere will still be allowed and its turning points will be located at the border of the ergosphere.

$k_{\rm axis}$ can be found if the $x$-equation for $E=m=0$ (\ref{eqn:charge-x-Gleichung}) is compared with the $x$-equation on the $\psi$-axis (see section \ref{sec:psi-axis}). Both equations have to be the same for $y=-1$ and $E=m=0$.
\end{itemize}

\begin{table}[h]
\begin{center}
\begin{tabular}{|lcll|}\hline
type &  zeros  & range of $y$ & orbit \\
\hline\hline
A  & 1 &
\begin{pspicture}(-2.5,-0.2)(3,0.2)
\psline[linewidth=0.5pt]{|->}(-2.5,0)(3,0)
\psline[linewidth=0.5pt,doubleline=true](1.0,-0.2)(1.,0.2)
\psline[linewidth=0.5pt,doubleline=true](-0.5,-0.2)(-0.5,0.2)
\psline[linewidth=1.2pt]{-*}(-2.5,0)(1.5,0)
\end{pspicture}
& TO 
\\ \hline
B & 2 & 
\begin{pspicture}(-2.5,-0.2)(3,0.2)
\psline[linewidth=0.5pt]{|->}(-2.5,0)(3,0)
\psline[linewidth=0.5pt,doubleline=true](1.0,-0.2)(1.0,0.2)
\psline[linewidth=0.5pt,doubleline=true](-0.5,-0.2)(-0.5,0.2)
\psline[linewidth=1.2pt]{*-*}(-1.0,0)(1.5,0)
\end{pspicture}
& MBO 
\\  \hline
C & 3 & 
\begin{pspicture}(-2.5,-0.2)(3,0.2)
\psline[linewidth=0.5pt]{|->}(-2.5,0)(3,0)
\psline[linewidth=0.5pt,doubleline=true](1.0,-0.2)(1.0,0.2)
\psline[linewidth=0.5pt,doubleline=true](-0.5,-0.2)(-0.5,0.2)
\psline[linewidth=1.2pt]{*-*}(-1.0,0)(1.5,0)
\psline[linewidth=1.2pt]{*}(2.7,0)(2.7,0)
\end{pspicture}
  & MBO, BO
\\  \hline
C$_0$ & 3 & 
\begin{pspicture}(-2.5,-0.2)(3,0.2)
\psline[linewidth=0.5pt]{|->}(-2.5,0)(3,0)
\psline[linewidth=0.5pt,doubleline=true](1.0,-0.2)(1.0,0.2)
\psline[linewidth=0.5pt,doubleline=true](-0.5,-0.2)(-0.5,0.2)
\psline[linewidth=1.2pt]{*-*}(-0.5,0)(1,0)
\psline[linewidth=1.2pt]{*}(2.7,0)(2.7,0)
\end{pspicture}
  & MBO, BO
\\ \hline\hline
\end{tabular}
\caption{Types of orbits of light in the (charged) doubly spinning black ring spacetime for $E=m=0$. The thick lines represent the range of $y$ and the turning points are shown by thick dots. The horizons are indicated by a vertical double line. The single vertical line at the left end is the singularity. The coordinate $y$ ranges from $-\infty$ to $-1$. Since the BO of type C is located at $y=-1$, the separation constant has to fulfill a certain condition (equation (\ref{eqn:carter})).}
\label{tab:typen-orbits}
\end{center}
\end{table}

\subsection{Solution of the $x$-equation}

Equation (\ref{eqn:charge-x-Gleichung}) can be written as 
\begin{equation}
 \left( \frac{\text{d}x}{\text{d}\gamma}\right) ^2 = X(x)=a_{x,4} x ^4 + a_{x,3} x ^3 + a_{x,2} x ^2 + a_{x,1} x + a_{x,0} \, ,
 \label{eqn:x-quadr}
\end{equation}
where $X$ is a polynomial of fourth order with the coefficients
\begin{eqnarray}
a_{x,4} &=& -(1-\nu)^2[\Phi ^2 \nu ^2 (1-\lambda ^2+\nu (1-\nu))-2\Phi\Psi\lambda\nu + \Psi ^2 \nu (1-\lambda^2)-k\nu] \\
a_{x,3} &=& -(1-\nu)^2[(\Phi^2-\Psi^2)\lambda\nu^2(3-2\nu) - \lambda k] \\
a_{x,2} &=& -(1-\nu)^2[\Psi^2\nu (2\lambda^2 + \nu (1-\nu)) + 2\Phi\Psi (\lambda^2 \nu + \lambda) + \Psi^2(2\lambda^2\nu+ \nu^2(1-\nu))  \nonumber\\
    &&  + k(\nu -1)] \\
a_{x,1} &=& -(1-\nu)^2[-\Phi^2\lambda (1+(1-\nu)^2) + 2\Phi\Psi\lambda\sqrt{\nu}(1-\nu^2) + \Psi^2\lambda\nu (2-3\nu) \nonumber\\
    && + k\lambda] \\
a_{x,0} &=& -(1-\nu)^2[-\Phi^2(1+\lambda^2) -2\Phi\Psi\lambda^2\sqrt{\nu} -\Psi^2\nu (\lambda^2-\nu^2+\nu+1)+k] \, .
\end{eqnarray}
The substitution $x = \pm \frac{1}{u} + x_0$, where $x_0$ is a zero of $X(x)$, reduces the polynomial $X(x)$ to third order:
\begin{equation}
\left( \frac{\mathrm{d}u}{\mathrm{d}\gamma} \right) ^ 2= b_{x,3} u^3 + b_{x,2} u^2 + b_{x,1} u+ b_{x,0} \, .
\end{equation}
The coefficients are
\begin{eqnarray}
b_{x,3}&=& \pm (4a_{x,4}x_0 ^3+3a_{x,3}x_0 ^2+2a_{x,2}x_0+a_{x,1})\\
b_{x,2}&=& 6a_{x,4}x_0^2+3a_{x,3}x_0+a_{x,2}\\
b_{x,1}&=& \pm (4a_{x,4}x_0+a_{x,3})\\
b_{x,0}&=& a_{x,4} \, .
\end{eqnarray}
A further substitution $u=\frac{1}{b_{x,3}}\left( 4v-\frac{b_{x,2}}{3} \right) $ transforms that into the standard Weierstra{\ss} form
\begin{equation}
\left( \frac{\mathrm{d}v}{\mathrm{d}\gamma} \right) ^ 2= 4v^3 - g_{x,2} v^2 -g_{x,3} :=P_{x,3}(v)
\label{eqn:weierstrass-form}
\end{equation}
where
\begin{equation}
g_{x,2}=\frac{b_{x,2}^2}{12} - \frac{b_{x,1} b_{x,3}}{4} \quad \text{and} \quad
g_{x,3}=\frac{b_{x,1} b_{x,2} b_{x,3}}{48}-\frac{b_{x,0} b_{x,3}^2}{16}-\frac{b_{x,2}^3}{216} \, .
\end{equation}
Equation (\ref{eqn:weierstrass-form}) is of elliptic type and is solved by the Weierstra{\ss} elliptic function \cite{Markushevich:1967}
\begin{equation}
v(\gamma )=\wp (\gamma - \gamma '_{\rm in},g_{x,2},g_{x,3}) \, ,
\end{equation}
where $\gamma' _{\rm in} = \gamma _{\rm in} + \int _{v_{x, \rm in}}^\infty \! \frac{1}{\sqrt{4v'^3 - g_{x,2} v'^2 -g_{x,3}}} \, \mathrm{d}v'$ and  $v_{x,\rm in}=\frac{b_{x,3}}{4}(x_{\rm in} - x_0)^{-1}+\frac{b_{x,2}}{12}$.

Then the solution of (\ref{eqn:charge-x-Gleichung}) takes the form
\begin{equation}
x (\gamma)=\pm\frac{b_{x,3}}{4\wp (\gamma - \gamma '_{\rm in},g_{x,2},g_{x,3}) - \frac{b_{x,2}}{3}} + x_0 \, .
\end{equation}

\subsection{Solution of the $y$-equation}

Equation (\ref{eqn:charge-y-Gleichung}) can be written as
\begin{equation}
 \left( \frac{\text{d}y}{\text{d}\gamma}\right) ^2 = Y(y) = a_{y,4} y ^4 + a_{y,3} y^3 + a_{y,2} y^2 + a_{y,1} y + a_{y,0} \, ,
 \label{eqn:y-quadr}
\end{equation}
where $Y$ is a polynomial of fourth order with the coefficients
\begin{eqnarray}
a_{y,4} &=& -(1-\nu)^2[\Psi ^2 \nu ^2 (1-\lambda ^2+\nu (1-\nu))-2\Psi\Phi\lambda\nu + \Phi ^2 \nu (1-\lambda^2)-c\nu] \\
a_{y,3} &=& -(1-\nu)^2[(\Psi^2-\Phi^2)\lambda\nu^2(3-2\nu) - \lambda c] \\
a_{y,2} &=& -(1-\nu)^2[\Psi^2\nu (2\lambda^2 + \nu (1-\nu)) + 2\Psi\Phi (\lambda^2 \nu + \lambda) + \Phi^2(2\lambda^2\nu+ \nu^2(1-\nu))  \nonumber\\
    &&  + c(\nu -1)] \\
a_{y,1} &=& -(1-\nu)^2[-\Psi^2\lambda (1+(1-\nu)^2) + 2\Psi\Phi\lambda\sqrt{\nu}(1-\nu^2) + \Phi^2\lambda\nu (2-3\nu) \nonumber\\
    && + c\lambda] \\
a_{y,0} &=& -(1-\nu)^2[-\Psi^2(1+\lambda^2) -2\Psi\Phi\lambda^2\sqrt{\nu} -\Phi^2\nu (\lambda^2-\nu^2+\nu+1)+c] \, .
\end{eqnarray}
The problem can be solved analogously to the $x$-equation. The polynomial $Y(y)$ is reduced to third order by the substitution $y = \pm \frac{1}{u} + y_0$ ($y_0$ is a zero of $Y(y)$):
\begin{equation}
\left( \frac{\mathrm{d}u}{\mathrm{d}\gamma} \right) ^ 2= b_{y,3} u^3 + b_{y,2} u^2 + b_{y,1} u+ b_{y,0} \, ,
\end{equation}
where
\begin{eqnarray}
b_{y,3}&=& \pm (4a_{y,4}y_0 ^3+3a_{y,3}y_0 ^2+2a_{y,2}y_0 +a_{y,1})\\
b_{y,2}&=& 6a_{y,4}y_0^2+3a_{y,3}y_0+a_{y,2}\\
b_{y,1}&=& \pm (4a_{y,4}y_0+a_{y,3})\\
b_{y,0}&=& a_{y,4} \, .
\end{eqnarray}
Next the polynomial is transformed into the Weierstra{\ss} form by substituting $u=\frac{1}{b_{y,3}}\left( 4v-\frac{b_{y,2}}{3} \right)$:
\begin{equation}
\left( \frac{\mathrm{d}v}{\mathrm{d}\gamma} \right) ^ 2= 4v^3 - g_{y,2} v^2 -g_{y,3} :=P_{y,3}(v) \, ,
\label{eqn:weierstrass-form-y}
\end{equation}
where
\begin{equation}
g_{y,2}=\frac{b_{y,2}^2}{12} - \frac{b_{y,1} b_{y,3}}{4} \quad \text{and} \quad
g_{y,3}=\frac{b_{y,1} b_{y,2} b_{y,3}}{48}-\frac{b_{y,0} b_{y,3}^2}{16}-\frac{b_{y,2}^3}{216} \, .
\end{equation}
Since (\ref{eqn:weierstrass-form-y}) is solved by the Weierstra{\ss} $\wp$-function, the solution of (\ref{eqn:charge-y-Gleichung}) yields
\begin{equation}
y (\gamma)=\pm \frac{b_{y,3}}{4\wp (\gamma - \gamma ''_{\rm in},g_{y,2},g_{y,3}) - \frac{b_{y,2}}{3}} + y_0 \, ,
\end{equation}
where $\gamma'' _{\rm in} = \gamma _{in} - \int _{v_{y,\rm in}}^\infty \! \frac{1}{\sqrt{4v'^3 - g_{y,2} v'^2 -g_{y,3}}} \, \mathrm{d}v'$ and $v_{y,\rm in}=\frac{b_{y,3}}{4}(y_{\rm in} - y_0)^{-1}+\frac{b_{y,2}}{12}$.

\subsection{Solution of the $\phi$-equation}
\label{sec:charge-ergo-phi}

To solve (\ref{eqn:charge-phi-Gleichung}) the $x$- and $y$-parts have to be separated. Inserting $A(x,y)$ and $L(x,y)$ from (\ref{eqn:A-L-kurz}) yields
\begin{equation}
 \frac{\mathrm{d}\phi}{\mathrm{d}\gamma}= \frac{\beta (x)\Phi +\delta (x)\Psi }{R^2G(x)} + \dfrac{\alpha (y)\Phi + \delta (y)\Psi}{R^2G(y)} \, .
\end{equation}
Using (\ref{eqn:charge-x-Gleichung}) and (\ref{eqn:charge-y-Gleichung}) we get
\begin{equation}
 \mathrm{d}\phi = \frac{\beta (x)\Phi +\delta (x)\Psi }{R^2G(x)}\frac{dx}{\sqrt{X(x)}} - \frac{\alpha (y)\Phi + \delta (y)\Psi}{R^2G(y)}\frac{dy}{\sqrt{Y (y)}} \, .
\end{equation}
So the $\phi$-equation consist of a $x$-integral and a $y$-integral:
\begin{eqnarray}
 \phi - \phi _{in} &=\phantom{:}& \int_{x_{\rm in}}^x \! \frac{\beta (x')\Phi +\delta (x')\Psi }{R^2G(x')}\frac{dx'}{\sqrt{X(x')}} - \int_{y_{\rm in}}^y \! \frac{\alpha (y')\Phi + \delta (y')\Psi}{R^2G(y')}\frac{dy'}{\sqrt{Y(y')}} \nonumber \\
   &=:& I_x + I_y \, .
\end{eqnarray}
Let us first consider $I_x$. Here we substitute $x=\pm \frac{b_{x,3}}{4u-\frac{b_{x,2}}{3}} + x_0$ to transform the polynonial $X(x)$ into the Weierstra{\ss} form $P_{x,3}$:
\begin{equation}
 I_x= \int_{u_{\rm in}}^u \! \frac{\beta \left(\pm \frac{b_{x,3}}{4u'-\frac{b_{x,2}}{3}}+ x_0\right)\Phi +\delta \left(\pm \frac{b_{x,3}}{4u'-\frac{b_{x,2}}{3}}+ x_0\right)\Psi }{R^2G\left(\pm \frac{b_{x,3}}{4u'-\frac{b_{x,2}}{3}}+ x_0\right)}\frac{du'}{\sqrt{P_{x,3}(u')}} \, .
\label{eqn:em0-Ix-subs-u}
\end{equation}
This integral has four poles $p_1$, $p_2$, $p_3$ and $p_4$ (zeros of the function $G$ with respect to $u$). We next apply a partial fractions decomposition upon (\ref{eqn:em0-Ix-subs-u}):
\begin{equation}
I_x=\int^u_{u_{\rm in}} \Biggl( K_0 + \sum^4_{j=1}\frac{K_j}{u-p_j} \Biggr) \frac{du'}{\sqrt{P_{x,3}(u')}} \, .
\label{eqn:Ix-partial}
\end{equation}
$K_j$ are constants which arise from the partial fractions decomposition and depend on the parameters of the metric and the test particle. Then we substitute $u = \wp (v, g_{x,2}, g_{x,3})$ with $\wp^\prime(v)=\sqrt{4 \wp^3(v)-g_{x,2}\wp(v)-g_{x,3}}$. Equation (\ref{eqn:Ix-partial}) now yields
\begin{equation}
 I_x= \int^{v_x}_{v_{x, \rm in}} \Biggl( K_0 + \sum^4_{j=1}\frac{K_j}{\wp_x(v)-p_j} \Biggr)dv
\label{eqn:Ix-dritterArt}
\end{equation}
with $v_x=v_x(\gamma)=\gamma-\gamma^\prime_{\rm in}$ and $v_{x,\rm in}=v_x(\gamma_{\rm in})$.

The second integral $I_y$ can be rewritten by canceling the factor $(1-y^2)$:
\begin{equation}
 I_y=-\int_{y_{\rm in}}^y \! \frac{\nu [-(1+\lambda^2)-\nu (1-\nu)+\lambda y' (2-3\nu) - (1-\lambda^2)y'^2]\cdot\Phi + \lambda\sqrt{\nu}[\lambda-(1-\nu^2)y' -\lambda\nu y'^2]\cdot\Psi}{R^2(1+\lambda y'+\nu y'^2)}\frac{dy'}{\sqrt{Y(y')}} \, .
\label{eqn:em0-Iy-subs-u}
\end{equation}
We substitute $y=\pm \frac{b_{y,3}}{4u-\frac{b_{y,2}}{3}} + y_0$ to transform the polynonial $Y(y)$ into the Weierstra{\ss} form $P_{y,3}$ and apply a partial fractions decomposition upon (\ref{eqn:em0-Iy-subs-u}) where the constants $L_j$ arise. Note that the two poles $p_1$ and $p_2$ of $I_y$ are also poles of $I_x$.
\begin{equation}
I_y=\int^u_{u_{\rm in}} \Biggl( L_0 + \sum^2_{j=1}\frac{L_j}{u-p_j} \Biggr) \frac{du'}{\sqrt{P_{y,3}(u')}} \, .
\label{eqn:Iy-partial}
\end{equation}
If we now substitute $u = \wp (v, g_{y,2}, g_{y,3})$ with $\wp^\prime(v)=\sqrt{4 \wp^3(v)-g_{y,2}\wp(v)-g_{y,3}}$, equation (\ref{eqn:Ix-partial}) yields
\begin{equation}
 I_y= \int^{v_y}_{v_{y, \rm in}} \Biggl( L_0 + \sum^2_{j=1}\frac{L_j}{\wp_y(v)-p_j} \Biggr)dv
\label{eqn:Iy-dritterArt}
\end{equation}
with $v_y=v_y(\gamma)=\gamma-\gamma''_{\rm in}$ and $v_{y,\rm in}=v_y(\gamma_{\rm in})$.\\

After solving the elliptic integrals of the third kind in (\ref{eqn:Ix-dritterArt}) and (\ref{eqn:Iy-dritterArt}) (see e.g. \cite{Enolski:2012}), the final solution is
\begin{equation}
\begin{split}
 \phi (\gamma) &= \Biggl\{K_0(v_x-v_{x,\rm in}) +  \sum^4_{j=1} \frac{K_j}{\wp^\prime_x(v_{j})}\Biggl( 2\zeta_x(v_{j})(v_x-v_{x,\rm in}) + \log\frac{\sigma_x(v_x-v_{j})}{\sigma_x(v_{x,\rm in}-v_{j})} -\log\frac{\sigma_x(v_x+v_{j})}{\sigma_x(v_{x,\rm in}+v_{j})} \Biggr) \Biggr\}  \\
 & + \Biggl\{L_0(v_y-v_{y,\rm in}) +  \sum^2_{j=1} \frac{L_j}{\wp^\prime_y(v_{j})}\Biggl( 2\zeta_y(v_{j})(v_y-v_{y,\rm in}) + \log\frac{\sigma_y(v_y-v_{j})}{\sigma_y(v_{y,\rm in}-v_{j})} -\log\frac{\sigma_y(v_y+v_{j})}{\sigma_y(v_{y,\rm in}+v_{j})}  \Biggr) \Biggr\}  \\
 & + \phi _{\rm in}
\end{split}
\end{equation}
with $p_j=\wp(v_j)$ and
\begin{eqnarray}
\wp_x(v) &= \wp (v, g_{x,2}, g_{x,3})\, , \qquad \wp_y(v) &= \wp (v, g_{y,2}, g_{y,3}) \, ,\nonumber\\
\zeta_x(v) &= \zeta (v, g_{x,2}, g_{x,3})\, , \qquad \zeta_y(v) &= \zeta (v, g_{y,2}, g_{y,3}) \, ,\\
\sigma_x(v) &= \sigma (v, g_{x,2}, g_{x,3})\, , \qquad \sigma_y(v) &= \sigma (v, g_{y,2}, g_{y,3}) \, .\nonumber
\end{eqnarray}

\subsection{Solution of the $\psi$-equation}

To solve (\ref{eqn:charge-psi-Gleichung}) the $x$- and $y$-parts have to be separated. Inserting $A(x,y)$ and $L(x,y)$ from (\ref{eqn:A-L-kurz}) yields
\begin{equation}
 \frac{\mathrm{d}\psi}{\mathrm{d}\gamma} = \frac{\delta(x)\Phi -\alpha (x)\Psi}{R^2G(x)}-\frac{\delta (y)\Phi +\beta (y)\Psi}{R^2G(y)} \, .
\end{equation}
Using (\ref{eqn:charge-x-Gleichung}) and (\ref{eqn:charge-y-Gleichung}) we get
\begin{equation}
 \mathrm{d}\psi = \frac{\delta(x)\Phi -\alpha (x)\Psi}{R^2G(x)}\frac{\mathrm{d}x}{\sqrt{X(x)}}+\frac{\delta (y)\Phi +\beta (y)\Psi}{R^2G(y)}\frac{\mathrm{d}y}{\sqrt{Y(y)}}
\end{equation}
or
\begin{equation}
 \psi - \psi_{\rm in}= \int_{x_{in}}^x \! \frac{\delta(x')\Phi -\alpha (x')\Psi}{R^2G(x')}\frac{\mathrm{d}x'}{\sqrt{X(x')}}+\int_{y_{in}}^y \!\frac{\delta (y')\Phi +\beta (y')\Psi}{R^2G(y')}\frac{\mathrm{d}y'}{\sqrt{Y(y')}} \, .
\end{equation}
The $\psi$-equation can be solved analogously to the $\phi$-equation from the previous section. Here the solution is
\begin{equation}
\begin{split}
 \psi (\gamma) &= \Biggl\{M_0(v_x-v_{x,\rm in}) +  \sum^2_{j=1} \frac{M_j}{\wp^\prime_x(v_{j})}\Biggl( 2\zeta_x(v_{j})(v_x-v_{x,\rm in}) + \log\frac{\sigma_x(v_x-v_{j})}{\sigma_x(v_{x,\rm in}-v_{j})} -\log\frac{\sigma_x(v_x+v_{j})}{\sigma_x(v_{x,\rm in}+v_{j})} \Biggr) \Biggr\}  \\
 & + \Biggl\{N_0(v_y-v_{y,\rm in}) +  \sum^4_{j=1} \frac{N_j}{\wp^\prime_y(v_{j})}\Biggl( 2\zeta_y(v_{j})(v_y-v_{y,\rm in}) + \log\frac{\sigma_y(v_y-v_{j})}{\sigma_y(v_{y,\rm in}-v_{j})} -\log\frac{\sigma_y(v_y+v_{j})}{\sigma_y(v_{y,\rm in}+v_{j})}  \Biggr) \Biggr\}  \\
 & + \psi _{\rm in}
\end{split}
\end{equation}
where $v=v(\gamma)=\gamma-\gamma''_{\rm in}$, $v_{\rm in}=v(\gamma_{\rm in})$, $p_j=\wp(v_j)$ and
\begin{eqnarray}
\wp_x(v) &= \wp (v, g_{x,2}, g_{x,3})\, , \qquad \wp_y(v) &= \wp (v, g_{y,2}, g_{y,3}) \, ,\nonumber\\
\zeta_x(v) &= \zeta (v, g_{x,2}, g_{x,3})\, , \qquad \zeta_y(v) &= \zeta (v, g_{y,2}, g_{y,3}) \, ,\\
\sigma_x(v) &= \sigma (v, g_{x,2}, g_{x,3})\, , \qquad \sigma_y(v) &= \sigma (v, g_{y,2}, g_{y,3}) \, .\nonumber
\end{eqnarray}
$M_j$ and $N_j$ are constants which arise from the partial fractions decomposition and depend on the para\-meters of the metric and the test particle.

\section{Geodesics on the $\psi$-axis}
\label{sec:psi-axis}

The surface $y=-1$ is the axis of $\psi$-rotation of the doubly spinning black ring. Here the Hamilton-Jacobi equation depends on the coordinate $x$ only. We set $y=-1$, $\Psi =0$ and $p_y=\frac{\partial S}{\partial y}=0$ in the Hamilton-Jacobi equation (\ref{eqn:hj-c-d-ring}):
\begin{eqnarray}
 0 &=& m^2-D^{2/3}(x,-1)\frac{H(x,-1)}{H(-1,x)}E^2 +D^{-1/3}(x,-1)\frac{(x+1)^2(1-\nu)^2}{R^2H(x,-1)}\left\lbrace G(x) \left(\frac{\partial S}{\partial x}\right) ^2\right. \nonumber \\
   && +\left. \frac{(\Phi+c\Omega_\phi E)^2}{(1-\nu)^2} \left[ \frac{\beta(x)}{G(x)}-\frac{\nu[2+\nu(1-\nu)+\lambda(2-3\nu)]}{1-\lambda+\nu}\right] \right\rbrace \, .
\end{eqnarray}
This can be rearranged to
\begin{eqnarray}
 \left(\frac{\partial S}{\partial x}\right) ^2 &=&\frac{D^{1/3}(x,-1)R^2 H(x,-1)}{(x+1)^2(1-\nu)^2G(x)}\left( D^{2/3}(x,-1)\frac{H(x,-1)}{H(-1,x)}E^2-m^2 \right) \nonumber \\
 && - \frac{(\Phi+c\Omega_\phi E)^2}{(1-\nu)^2G(x)} \left( \frac{\beta(x)}{G(x)}-\frac{\nu[2+\nu(1-\nu)+\lambda(2-3\nu)]}{1-\lambda+\nu}\right) =: X_S \, . \nonumber \\
\end{eqnarray}
Then we have
\begin{equation}
 S = \frac{1}{2}m^2\tau -Et + \Phi\phi + \int \! \sqrt{X_S} \, \mathrm{d}x \, .
\end{equation}
Now we set the partial derivatives of $S$ with respect to the constants $m^2$, $E$ and $\Phi$ to zero in order to obtain the equations of motion.
With the Mino-time \cite{Mino:2003yg} $d\gamma = \frac{(x+1)}{D(x,-1)^{1/3}R^2 H(x,-1) H(-1,x)} d\tau$ the equations of motion take the form
\begin{eqnarray}
 \frac{\mathrm{d}x}{\mathrm{d}\gamma} &=& \sqrt{X(x)} \label{eqn:charge-psi-x-gleichung}\\
 \frac{\mathrm{d}\phi}{\mathrm{d}\gamma} &=& \frac{(x+1)H(x,-1)H^2(-1,x)}{(1+\nu-\lambda)^2G(x)}(\Phi +c\Omega_\phi E) \label{eqn:charge-psi-phi-gleichung}\\
 \frac{\mathrm{d}t}{\mathrm{d}\gamma} &=& R^2 E\frac{D(x,-1)H^2(x,-1)}{(x+1)}- \frac{(x+1)H(x,-1)H^2(-1,x)}{(1+\nu-\lambda)^2G(x)}c\Omega_\phi(\Phi +c\Omega_\phi E) \label{eqn:charge-psi-t-gleichung}
\end{eqnarray}
where
\begin{eqnarray}
 X(x) &=& (1-\nu)^2 H(x,-1) H(-1,x) \left\lbrace R^2 G(x) \left[D(x,-1) H(x,-1)E^2 - D^{1/3}(x,-1) H(-1,x)m^2\right] \right. \nonumber \\
 && \left. - \frac{(x+1)^2}{(1-\lambda +\nu)^2} \left[ H(-1,x)\Phi +cR\lambda \sqrt{\nu}\sqrt{2((1+\nu)^2-\lambda^2)}(1-x^2)E \right] ^2 \right\rbrace \nonumber \\
\end{eqnarray}
and
\begin{eqnarray}
 H(-1,x) &=& (1-\lambda)^2-\nu ^2 + \nu x^2 (1-\lambda ^2-\nu ^2+2\lambda\nu) \nonumber\\
 H(x,-1) &=& 1+\lambda ^2 -\nu ^2-2\lambda\nu (1-x^2)+2\lambda x (1-\nu ^2) + x^2\nu(1-\lambda ^2-\nu ^2) \nonumber\\
 D(x,-1) &=& 1+\frac{s^2}{H(x,-1)}[2\lambda(1-\nu)(x+1)(1+\nu x)] \nonumber\\
 \Omega _\phi &=& \frac{R\lambda\sqrt{2((1+\nu)^2-\lambda^2)}}{H(-1,x)}(1-x^2)\sqrt{\nu} \, .
\end{eqnarray}

Solving the equations of motion (\ref{eqn:charge-psi-x-gleichung})-(\ref{eqn:charge-psi-t-gleichung}) analytically is only possible if $X(x)$ is a polynomial. This happens in two cases:
\begin{enumerate}
 \item $D(x,-1)=1$ (which implies $c=1$ and $s=0$): uncharged doubly spinning black ring
 \item $m=0$: charged doubly spinning black ring with photons
\end{enumerate}
In both cases $X(x)$ is a polynomial of 10th order, so that the equations of motion are of hyperelliptic type.

\subsection{Classification of Geodesics}

From (\ref{eqn:charge-psi-x-gleichung}) we can read off the effective potential consisting of two parts $U_+(x)$ and $U_-(x)$:
\begin{equation}
 X=a(x)(E-U_+)(E-U_-) \, .
\end{equation}
Since $X(x)$ can be written as $X(x)=a(x)E^2+b(x)E+c(x)$ the effective potential takes the form
\begin{equation}
 U_\pm (x) = \frac{-b(x)\pm\sqrt{b(x)^2-4a(x)c(x)}}{2a(x)}\, ,
\end{equation}
where
\begin{equation}
 \begin{split}
  a(x) &= (1-\nu)^2 H(x,-1) H(-1,x) \left(R^2 G(x) D(x,-1) H(x,-1)-\frac{c^2\Omega_\phi^2 H^2(-1,x)(x+1)^2}{(1-\lambda+\nu)^2}\right)\\
  b(x)  &= -2(1-\nu)^2\Phi\frac{ c \Omega_\phi H^3(-1,x) H(x,-1) (x+1)^2  }{(1-\lambda+\nu)^2}\\
  c(x)  &= -(1-\nu)^2 H(x,-1) H(-1,x)\left(R^2G(x) D^{1/3}(x,-1) H(-1,x) m^2+\frac{(x+1)^2 H^2(-1,x)\Phi^2}{(1-\lambda+\nu)^2}\right) \, .
 \end{split}
\end{equation}

\begin{figure}[h]
 \centering
 \subfigure[$\Phi=0$ \newline Examples of orbits of type A and B.]{
   \includegraphics[width=4.8cm]{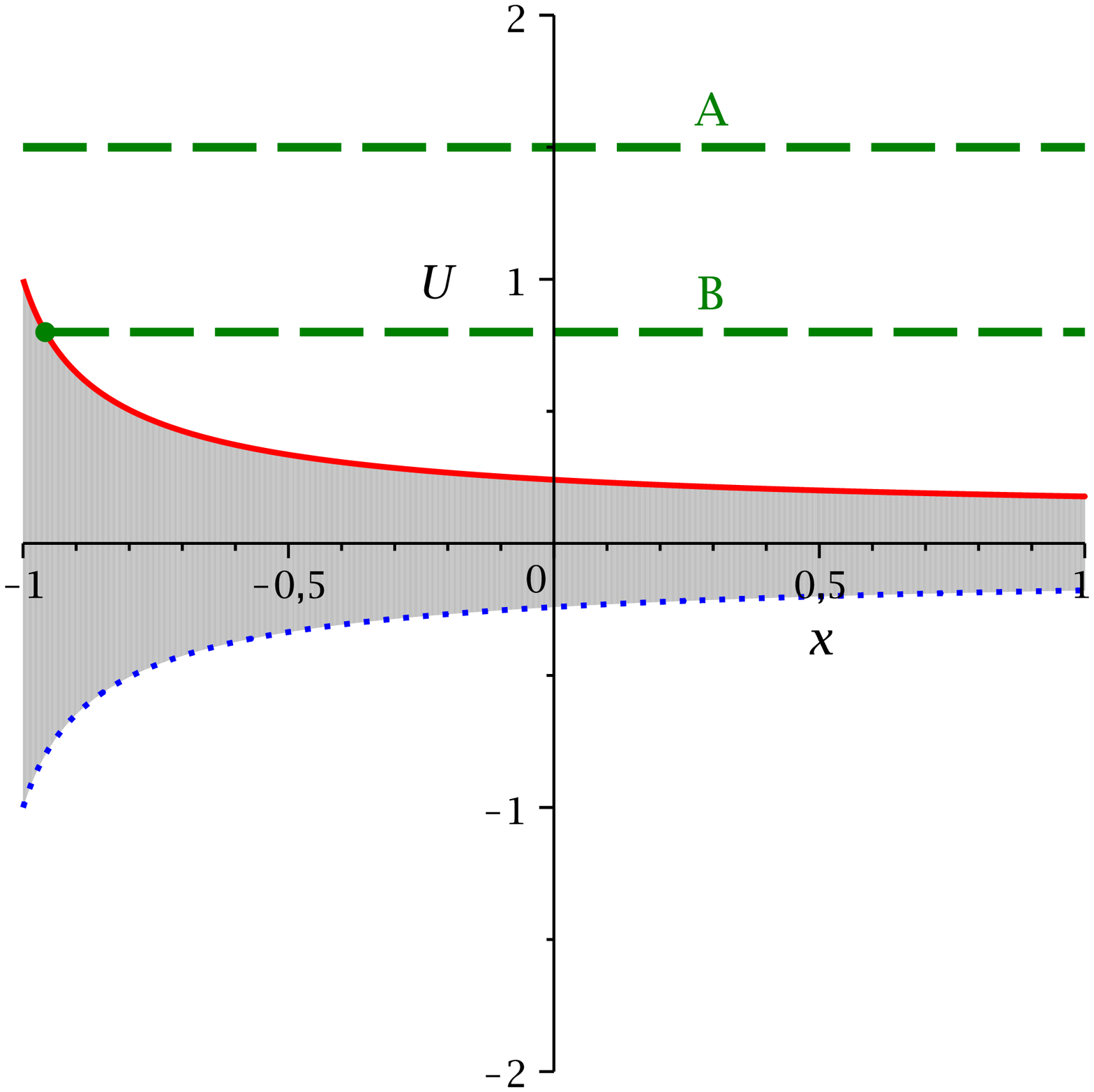}
 }
 \subfigure[$\Phi=1$ \newline Examples of orbits of type C and D. If $|\Phi|>0$ there is a potential barrier which prevents the geodesics from reaching $x=+1$. $X(x)$ can have one or two zeros.]{
   \includegraphics[width=4.8cm]{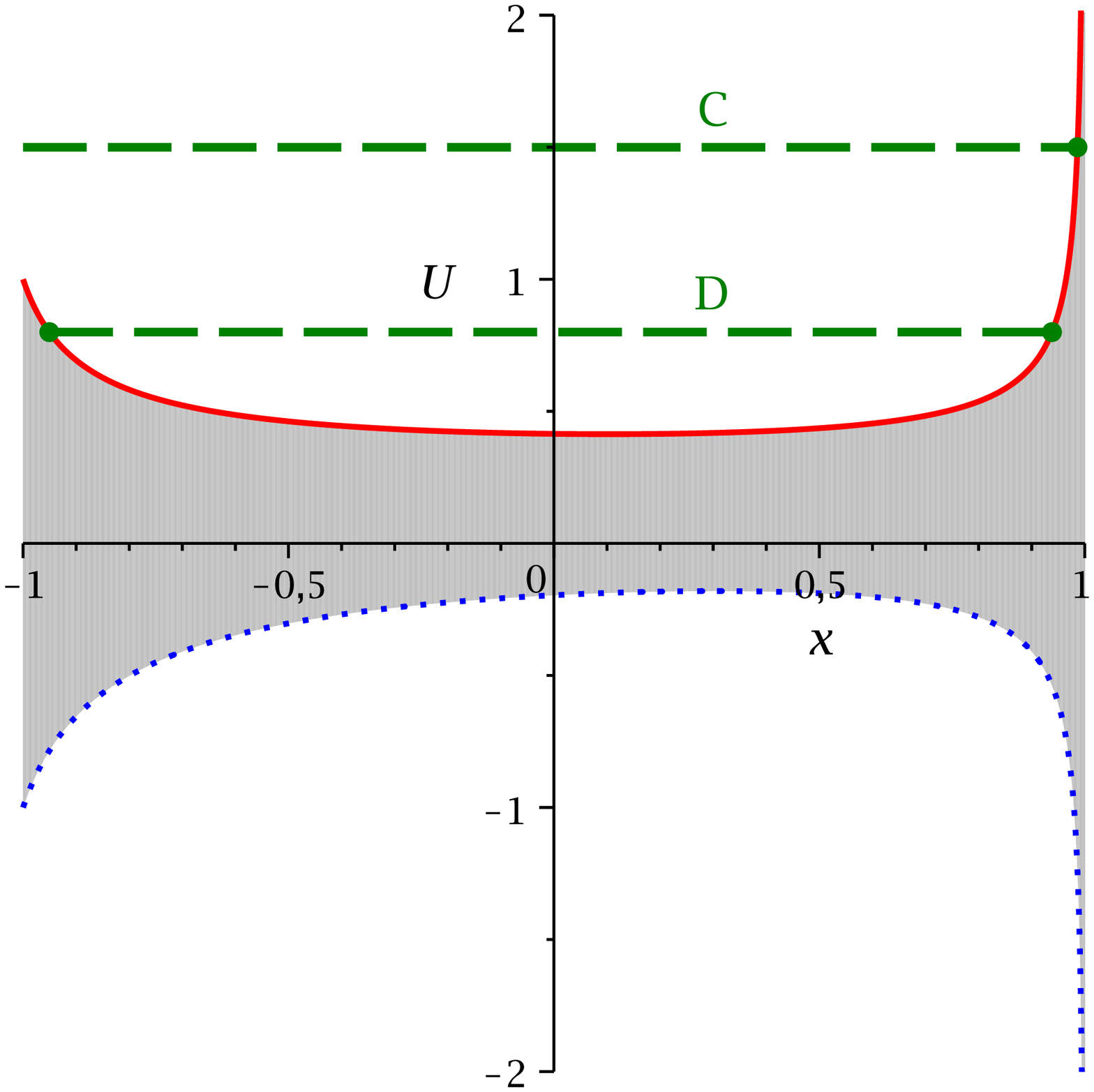}
 }
 \subfigure[$\Phi=8$ \newline Example of an orbit of type E. The potential can have local extrema which lead to three zeros of $X(x)$]{
   \includegraphics[width=4.8cm]{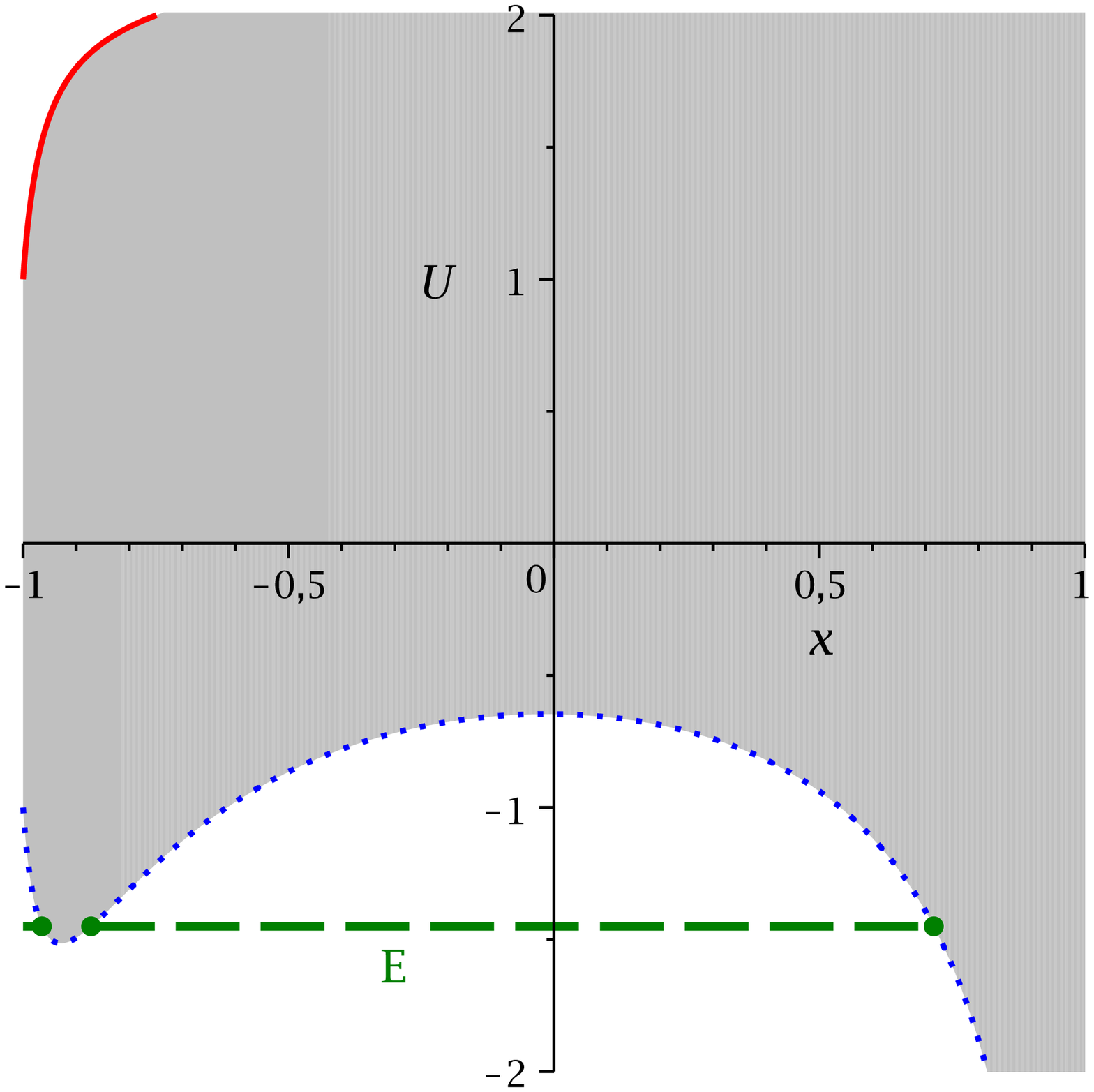}
 }
 \caption{$R=1$, $m=1$, $s=1$, $\nu =0.1$ and $\lambda=0.7$ \newline
 Effective potentials $U_+(x)$ (red solid line) and $U_-(x)$ (blue dotted line) on the $\psi$-axis ($y=-1$) of the black ring. The grey area is a forbidden zone, where no motion is possible. Green dashed lines represent energies and green points mark the turning points.}
 \label{pic:c-psi-orbits1}
\end{figure}

Figure \ref{pic:c-psi-orbits1} shows the effective potential for the motion on the $\psi$ axis. $U_+$ is plotted in red (solid line) while $U_-$ is plotted in blue (dotted line). The grey area between the two parts of the potential is a forbidden zone where no motion is possible because $X(x)$ becomes negative there. If $\Phi=0$ then $X(x)$ has none or one zeros and $U_+$ and $U_-$ are symmetric ($U_+=-U_-$). If $|\Phi|>0$ there is a potential barrier which prevents the geodesics from reaching $x=+1$. $X(x)$ can have up to three zeros.

Possible orbits are Bound Orbits (BO), where light or test particles circle the black ring, and Escape Orbits (EO), where light or a test particle approaches the black ring, turns around at a certain point and escapes the gravitational field.

There are five different types of orbits (see table \ref{tab:c-psi-typen-orbits1}), which exist in the charged as well as in the uncharged black ring spacetime:
\begin{itemize}
 \item Type A:\\
  $X(x)$ has no zero. EOs without a turning point exist. The orbit crosses the equatorial plane ($x=+1$) and reaches infinity ($x=-1$ and $y=-1$).
 \item Type B:\\
  $X(x)$ has one zero. BOs with a turning point on each side of the ring exist, so that the orbit crosses the equatorial plane ($x=+1$).
 \item Type C:\\
  $X(x)$ has one zero. EOs with a turning point exist.
 \item Type D:\\
  $X(x)$ has two zeros. BOs which do not cross the equatorial plane exist.
 \item Type E:\\
  $X(x)$ has three zeros. BOs which do not cross the equatorial plane and EOs exist.
\end{itemize}

\begin{table}[h]
\begin{center}
\begin{tabular}{|lcll|}\hline
type &  zeros  & range of $x$ & orbit \\
\hline\hline
A & 0 & 
\begin{pspicture}(-2.5,-0.2)(3,0.2)
\psline[linewidth=0.5pt]{->}(-2.5,0)(3,0)
\psline[linewidth=1.2pt]{-}(-2.5,0)(3,0)
\end{pspicture}
  & EO
\\  \hline
B & 1 & 
\begin{pspicture}(-2.5,-0.2)(3,0.2)
\psline[linewidth=0.5pt]{->}(-2.5,0)(3,0)
\psline[linewidth=1.2pt]{*-}(-1.5,0)(3,0)
\end{pspicture}
  & BO 
\\ \hline
C  & 1 &
\begin{pspicture}(-2.5,-0.2)(3,0.2)
\psline[linewidth=0.5pt]{->}(-2.5,0)(3,0)
\psline[linewidth=1.2pt]{-*}(-2.5,0)(2,0)
\end{pspicture}
& EO 
\\ \hline
D & 2 & 
\begin{pspicture}(-2.5,-0.2)(3,0.2)
\psline[linewidth=0.5pt]{->}(-2.5,0)(3,0)
\psline[linewidth=1.2pt]{*-*}(-1,0)(2,0)
\end{pspicture}
  & BO
\\ \hline
E & 3 & 
\begin{pspicture}(-2.5,-0.2)(3,0.2)
\psline[linewidth=0.5pt]{->}(-2.5,0)(3,0)
\psline[linewidth=1.2pt]{*-*}(-0,0)(2,0)
\psline[linewidth=1.2pt]{-*}(-2.5,0)(-1.0,0)
\end{pspicture}
& EO, BO 
\\ \hline\hline
\end{tabular}
\caption{Types of orbits of light and particles in the charged doubly spinning black ring spacetime for $y=-1$, $\Psi =0$. The thick lines represent the range of the orbits. The turning points are shown by thick dots. The coordinate $x$ ranges from $-1$ to $+1$. An orbit of type B crosses the equatorial plane.}
\label{tab:c-psi-typen-orbits1}
\end{center}
\end{table}

\subsection{Solution of the $x$-equation}
\label{sec:charge-psi-xsolution}

In the two cases $D(x,-1)=1$ (i.e. $s=0$) and $m=0$ equation (\ref{eqn:charge-psi-x-gleichung}) can be written in the form
\begin{equation}
 \left( \frac{\mathrm{d}x}{\mathrm{d}\gamma}\right)^2 = X(x) = \sum _{i=0}^{10} a_{i} x^{i} \, .
 \label{eqn:charge-psi-x-quadrgleichung}
\end{equation}
A separation of variables gives the hyperelliptic integral
\begin{equation}
 \gamma - \gamma_{\rm in} = \int _{x_{\rm in}}^x \! \frac{\mathrm{d}x'}{\sqrt{X(x')}}.
 \label{eqn:charge-psi-x-intgleichung}
\end{equation}
Since $X(x)$ is a polynomial of 10th order, the genus of the Riemann surface is $g=4$. A canonical basis of holomorphic ($\mathrm{d}u_i$) and meromorphic ($\mathrm{d}r_i$) differentials associated with the hyperelliptic curve $w^2=X(x)$ is given by (see \cite{Hackmann:2008tu} or \cite{Enolski:2012})
\begin{equation}
 \mathrm{d}u_i := \frac{x^{i-1}}{w}\mathrm{d}x\, ,
\end{equation}
\begin{equation}
 \mathrm{d}r_i := \sum _{k=i}^{2g+1-i} (k+1-i)\lambda _{k+1+i}\frac{x^{k}}{4w}\mathrm{d}x
\end{equation}
with $i=1,...,g$ and $\lambda_i$ being the coefficients of the hyperelliptic curve written as
\begin{equation}
w^2=\lambda_{2g+2}x^{2g+2}+\lambda_{2g+1}x^{2g+1}+\lambda_{2g}x^{2g}+\ldots+\lambda_0 \label{curve} \, .
\end{equation}
Furthermore we introduce the holomorphic and meromorphic period matrices $(2\omega, 2\omega ')$ and $(2\eta, 2\eta ')$:
\begin{equation}
 \begin{split}
  2\omega_{ij} := \oint_{\mathfrak{a}_j} \mathrm{d}u_i, \qquad 2\omega'_{ij} := \oint_{\mathfrak{b}_j} \mathrm{d}u_i, \\
  2\eta_{ij} := -\oint_{\mathfrak{a}_j} \mathrm{d}r_i, \qquad 2\eta'_{ij} := -\oint_{\mathfrak{b}_j} \mathrm{d}r_i \,
 \end{split}
\end{equation}
with $i,j=1,...,g$, where $\{ {\mathfrak a}_i, {\mathfrak b}_i|i=1,...,g \}$ is the canonical basis of closed paths.

The solution of equation (\ref{eqn:charge-psi-x-intgleichung}) is extensively discussed in \cite{Hackmann:2008zz}, \cite{Enolski:2010if},  \cite{Hackmann:2008tu}, \cite{Enolski:2012} and is given in terms of the derivatives
\begin{equation}
 \begin{split}
 \sigma_{i}(\boldsymbol{u}) &= -\frac{\partial}{\partial u_i}\sigma(\boldsymbol{u}) \\
 \sigma_{ij}(\boldsymbol{u}) &= -\frac{\partial}{\partial u_i}\frac{\partial}{\partial u_j}\sigma(\boldsymbol{u}) \\
 \vdots
 \end{split}
\end{equation}
of the Kleinian sigma function $\sigma (\boldsymbol{u}) = k e^{-(1/2)\boldsymbol{u}^t\eta\omega^{-1}\boldsymbol{u}} \vartheta ((2\omega)^{-1}\boldsymbol{u} + \boldsymbol{K}_{x_{\rm in}};\tau)$: 
\begin{equation}
 x(\gamma)=-\frac{\sigma_{144}(\boldsymbol{\gamma}_\Theta)}{\sigma_{244}(\boldsymbol{\gamma}_\Theta)} \, ,
\end{equation}
where
\begin{equation}
 \boldsymbol{\gamma}_\Theta :=
 \left(
 \begin{array}{c}
 \gamma - \gamma_{\rm in}' \\
 \gamma_2 \\
 \gamma_3 \\
 \gamma_4
 \end{array}
 \right).
\end{equation}
The constant $\gamma_{\rm in}' = \gamma_{\rm in} +\int_{x_{\rm in}}^{\infty}\! \mathrm{d}u_1$ depends on $ \gamma_{\rm in}$ and $x_{\rm in}$ only. $\gamma_2$, $\gamma_3$ and $\gamma_4$ are defined by the vanishing condition of the Kleinian sigma function: $\sigma (\boldsymbol{\gamma}_\Theta) = 0$, $\sigma_4 (\boldsymbol{\gamma}_\Theta) = 0$ and $\sigma_{44} (\boldsymbol{\gamma}_\Theta) = 0$ so that $(2\omega )^{-1}\boldsymbol{\gamma}_\Theta$ is an element of the theta divisor (the set of zeros of the theta function).

\subsection{Solution of the $\phi$-equation}
\label{sec:charge-psi-phisolution}

With (\ref{eqn:charge-psi-x-gleichung}) equation (\ref{eqn:charge-psi-phi-gleichung}) yields
\begin{equation}
\mathrm{d}\phi = \frac{(x+1)H(x,-1)H^2(-1,x)}{(1+\nu-\lambda)^2G(x)}(\Phi +c\Omega_\phi E) \frac{\mathrm{d}x}{\sqrt{X(x)}} \, .
\end{equation}
This can be rewritten as
\begin{equation}
 \mathrm{d}\phi=\frac{P_6(x)}{P_3(x)} \frac{\mathrm{d}x}{\sqrt{X(x)}}\, ,
\end{equation}
where $P_6(x)$ is a polynomial of sixth order and $P_3(x)$ is a polynomial of third order. So we have to solve the integral
\begin{equation}
 \phi - \phi_{\rm in} = \int _{x_{\rm in}}^x \! \frac{P_6(x')}{P_3(x')} \, \frac{\mathrm{d}x'}{\sqrt{X(x')}}
\label{eqn:psi-phiint}
\end{equation}
which has poles at $p_{1,2}=\frac{-\lambda\pm\sqrt{\lambda^2-4\nu}}{2\nu}$ and $p_3=1$. We apply a partial fractions decomposition upon (\ref{eqn:psi-phiint}):
\begin{equation}
 \phi - \phi_{\rm in} = \int _{x_{\rm in}}^x \! \left( \sum _{i=1}^3 \frac{K_i}{x'-p_i} + \sum _{i=4}^7 K_i x'\, ^{i-4} \right) \,  \frac{\mathrm{d}x'}{\sqrt{X(x')}}\, ,
 \label{eqn:charge-phi-pbz}
\end{equation}
where $K_i$ are constants which arise from the partial fractions decomposition. 

If $D(x,-1)=1$ (i.e. $s=0$) or $m=0$ then $X(x)$ is a polynomial of tenth order and equation (\ref{eqn:charge-phi-pbz}) consists of hyperelliptic integrals of the first and third kind.\\

First we will solve the holomorphic integrals of the first kind $\int _{x_{\rm in}}^x\! \frac{x'\,^i}{\sqrt{X(x')}}\mathrm{d}x'$ (in this case $i=0,1,2,3$). 
We introduce a variable $v$ so that $v-v_{0} =  \int _{x_{\rm in}}^x\! \frac{\mathrm{d}x'}{\sqrt{X(x')}}$. The inversion of this integral yields $x(v)=-\frac{\sigma_{144}(\boldsymbol{u})}{\sigma_{244}(\boldsymbol{u})}$ (see section \ref{sec:charge-psi-xsolution} or \cite{Enolski:2012}), where
\begin{equation}
\boldsymbol{u}=\boldsymbol{\mathfrak A}_i+\left( 
\begin{array}{c} 
 v - v_0 \\ 
 f_1(v - v_0) \\
 f_2(v - v_0) \\
 f_3(v - v_0)
 \end{array}   \right) ,\quad f_1(0)=f_2(0)=f_3(0)=0 \, .
\end{equation}
The functions $f_1(v - v_0)$, $f_2(v - v_0)$ and $f_3(v - v_0)$ can be found from the conditions $\sigma(\boldsymbol{u})=0$, $\sigma_4(\boldsymbol{u})=0$ and $\sigma_{44}(\boldsymbol{u})=0$.

The vector $\boldsymbol{\mathfrak A}_i$ identified with each branch point $e_i$ is defined as \cite{Enolski:2012}
\begin{equation}
\boldsymbol{\mathfrak{A}}_i=\int_{\infty}^{(e_i,0)} \mathrm{d}\boldsymbol{u}= 2\omega \boldsymbol{\varepsilon}_i+2\omega' \boldsymbol{\varepsilon}_i', \quad i=1,\ldots,10 \,,
\label{eqn:characteristics}
\end{equation}
with the vectors $\boldsymbol{\varepsilon}_i$ and $\boldsymbol{\varepsilon}_i'$ $\in\mathbb{R}^4$ with entries equal $\frac{1}{2}$ or $0$. The matrix
\begin{equation}
 [\boldsymbol{\mathfrak{A}}_i] = \left(
 \begin{array}{c}
  \boldsymbol{\varepsilon}_i'\\
  \boldsymbol{\varepsilon}_i
 \end{array}
\right) =
\left(
 \begin{array}{c}
  \varepsilon_{i1}' \ldots \varepsilon_{i4}'\\
  \varepsilon_{i1}  \ldots \varepsilon_{i4}
 \end{array}
\right) 
\end{equation}
is called the characteristic of a branch point $e_i$.\\

Equation (\ref{eqn:charge-phi-pbz}) now reads
\begin{equation}
 \phi - \phi_{\rm in} = \sum _{i=1}^3 \int _{x_{\rm in}}^x \!\frac{K_i}{x'-p_i} \frac{\mathrm{d}x'}{\sqrt{X(x')}} + K_4(v - v_0) + K_5f_1(v - v_0) + K_6f_2(v - v_0)  + K_7f_3(v - v_0) \, .
\end{equation}\\

The remaining hyperelliptic integrals of the third kind can be solved with the help of the following equation (see \cite{Enolski:2012}). 
\begin{align}\begin{split}
W\int_{P'}^P\frac{1}{x-Z}\frac{\mathrm{d}x}{w} = & 2\int_{P'}^P \mathrm{d}\boldsymbol{u}^T(x,y) \left[ \boldsymbol{\zeta} \left( \int_{(e_2,0)}^{(Z,W)} \mathrm{d} \boldsymbol{u} + \boldsymbol{K}_\infty  \right) - 2( \boldsymbol{\eta}^{\prime}\boldsymbol{\varepsilon}^\prime + \boldsymbol{\eta}\boldsymbol{\varepsilon} )  - \frac12 \boldsymbol{\mathfrak{Z}}(Z,W)    \right]\\
& +\ln \frac{\sigma\left(\int_{\infty}^P \mathrm{d}\boldsymbol{u}- \int_{(e_2,0)}^{(Z,W)} \mathrm{d}\boldsymbol{u} - \boldsymbol{K}_\infty  \right)}{\sigma\left(\int_{\infty}^P \mathrm{d}\boldsymbol{u}+ \int_{(e_2,0)}^{(Z,W)} \mathrm{d}\boldsymbol{u} - \boldsymbol{K}_\infty \right)}
- \mathrm{ln} \frac{\sigma\left(\int_{\infty}^{P'} \mathrm{d}\boldsymbol{u} - \int_{(e_2,0)}^{(Z,W)} \mathrm{d}\boldsymbol{u} - \boldsymbol{K}_\infty  \right)}{\sigma\left(\int_{\infty}^{P'} \mathrm{d}\boldsymbol{u} + \int_{(e_2,0)}^{(Z,W)} \mathrm{d}\boldsymbol{u} - \boldsymbol{K}_\infty  \right)}.\end{split} \label{main1-2}
\end{align}
$P$ and $P'$ are points on the hyperelliptic curve, $Z$ is a pole, $W=w(Z)$ and $w^2=X(x)$. The zeros $e_i$ of $w^2(x)$ are the branch points of the curve $w^2$. $\mathrm{d}\boldsymbol{u}$ is the vector of the holomorphic differentials of the first kind $\mathrm{d}u_i=\frac{x^{i-1}}{w}\mathrm{d}x$ with $i=1,...,g$. $\zeta$ and $\sigma$ are Kleinian functions and $\boldsymbol{K}_\infty$ is the vector of Riemann constants. 

The $g$th component (in this case genus $g=4$) of the vector $\boldsymbol{\mathfrak{Z}}(Z,W)$ is $\mathfrak{Z}_g(Z,W)=0$ and for $1\leq j<g$ we have
\begin{equation}
\mathfrak{Z}_j(Z,W)=\frac{W}{\prod_{k=2}^{g} (Z-e_{2k})}\sum_{k=0}^{g-j-1}(-1)^{g-k+j+1}Z^kS_{g-k-j-1}(\boldsymbol{e}) \, .
 \end{equation}
The $S_k(\boldsymbol{e})$ are elementary symmetric functions of order $k$ built on  $g-1$ branch points $e_4,\ldots, e_{2g}$: $S_0=1$, $S_1=e_4+\ldots+e_{2g}$, etc.\\

Finally the solution of the $\phi$-equation (\ref{eqn:charge-psi-phi-gleichung}) is
\begin{equation}
\begin{split}
 \phi &=\phi_{\rm in} +  \sum _{i=1}^3 K_i \left[ \frac{2}{W_i} \left(\int _{x_{\rm in}}^x d\boldsymbol{u}\right)^T \left( \boldsymbol{\zeta} \left( \int_{(e_2,0)}^{(p_i,W_i)} \mathrm{d} \boldsymbol{u} + \boldsymbol{K}_\infty  \right) - 2( \boldsymbol{\eta}^{\prime}\boldsymbol{\varepsilon}^\prime + \boldsymbol{\eta}\boldsymbol{\varepsilon} )  - \frac12 \boldsymbol{\mathfrak{Z}}(p_i,W_i)  \right) \right. \\ 
& \left. + \ln\frac{\sigma\left(  W^2(x)  \right)}{\sigma\left( W^1(x) \right)}
-  \ln \frac{\sigma\left(  W^2(x_{\rm in})  \right)}{\sigma\left( W^1(x_{\rm in}) \right)}  \right] + K_4(v - v_0) + K_5f_1(v - v_0) + K_6f_2(v - v_0)  + K_7f_3(v - v_0) \, ,
\end{split}
\end{equation}
where $W_i=\sqrt{X(p_i)}$ and $W^{1,2}(x) = \int^{x}_{\infty}{d\boldsymbol{u}} \pm  \int_{(e_2,0)}^{(p_i,W_i)} \mathrm{d} \boldsymbol{u} - \boldsymbol{K}_\infty $.

\subsection{Solution of the $t$-equation}

With (\ref{eqn:charge-psi-x-gleichung}) equation (\ref{eqn:charge-psi-t-gleichung}) yields
\begin{equation}
 \mathrm{d}t= \left( R^2 E\frac{D(x,-1)H^2(x,-1)}{(x+1)}- \frac{(x+1)H(x,-1)H^2(-1,x)}{(1+\nu-\lambda)^2G(x)}c\Omega_\phi(\Phi +c\Omega_\phi E) \right) \frac{\mathrm{d}x}{\sqrt{X(x)}} \, .
\end{equation}
This can be rewritten as
\begin{equation}
 \mathrm{d}t = \left(\frac{P_4(x)}{x+1} + \frac{P_5(x)}{P_2(x)} \right)\frac{\mathrm{d}x}{\sqrt{X(x)}}\, ,
\end{equation}
where $P_d(x)$ are polynomials of order $d$. $P_2(x)$ has the zeros $p_{1,2}=\frac{-\lambda\pm\sqrt{\lambda^2-4\nu}}{2\nu}$.
So we have to solve the integral
\begin{equation}
 t - t_{\rm in} = \int _{x_{\rm in}}^x \! \left(\frac{P_4(x')}{x'+1} + \frac{P_5(x')}{P_2(x')}\right) \, \frac{\mathrm{d}x'}{\sqrt{X(x')}} \, .
\end{equation}
We apply a partial fractions decomposition where the constants $M_j$ and $N_j$ arise:
\begin{equation}
 t - t_{\rm in} = \int _{x_{\rm in}}^x \! \left( \frac{M_1}{x'+1} + \sum _{i=2}^5 M_i x'\, ^{i-2}  +  \sum _{i=1}^2 \frac{N_i}{x'-p_i} + \sum _{i=3}^6 N_i x'\, ^{i-3} \right)  \, \frac{\mathrm{d}x'}{\sqrt{X(x')}} \, .
\label{eqn:psi-tint}
\end{equation}
If $D(x,-1)=1$ (i.e. $s=0$) or $m=0$ then $X(x)$ is a polynomial of tenth order and the equation (\ref{eqn:psi-tint}) can be integrated analytically.
The hyperelliptic integrals of the first and third kind can be solved as shown in section \ref{sec:charge-psi-phisolution}. The solution of the $t$-equation (\ref{eqn:charge-psi-t-gleichung}) is
\begin{equation}
\begin{split}
 t &= \left\lbrace  M_1 \left[ \frac{2}{\sqrt{X(-1)}} \left(\int _{x_{\rm in}}^x d\boldsymbol{u}\right)^T \left( \boldsymbol{\zeta} \left( \int_{(e_2,0)}^{(-1,\sqrt{X(-1)})} \mathrm{d} \boldsymbol{u} + \boldsymbol{K}_\infty  \right) - 2( \boldsymbol{\eta}^{\prime}\boldsymbol{\varepsilon}^\prime + \boldsymbol{\eta}\boldsymbol{\varepsilon} )  - \frac12 \boldsymbol{\mathfrak{Z}}(-1,\sqrt{X(-1)})  \right) \right.  \right.\\ 
& \left.\left. + \ln\frac{\sigma\left(  W^2(x)  \right)}{\sigma\left( W^1(x) \right)}
-  \ln \frac{\sigma\left(  W^2(x_{\rm in})  \right)}{\sigma\left( W^1(x_{\rm in}) \right)}  \right] + M_2(v - v_0) + M_3f_1(v - v_0) +M_4f_2(v - v_0)  + M_5f_3(v - v_0) \right\rbrace \\
& + \left\lbrace  \sum _{i=1}^2 N_i \left[ \frac{2}{W_i} \left(\int _{x_{\rm in}}^x d\boldsymbol{u}\right)^T \left( \boldsymbol{\zeta} \left( \int_{(e_2,0)}^{(p_i,W_i)} \mathrm{d} \boldsymbol{u} + \boldsymbol{K}_\infty  \right) - 2( \boldsymbol{\eta}^{\prime}\boldsymbol{\varepsilon}^\prime + \boldsymbol{\eta}\boldsymbol{\varepsilon} )  - \frac12 \boldsymbol{\mathfrak{Z}}(p_i,W_i)  \right) \right.  \right.\\ 
& \left.\left. + \ln\frac{\sigma\left(  W^2(x)  \right)}{\sigma\left( W^1(x) \right)}
-  \ln \frac{\sigma\left(  W^2(x_{\rm in})  \right)}{\sigma\left( W^1(x_{\rm in}) \right)}  \right] + N_3(v - v_0) + N_4f_1(v - v_0) + N_5f_2(v - v_0)  + N_6f_3(v - v_0) \right\rbrace \\
& + t_{\rm in} \, ,
\end{split}
\end{equation}
where $W_i=\sqrt{X(p_i)}$ and $W^{1,2}(x) = \int^{x}_{\infty}{d\boldsymbol{u}} \pm  \int_{(e_2,0)}^{(p_i,W_i)} \mathrm{d} \boldsymbol{u} - \boldsymbol{K}_\infty $.

\section{Geodesics on the $\phi$-axis}

The surface $x=\pm 1$ is the axis of $\phi$ rotation. It can be seen as the equatorial plane of the black ring, which is divided into two parts. The first part $x=+1$ is the plane enclosed by the ring (or more precisely: enclosed by the singularity), which we will refer to as ``inside'' the ring. The second part $x=-1$ describes the equatorial plane around the black ring (or more precisely: around the singularity), which we will refer to as ``outside'' the ring.\\

If we set $x=\pm 1$, $\Phi =0$ and $p_x=\frac{\partial S}{\partial x}=0$ in the Hamilton-Jacobi equation (\ref{eqn:hj-c-d-ring}), it depends on the coordinate $y$ only:
\begin{eqnarray}
 0 &=& m^2-D^{2/3}(\pm 1,y)\frac{H(\pm 1,y)}{H(y,\pm 1)}E^2 -D^{-1/3}(\pm 1,y)\frac{(\pm 1-y)^2(1-\nu)^2}{R^2H(\pm 1,y)}\left\lbrace G(y) \left(\frac{\partial S}{\partial y}\right) ^2\right. \nonumber \\
   && +\left. \frac{(\Psi+c\Omega_\psi E)^2}{(1-\nu)^2} \left[ \frac{\beta(y)}{G(y)}-\frac{\nu[2+\nu(1-\nu)\mp\lambda(2-3\nu)]}{1\pm\lambda+\nu}\right] \right\rbrace \, .
\end{eqnarray}
This can be rearranged to
\begin{eqnarray}
 \left(\frac{\partial S}{\partial y}\right) ^2 &=&\frac{D^{1/3}(\pm 1,y)R^2 H(\pm1,y)}{(\pm 1-y)^2(1-\nu)^2G(y)}\left( m^2-D^{2/3}(\pm 1,y)\frac{H(\pm 1,y)}{H(y,\pm 1)}E^2 \right) \nonumber \\
 && - \frac{(\Psi+c\Omega_\psi E)^2}{(1-\nu)^2G(y)} \left( \frac{\beta(y)}{G(y)}-\frac{\nu[2+\nu(1-\nu)\mp\lambda(2-3\nu)]}{1\pm\lambda+\nu}\right) =: Y_S \, . \nonumber \\
\end{eqnarray}
Then we have
\begin{equation}
 S = \frac{1}{2}m^2\tau -Et + \Psi\psi + \int \! \sqrt{Y_S} \, \mathrm{d}y \, .
\end{equation}
Now we set the derivatives of $S$ with respect to the constants $m^2$, $E$ and $\Psi$ to zero in order to obtain the equations of motion.
With the Mino-time \cite{Mino:2003yg} $d\gamma = \frac{(\pm1-y)}{D(\pm1,y)^{1/3}R^2 H(\pm1,y)} d\tau$ the equations of motion take the form
\begin{eqnarray}
 \frac{\mathrm{d}y}{\mathrm{d}\gamma} &=& -\sqrt{Y(y)} \label{eqn:charge-phi-y-gleichung}\\
 \frac{\mathrm{d}\psi}{\mathrm{d}\gamma} &=& -\frac{(\pm1-y)H(\pm1,y)H(y,\pm1)}{(1+\nu\pm\lambda)^2G(y)}(\Psi+c\Omega_\psi E) \label{eqn:charge-phi-psi-gleichung}\\
 \frac{\mathrm{d}t}{\mathrm{d}\gamma} &=&  R^2 E \frac{D(\pm1,y)H^2(\pm 1,y)}{(\pm 1-y)H(y,\pm1)} - \frac{(\pm1-y)H(\pm1,y)H(y,\pm1)}{(1+\nu\pm\lambda)^2G(y)}c\Omega_\psi(\Psi+c\Omega_\psi E)\, , \label{eqn:charge-phi-t-gleichung}
\end{eqnarray} 
where
\begin{eqnarray}
 Y(y) &=& (1-\nu)^2 \frac{H(\pm 1,y)}{H(y,\pm 1)}  \bigg\lbrace R^2 G(y) [D^{1/3}H(y,\pm 1)m^2-DH(\pm 1,y)E^2] \nonumber \\
 && - \frac{(\pm 1-y)^2}{(1\pm\lambda +\nu)^2} \bigg[ H(y,\pm 1)\Psi -cR\lambda \sqrt{2((1+\nu)^2-\lambda^2)}\frac{1+y}{(1-\lambda +\nu)} \nonumber \\
 && \cdot (1+\lambda -\nu +y\nu(1-\lambda -\nu)\pm 2\nu(1-y)) E \bigg] ^2 \bigg\rbrace
\end{eqnarray}
and
\begin{eqnarray}
  H(\pm 1,y) &=& (1\pm\lambda)^2 -\nu ^2+y^2\nu(1-\lambda ^2-\nu ^2\mp 2\lambda\nu) \nonumber \\
  H(y,\pm 1) &=& 1+\lambda ^2 -\nu ^2 \pm 2\lambda\nu(1-y^2)+2\lambda y(1-\nu ^2) + y^2\nu(1-\lambda^2-\nu^2)\nonumber \\
  D(\pm1,y) &=& 1+\frac{s^2}{H(\pm 1,y)}[2\lambda(1-\nu)(\pm1-y)(1\mp\nu y)] \nonumber \\
 \Omega_\psi &=& -\frac{R\lambda\sqrt{2((1+\nu)^2-\lambda ^2)}}{H(y,\pm 1)} \frac{1+y}{1-\lambda +\nu} (1+\lambda -\nu +y\nu(1-\lambda -\nu)\pm 2\nu(1-y)) \, .
\end{eqnarray}
Solving the equations of motion (\ref{eqn:charge-phi-y-gleichung})-(\ref{eqn:charge-phi-t-gleichung}) analytically is only possible if $Y(y)$ is a polynomial. This happens in two cases:
\begin{enumerate}
 \item $D(\pm1,y)=1$ (which implies $c=1$ and $s=0$): uncharged doubly spinning black ring
 \item $m=0$: charged doubly spinning black ring with photons
\end{enumerate}
In both cases $Y(y)$ is a polynomial of 6th order, so that the equations of motion are of hyperelliptic type (genus $g=2$).

\subsection{Classification of Geodesics}

From (\ref{eqn:charge-phi-y-gleichung}) we can read off an effective potential consisting of the two parts $V_+(y)$ and $V_-(y)$:
\begin{equation}
 Y=a(y)(E-V_+)(E-V_-) \, .
\end{equation}
Since $Y(y)$ can be written as $Y(y)=a(y)E^2+b(y)E+c(y)$ the effective potential takes the form
\begin{equation}
 V_\pm (y) = \frac{-b(y)\pm\sqrt{b(y)^2-4a(y)c(y)}}{2a(y)} \, ,
\end{equation}
where
\begin{equation}
 \begin{split}
  a(x) &= (1-\nu)^2 H(\pm1,y) H(y,\pm1) \left(-R^2 G(y) D(\pm1,y) H(\pm1,y)-\frac{c^2\Omega_\psi^2 H^2(y,\pm1)(\pm1-y)^2}{(1\pm\lambda+\nu)^2}\right)\\
  b(x)  &= -2(1-\nu)^2\Psi\frac{ c \Omega_\psi H^3(y,\pm1) H(\pm1,y) (\pm1-y)^2  }{(1\pm\lambda+\nu)^2}\\
  c(x)  &= (1-\nu)^2 H(\pm1,y) H(y,\pm1)\left(R^2G(y) D^{1/3}(\pm1,y) H(y,\pm1) m^2-\frac{(\pm1-y)^2 H^2(y,\pm1)\Psi^2}{(1\pm\lambda+\nu)^2}\right) \, .
 \end{split}
\end{equation}
The two cases $x=+1$ (geodesics inside the ring) and $x=-1$ (geodesics outside the ring) have to be discussed separately.

\subsubsection{Geodesics outside the ring}

\begin{figure}
 \centering
 \subfigure[$s=0,5$ and $\Psi=5$ \newline Examples of orbits of type A and B.]{
   \includegraphics[width=4.5cm]{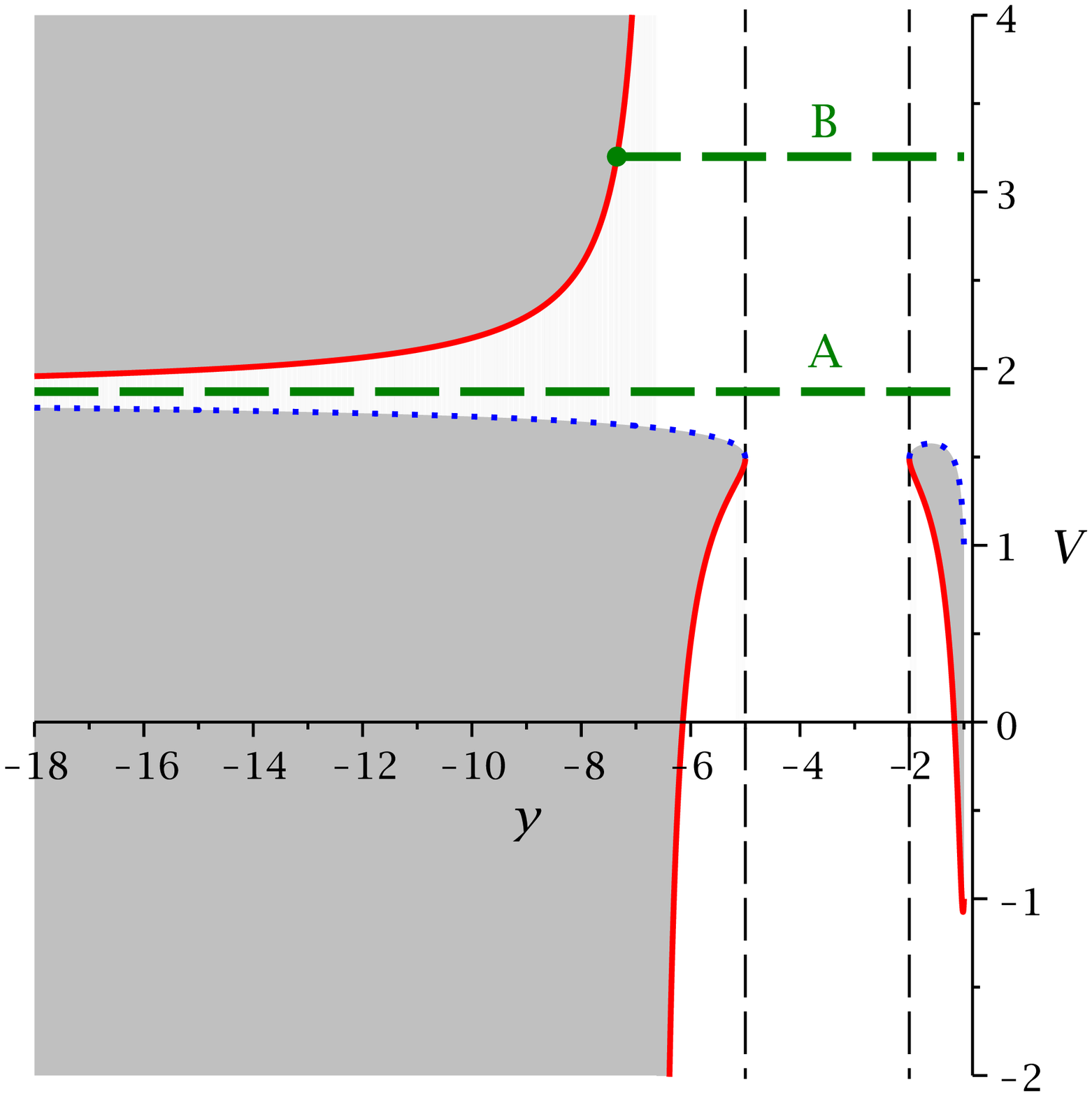}
 }
 \subfigure[$s=1$ and $\Psi=0$ \newline Examples of orbits of type C and C$_0$]{
   \includegraphics[width=4.5cm]{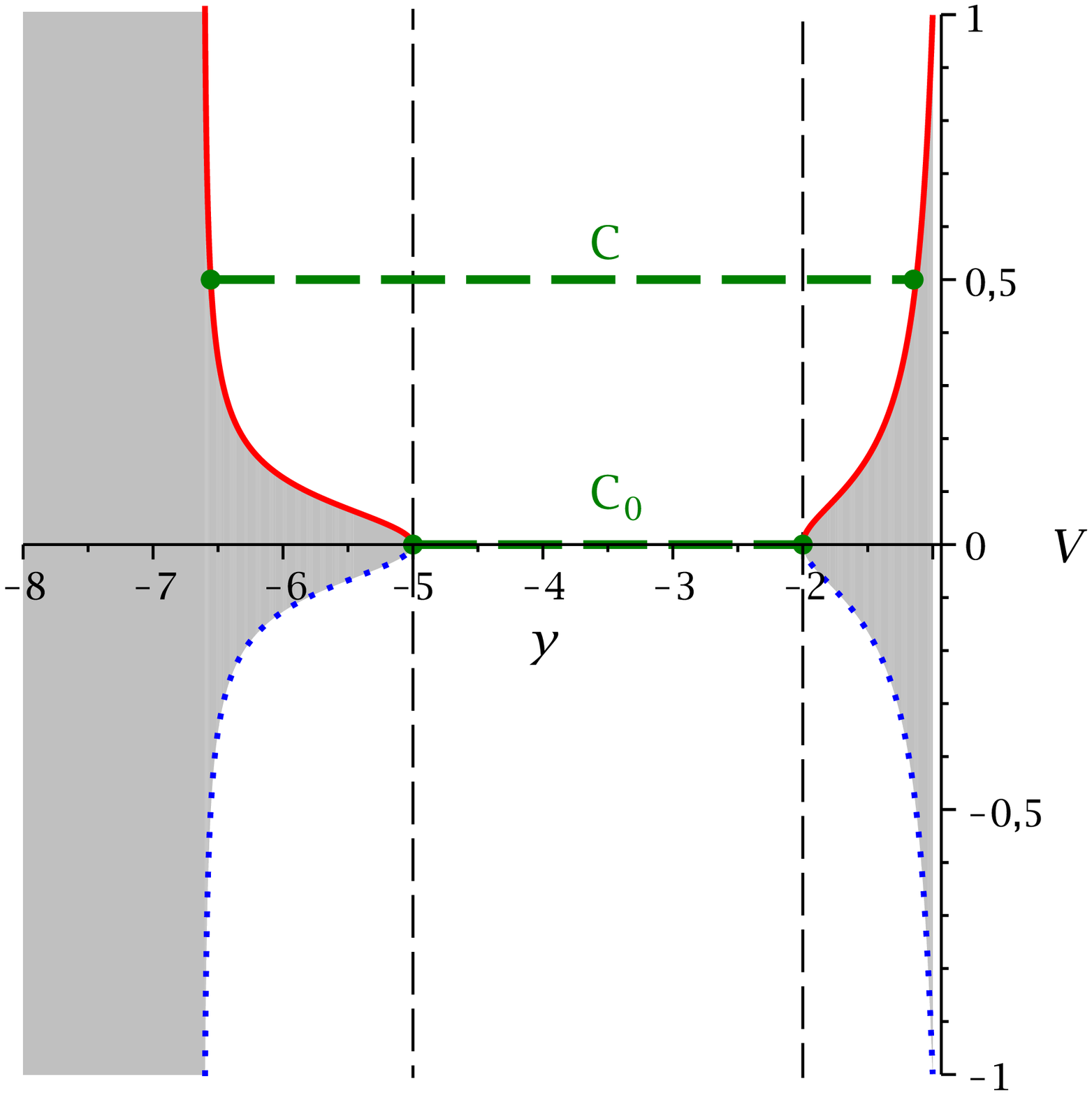}
 }
 \subfigure[$s=1$ and $\Psi=10$ \newline Examples of orbits of type D and D$_0$ ]{
   \includegraphics[width=4.5cm]{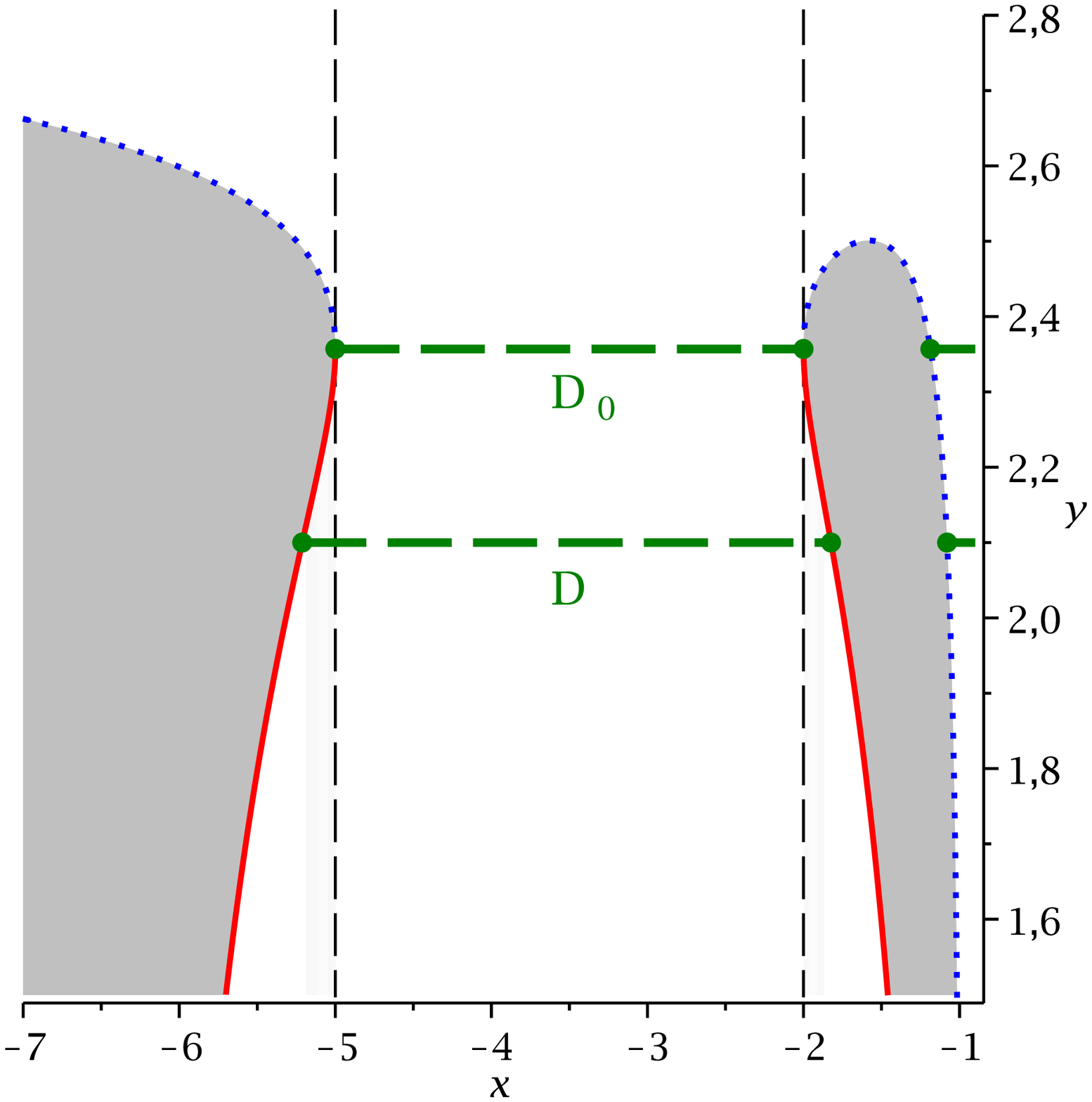}
 }
 \caption{$R=1$, $m=1$, $\nu =0.1$ and $\lambda=0.7$ \newline
 Effective potentials $V_+(y)$ (red solid line) and $V_-(y)$ (blue dotted line) on the $\phi$-axis outside the (charged) doubly spinning black ring. The grey area is a forbidden zone, where no motion is possible. The horizons are marked by vertical dashed lines. Horizontal green dashed lines represent energies and green points mark the turning points.}
 \label{pic:charge-phi-orbits1}
\end{figure}

Let us first take a look at the motion on the surface outside the black ring. Here we have $x=-1$. Figure \ref{pic:charge-phi-orbits1} shows the effective potential $V(y)$ for different values of the parameters. $V_+$ and $V_-$ meet at the horizons. Mainly the angular momentum $\Psi$ defines the shape of the effective potential, the charge parameter $s$ has less influence. For $\Psi =0$ the potential is symmetric with respect to the $y$-axis ($V(y)=0$) and $Y(y)$ has one or two zeros. A potential barrier prevents the test particles and light from falling into the singularity

If $|\Psi|>0$ the potential is no longer symmetric and up to three zeros of $Y(y)$ are possible. If $|\Psi|$ is large enough a gap in the potential barrier appears which allows test particles and light to reach the singularity.\\

Possible orbits outside the black ring are:
\begin{enumerate}
 \item \textit{Terminating Orbits} (TO):\\
  Light or test particles approach the black ring, cross both horizons and fall into the singularity.
 \item \textit{Many-World Bound Orbits} (MBO):\\
  Light or test particles circle the black ring on a periodic bound orbit, but cross both horizons several times on their flight. Everytime both horizons are traversed twice light or the test particles emerge into another universe.
 \item \textit{Escape Orbits} (EO):\\
  Light or test particles approach the black ring, turn around at a certain point and escape the gravitational field.
 \item \textit{Two-World Escape Orbits} (TEO):\\
  Light or test particles approach the black ring, cross both horizons turn around at a certain point, cross the horizons a second time and escape the gravitational field. Since both horizons are traversed twice light or the test particles emerge into another universe.
\end{enumerate}

There are four different types of orbits (see table \ref{tab:charge-phi-typen-orbits1}).
\begin{itemize}
 \item Type A:\\
  $Y(y)$ has no zeros and only TOs exist.
 \item Type B:\\
  $Y(y)$ has one zero and only TEOs exist.
 \item Type C:\\
  $Y(y)$ has two zeros and only MBOs exist. In a special case the two turning points lie on the horizons.
 \item Type D:\\
  $Y(y)$ has three zeros. MBOs and EOs exist. In a special case the two turning points of the MBO lie on the horizons.
\end{itemize}

\begin{table}[h!]
\begin{center}
\begin{tabular}{|lcll|}\hline
type &  zeros  & range of $y$ & orbit \\
\hline\hline
A & 0 & 
\begin{pspicture}(-2.5,-0.2)(3,0.2)
\psline[linewidth=0.5pt]{|->}(-2.5,0)(3,0)
\psline[linewidth=0.5pt,doubleline=true](1.0,-0.2)(1.0,0.2)
\psline[linewidth=0.5pt,doubleline=true](-0.5,-0.2)(-0.5,0.2)
\psline[linewidth=1.2pt]{-}(-2.5,0)(3,0)
\end{pspicture}
  & TO
\\  \hline
B  & 1 &
\begin{pspicture}(-2.5,-0.2)(3,0.2)
\psline[linewidth=0.5pt]{|->}(-2.5,0)(3,0)
\psline[linewidth=0.5pt,doubleline=true](1.0,-0.2)(1.,0.2)
\psline[linewidth=0.5pt,doubleline=true](-0.5,-0.2)(-0.5,0.2)
\psline[linewidth=1.2pt]{*-}(-1,0)(3,0)
\end{pspicture}
& TEO 
\\ \hline
C & 2 & 
\begin{pspicture}(-2.5,-0.2)(3,0.2)
\psline[linewidth=0.5pt]{|->}(-2.5,0)(3,0)
\psline[linewidth=0.5pt,doubleline=true](1.0,-0.2)(1.0,0.2)
\psline[linewidth=0.5pt,doubleline=true](-0.5,-0.2)(-0.5,0.2)
\psline[linewidth=1.2pt]{*-*}(-1.0,0)(1.5,0)
\end{pspicture}
  & MBO 
\\ \hline
C$_0$ & 2 & 
\begin{pspicture}(-2.5,-0.2)(3,0.2)
\psline[linewidth=0.5pt]{|->}(-2.5,0)(3,0)
\psline[linewidth=0.5pt,doubleline=true](1.0,-0.2)(1.0,0.2)
\psline[linewidth=0.5pt,doubleline=true](-0.5,-0.2)(-0.5,0.2)
\psline[linewidth=1.2pt]{*-*}(-0.5,0)(1.0,0)
\end{pspicture}
  & MBO
\\ \hline
D & 3 & 
\begin{pspicture}(-2.5,-0.2)(3,0.2)
\psline[linewidth=0.5pt]{|->}(-2.5,0)(3,0)
\psline[linewidth=0.5pt,doubleline=true](1.0,-0.2)(1.0,0.2)
\psline[linewidth=0.5pt,doubleline=true](-0.5,-0.2)(-0.5,0.2)
\psline[linewidth=1.2pt]{*-*}(-1.0,0)(1.5,0)
\psline[linewidth=1.2pt]{*-}(2.0,0)(3,0)
\end{pspicture}
& MBO, EO 
\\  \hline
D$_0$ & 3 & 
\begin{pspicture}(-2.5,-0.2)(3,0.2)
\psline[linewidth=0.5pt]{|->}(-2.5,0)(3,0)
\psline[linewidth=0.5pt,doubleline=true](1.0,-0.2)(1.0,0.2)
\psline[linewidth=0.5pt,doubleline=true](-0.5,-0.2)(-0.5,0.2)
\psline[linewidth=1.2pt]{*-*}(-0.5,0)(1.0,0)
\psline[linewidth=1.2pt]{*-}(2.0,0)(3,0)
\end{pspicture}
& MBO, EO 
\\ \hline\hline
\end{tabular}
\caption{Types of orbits of light and particles in the (charged) doubly spinning black ring spacetime for $x=-1$ and $\Phi =0$. The thick lines represent the range of $y$ and the turning points are shown by thick dots. The horizons are indicated by a vertical double line. The single vertical line at the left end is the singularity. The coordinate $y$ ranges from $-\infty$ to $-1$.}
\label{tab:charge-phi-typen-orbits1}
\end{center}
\end{table}

\subsubsection{Geodesics inside the ring}

\begin{figure}
 \centering
 \subfigure[$\nu=0,1$, $\lambda=0.7$, $m=1$ and $\Psi=0$ \newline Examples of orbits of type A,C, D and D$_0$.]{
   \includegraphics[width=4.5cm]{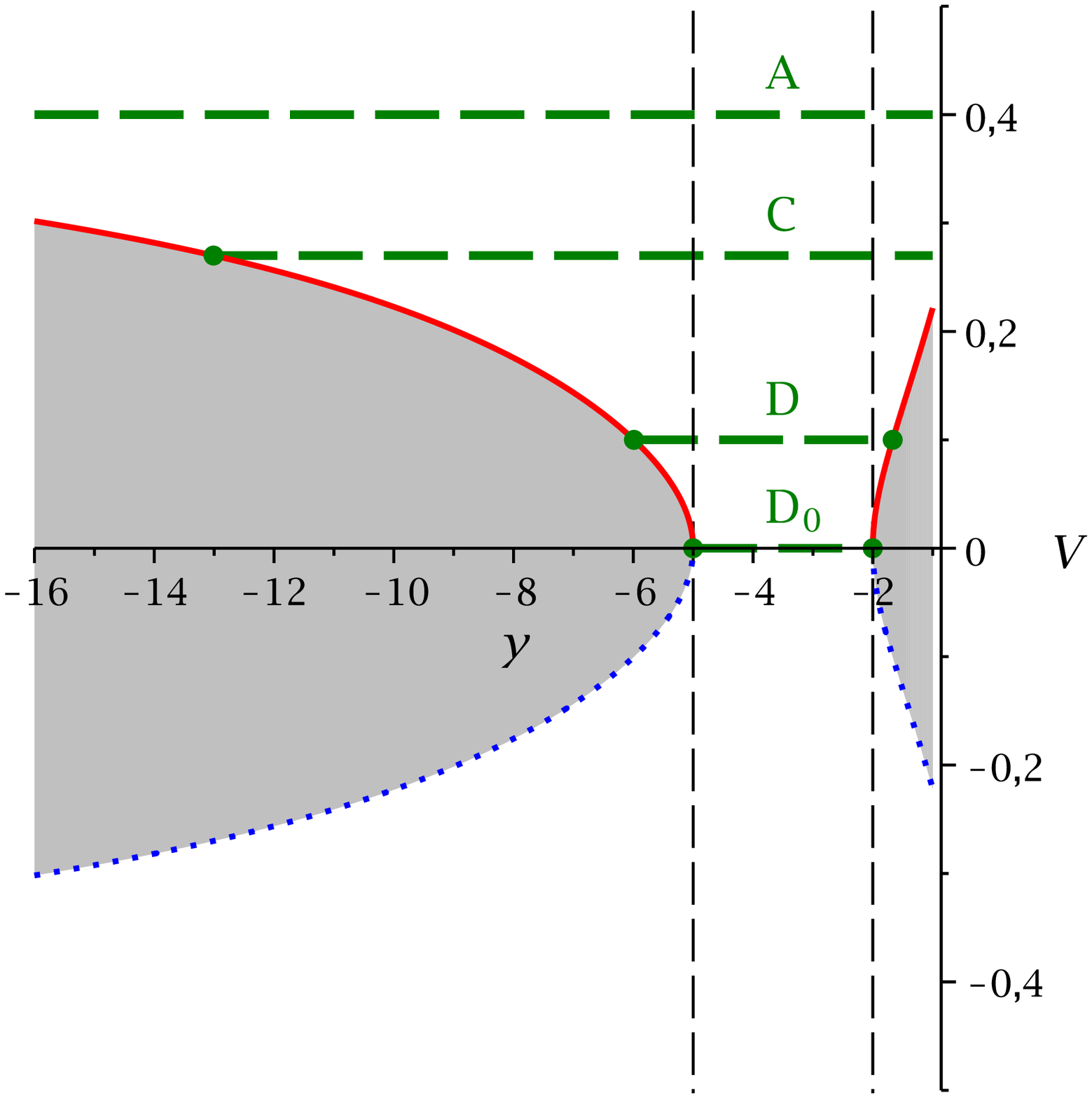}
 }
 \subfigure[$\nu=0,2$, $\lambda=0.9$, $m=0$ and $\Psi=5$  \newline Examples of orbits of type B,E and E$_0$]{
   \includegraphics[width=4.5cm]{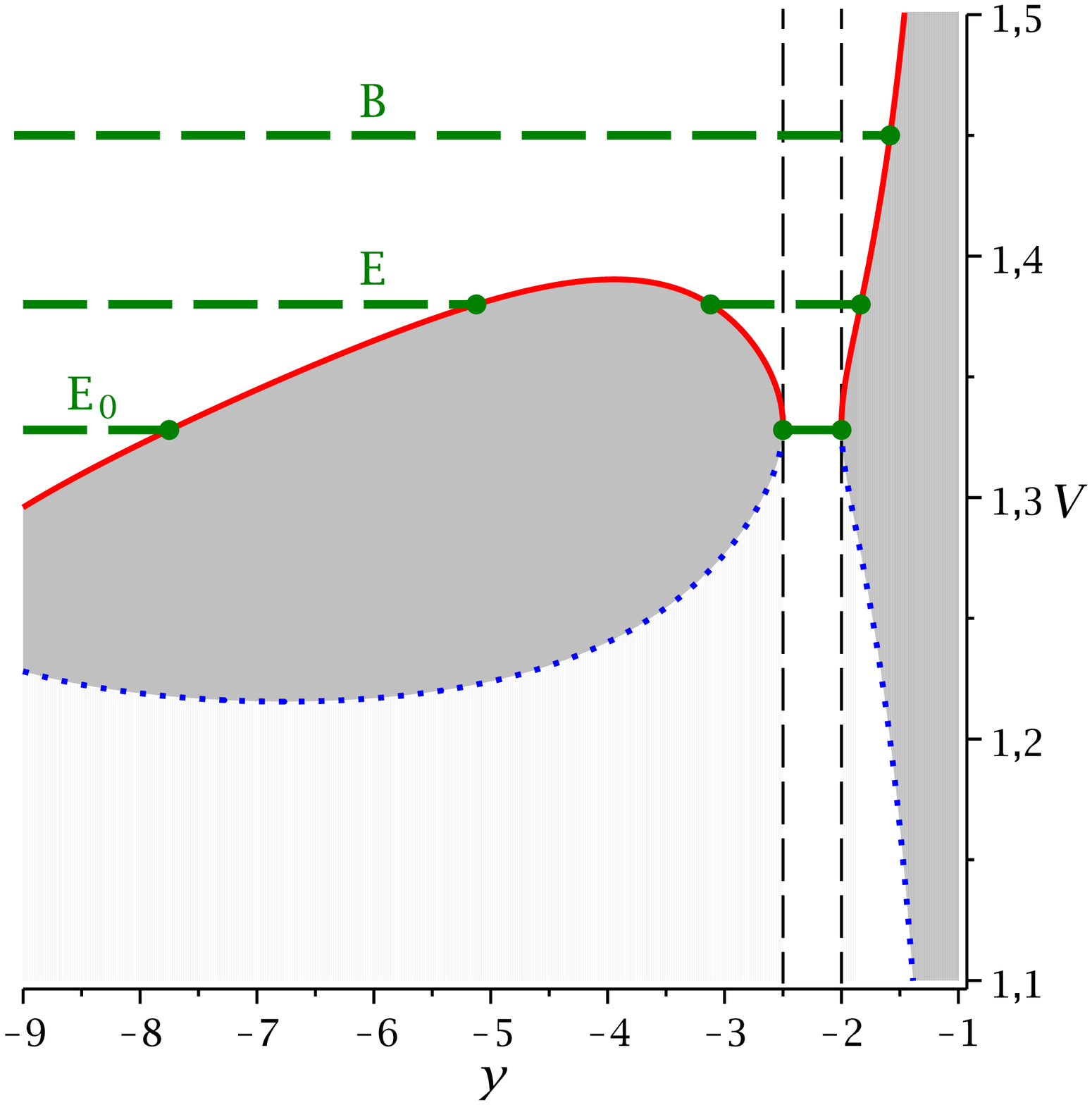}
 }
 \subfigure[$\nu=0,2$, $\lambda=0.9$, $m=1$ and $\Psi=20$ \newline Examples of orbits of type F.]{
   \includegraphics[width=4.5cm]{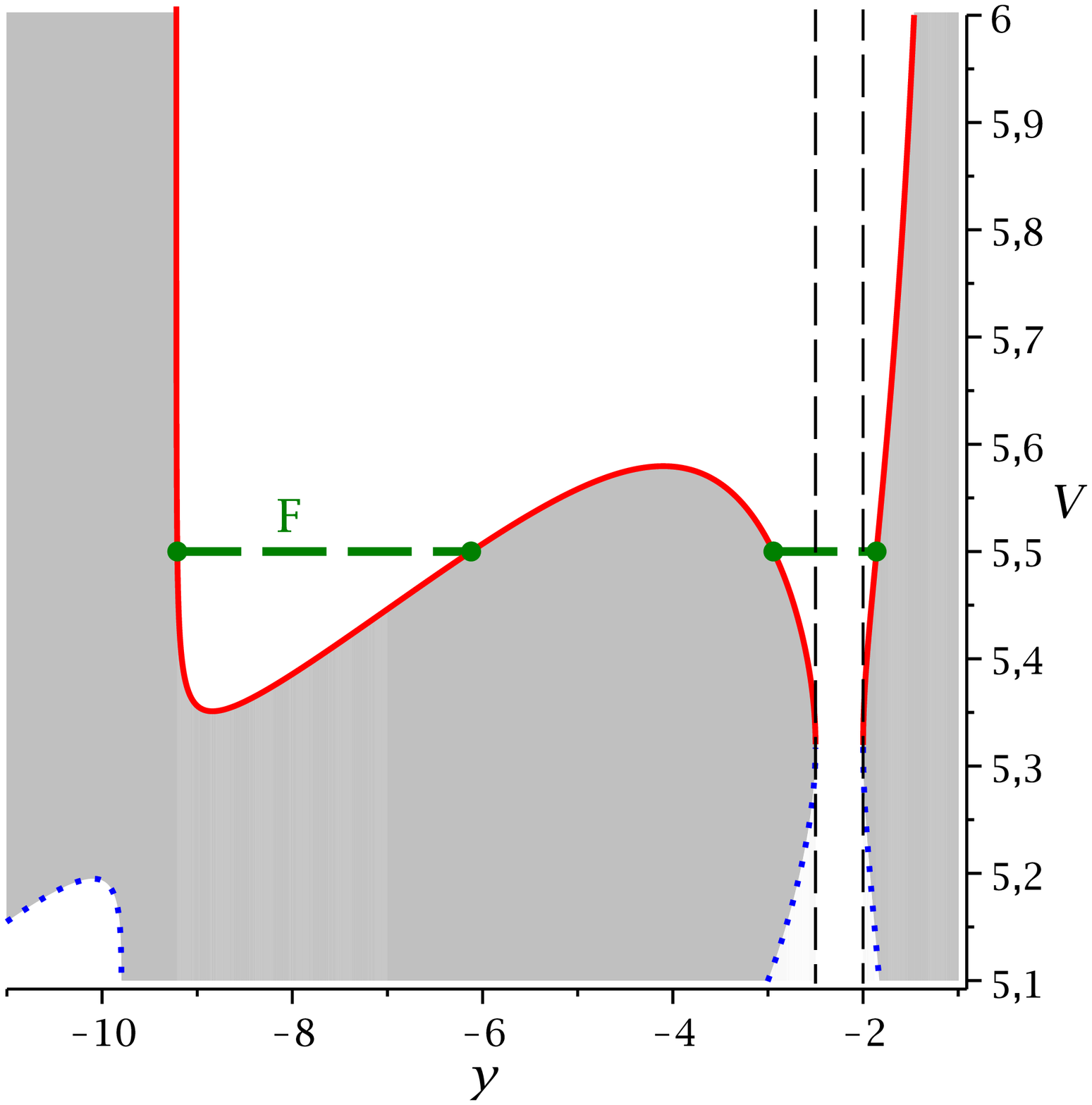}
 }
 \caption{$R=1$ and $s=0,1$ \newline
 Effective potentials $V_+(y)$ (red solid line) and $V_-(y)$ (blue dotted line) on the $\phi$-axis inside the (charged) doubly spinning black ring ($x=+1$). The grey area is a forbidden zone, where no motion is possible. The horizons are marked by vertical dashed lines. Horizontal green dashed lines represent energies and green points mark the turning points.}
 \label{pic:charge-phi-orbits2}
\end{figure}

The effective potential for geodesics on the surface enclosed by the black ring ($x=+1$) is shown in figure \ref{pic:charge-phi-orbits2}. Again, if we have $\Psi =0$ the potential is symmetric and $Y(y)$ has none or one zeros. In the case $|\Psi|>0$ a potential barrier appears which prevents test particles and light from reaching $y=-1$. In contrast to the motion of light and test particles outside the black ring, the charge parameter $s$ has now significant influence on the effective potential and the possible orbits. If $s>0$ up to four zeros of $Y(y)$ are possible, if $s=0$ a maximum of three zeros is possible. In the case of four zeros, bound orbits of test particles (not light) behind the inner horizon are possible.\\

Possible orbits are Terminating Orbits (TO), Many-World Bound Orbits (MBO) and Bound Orbits (BO). The bound orbits, which are only possible in the charged case, exist behind the inner horizon of the charged doubly spinning black ring.

There are six different types of orbits (see table \ref{tab:charge-phi-typen-orbits2}).
\begin{itemize}
 \item Type A:\\
  $Y(y)$ has no zeros and only TOs exist. Test particles or light come from infinity ($y=-1$) and fall into the singularity ($y=-\infty$).
 \item Type B:\\
  $Y(y)$ has one zero and only TOs exist. The orbits starts at a certain point and ends in the singularity.
 \item Type C:\\
  $Y(y)$ has one zero and only MBOs exist. Light or a test particle crosses the center of the ring ($x=+1, y=-1$). 
 \item Type D:\\
  $Y(y)$ has two zeros and only MBOs exist. In a special case the two turning points lie on the horizons.
 \item Type E:\\
  $Y(y)$ has three zeros. MBOs and TOs exist. In a special case the two turning points of the MBO lie on the horizons.
 \item Typ F:\\
  $Y(y)$ has four zeros. MBOs and BOs exist. The BOs lie behind the inner horizon and are only possible if the doubly spinning black ring is charged and $m\neq0$.
\end{itemize}

\begin{table}[h!]
\begin{center}
\begin{tabular}{|lcll|}\hline
type &  zeros  & range of $y$ & orbit \\
\hline\hline
A & 0 & 
\begin{pspicture}(-2.5,-0.2)(3,0.2)
\psline[linewidth=0.5pt]{|->}(-2.5,0)(3,0)
\psline[linewidth=0.5pt,doubleline=true](1.0,-0.2)(1.0,0.2)
\psline[linewidth=0.5pt,doubleline=true](-0.5,-0.2)(-0.5,0.2)
\psline[linewidth=1.2pt]{-}(-2.5,0)(3,0)
\end{pspicture}
  & TO
\\ \hline
B  & 1 &
\begin{pspicture}(-2.5,-0.2)(3,0.2)
\psline[linewidth=0.5pt]{|->}(-2.5,0)(3,0)
\psline[linewidth=0.5pt,doubleline=true](1.0,-0.2)(1.,0.2)
\psline[linewidth=0.5pt,doubleline=true](-0.5,-0.2)(-0.5,0.2)
\psline[linewidth=1.2pt]{-*}(-2.5,0)(1.5,0)
\end{pspicture}
& TO 
\\  \hline
C  & 1 &
\begin{pspicture}(-2.5,-0.2)(3,0.2)
\psline[linewidth=0.5pt]{|->}(-2.5,0)(3,0)
\psline[linewidth=0.5pt,doubleline=true](1.0,-0.2)(1.,0.2)
\psline[linewidth=0.5pt,doubleline=true](-0.5,-0.2)(-0.5,0.2)
\psline[linewidth=1.2pt]{*-}(-1,0)(3,0)
\end{pspicture}
& MBO 
\\ \hline
D & 2 & 
\begin{pspicture}(-2.5,-0.2)(3,0.2)
\psline[linewidth=0.5pt]{|->}(-2.5,0)(3,0)
\psline[linewidth=0.5pt,doubleline=true](1.0,-0.2)(1.0,0.2)
\psline[linewidth=0.5pt,doubleline=true](-0.5,-0.2)(-0.5,0.2)
\psline[linewidth=1.2pt]{*-*}(-1.0,0)(1.5,0)
\end{pspicture}
  & MBO 
\\ \hline
D$_0$ & 2 & 
\begin{pspicture}(-2.5,-0.2)(3,0.2)
\psline[linewidth=0.5pt]{|->}(-2.5,0)(3,0)
\psline[linewidth=0.5pt,doubleline=true](1.0,-0.2)(1.0,0.2)
\psline[linewidth=0.5pt,doubleline=true](-0.5,-0.2)(-0.5,0.2)
\psline[linewidth=1.2pt]{*-*}(-0.5,0)(1.0,0)
\end{pspicture}
  & MBO
\\ \hline
E & 3 & 
\begin{pspicture}(-2.5,-0.2)(3,0.2)
\psline[linewidth=0.5pt]{|->}(-2.5,0)(3,0)
\psline[linewidth=0.5pt,doubleline=true](1.0,-0.2)(1.0,0.2)
\psline[linewidth=0.5pt,doubleline=true](-0.5,-0.2)(-0.5,0.2)
\psline[linewidth=1.2pt]{*-*}(-1.0,0)(1.5,0)
\psline[linewidth=1.2pt]{-*}(-2.5,0)(-1.5,0)
\end{pspicture}
& TO, MBO
\\  \hline
E$_0$ & 3 & 
\begin{pspicture}(-2.5,-0.2)(3,0.2)
\psline[linewidth=0.5pt]{|->}(-2.5,0)(3,0)
\psline[linewidth=0.5pt,doubleline=true](1.0,-0.2)(1.0,0.2)
\psline[linewidth=0.5pt,doubleline=true](-0.5,-0.2)(-0.5,0.2)
\psline[linewidth=1.2pt]{*-*}(-0.5,0)(1.0,0)
\psline[linewidth=1.2pt]{-*}(-2.5,0)(-1.5,0)
\end{pspicture}
& TO, MBO
\\ \hline
F & 4 & 
\begin{pspicture}(-2.5,-0.2)(3,0.2)
\psline[linewidth=0.5pt]{|->}(-2.5,0)(3,0)
\psline[linewidth=0.5pt,doubleline=true](1.0,-0.2)(1.0,0.2)
\psline[linewidth=0.5pt,doubleline=true](-0.5,-0.2)(-0.5,0.2)
\psline[linewidth=1.2pt]{*-*}(-1.0,0)(1.5,0)
\psline[linewidth=1.2pt]{*-*}(-2,0)(-1.5,0)
\end{pspicture}
& BO, MBO
\\ \hline\hline
\end{tabular}
\caption{Types of orbits of light and particles in the (charged) doubly spinning black ring spacetime for $x=+1$ and $\Phi =0$. The thick lines represent the range of $y$ and the turning points are shown by thick dots. The horizons are indicated by a vertical double line. The single vertical line at the left end is the singularity. The coordinate $y$ ranges from $-\infty$ to $-1$.}
\label{tab:charge-phi-typen-orbits2}
\end{center}
\end{table}

\subsection{Solution of the $y$-equation}
\label{sec:charge-phi-y-solution}

If $D(\pm1,y)=1$ (i.e. $s=0$) or $m=0$ then $Y(y)$ is a polynomial of sixth order and equation (\ref{eqn:charge-phi-y-gleichung}) can be solved analogously to (\ref{eqn:charge-psi-x-gleichung}). A separation of variables yields 
\begin{equation}
 \gamma - \gamma_{\rm in} = -\int _{y_{\rm in}}^y \! \frac{\mathrm{d}y'}{\sqrt{Y(y')}} \, .
\end{equation}
The solution is
\begin{equation}
 y(\gamma)=-\frac{\sigma_{1}(\boldsymbol{\gamma}_\Theta)}{\sigma_{2}(\boldsymbol{\gamma}_\Theta)}.
\end{equation}
where
\begin{equation}
 \boldsymbol{\gamma}_\Theta :=
 \left(
 \begin{array}{c}
 -\gamma + \gamma_{\rm in}' \\
 \gamma_2 
 \end{array}
 \right) = \boldsymbol{\varphi}_\infty .
\end{equation}
The constant $\gamma_{\rm in}' = \gamma_{\rm in} -\int_{y_{\rm in}}^{\infty}\! \frac{\mathrm{d}y'}{\sqrt{Y(y')}}$ depends on $ \gamma_{\rm in}$ and $y_{\rm in}$ only. $\gamma_2$ is defined by the vanishing condition of the Kleinian sigma function: $\sigma (\boldsymbol{\gamma}_\Theta) = 0$ so that $(2\omega )^{-1}\boldsymbol{\gamma}_\Theta$ is an element of the theta divisor (the set of zeros of the theta function).

\subsection{Solution of the $\psi$-equation}
\label{sec:charge-phi-psisolution}

With (\ref{eqn:charge-phi-y-gleichung}) equation (\ref{eqn:charge-phi-psi-gleichung}) yields
\begin{equation}
 \mathrm{d}\psi = -\frac{(\pm1-y)H(\pm1,y)H(y,\pm1)}{(1+\nu\pm\lambda)^2G(y)}(\Psi+c\Omega_\psi E) \frac{\mathrm{d}y}{\sqrt{Y(y)}}\, .
\end{equation}
This can be rewritten as
\begin{equation}
 \mathrm{d}\psi =\frac{P_4(y)}{P_3(y)} \frac{\mathrm{d}y}{\sqrt{Y(y)}}\, ,
\end{equation}
where $P_4(y)$ is a polynomial of sixth order and $P_3(y)$ is a polynomial of third order. So we have to solve the integral
\begin{equation}
 \psi - \psi_{\rm in} = \int _{y_{\rm in}}^y \! \frac{P_4(y')}{P_3(y')} \, \frac{\mathrm{d}y'}{\sqrt{Y(y')}}
\label{eqn:phi-psiint}
\end{equation}
which has poles at $p_{1,2}=\frac{-\lambda\pm\sqrt{\lambda^2-4\nu}}{2\nu}=y_{h\pm}$ and $p_3=-1$ (if $x=+1$) or $p_3=1$ (if $x=-1$). We apply a partial fractions decomposition upon (\ref{eqn:phi-psiint}):
\begin{equation}
 \psi - \psi_{\rm in} = \int _{y_{\rm in}}^y \! \left( \sum _{i=1}^3 \frac{K_i}{y'-p_i} + K_4 + K_5 y' \right) \, \frac{\mathrm{d}y'}{\sqrt{Y(y')}}\, ,
 \label{eqn:charge-psi-pbz}
\end{equation}
where $K_i$ are constants which arise from the partial fractions decomposition.

If $D(\pm1,y)=1$ (i.e. $s=0$) or $m=0$ then $Y(y)$ is a polynomial of sixth order and equation (\ref{eqn:charge-psi-pbz}) consists of hyperelliptic integrals of the first and third kind. The solution can be found in the same way as in section \ref{sec:charge-ergo-phi} or \ref{sec:charge-psi-phisolution}:
\begin{equation}
\begin{split}
 \psi &=\psi_{\rm in} +  \sum _{i=1}^3 K_i \left[ \frac{2}{W_i} \left(\int _{y_{\rm in}}^y d\boldsymbol{u}\right)^T \left( \boldsymbol{\zeta} \left( \int_{(e_2,0)}^{(p_i,W_i)} \mathrm{d} \boldsymbol{u} + \boldsymbol{K}_\infty  \right) - 2( \boldsymbol{\eta}^{\prime}\boldsymbol{\varepsilon}^\prime + \boldsymbol{\eta}\boldsymbol{\varepsilon} )  - \frac12 \boldsymbol{\mathfrak{Z}}(p_i,W_i)  \right) \right. \\ 
& \left. + \ln\frac{\sigma\left(  W^2(y)  \right)}{\sigma\left( W^1(y) \right)}
-  \ln \frac{\sigma\left(  W^2(y_{\rm in})  \right)}{\sigma\left( W^1(y_{\rm in}) \right)}  \right] + K_4(v - v_0) + K_5f_1(v - v_0) \, ,
\end{split}
\end{equation}
where $W_i=\sqrt{Y(p_i)}$ and $W^{1,2}(y) = \int^{y}_{\infty}{d\boldsymbol{u}} \pm  \int_{(e_2,0)}^{(p_i,W_i)} \mathrm{d} \boldsymbol{u} - \boldsymbol{K}_\infty $.

\subsection{Solution of the $t$-equation}

With (\ref{eqn:charge-phi-y-gleichung}) equation (\ref{eqn:charge-phi-t-gleichung}) yields
\begin{equation}
 \mathrm{d}t= \left( R^2 E\frac{D(\pm1,y)H^2(\pm1,y)}{(\pm1-y)H(y,\pm1)}- \frac{(\pm1-y)H(\pm1,y)H(y,\pm1)}{(1+\nu\pm\lambda)^2G(y)}c\Omega_\psi(\Psi +c\Omega_\psi E) \right) \frac{\mathrm{d}y}{\sqrt{Y(y)}} \, .
\end{equation}
This can be rewritten as
\begin{equation}
 \mathrm{d}t = \frac{P_7(y)}{P_6(y)} \frac{\mathrm{d}y}{\sqrt{Y(y)}} \, ,
\end{equation}
where $P_d(y)$ are polynomials of order $d$. $P_6(y)$ has the zeros $p_i$ ($i=1..6$).
So we have to solve the integral
\begin{equation}
 t - t_{\rm in} = \int _{y_{\rm in}}^y \! \frac{P_7(y')}{P_6(y')} \, \frac{\mathrm{d}y'}{\sqrt{Y(y')}} \, .
\end{equation}
We apply a partial fractions decomposition where the constants $M_i$ arise:
\begin{equation}
 t - t_{\rm in} = \left( \sum _{i=1}^6 \frac{M_i}{y'-p_i} + M_7 + M_8 y' \right)  \, \frac{\mathrm{d}y'}{\sqrt{Y(y')}} \, .
\end{equation}
If $D(\pm1,y)=1$ (i.e. $s=0$) or $m=0$ then $Y(y)$ is a polynomial of sixth order and the equation can be solved.
The hyperelliptic integrals of the first and third kind can be solved as shown in section \ref{sec:charge-ergo-phi} or \ref{sec:charge-psi-phisolution}. The solution of the $t$-equation (\ref{eqn:charge-phi-t-gleichung}) is
\begin{equation}
\begin{split}
 t &=t_{\rm in} + \left\lbrace  \sum _{i=1}^6 M_i \left[ \frac{2}{W_i} \left(\int _{y_{\rm in}}^y d\boldsymbol{u}\right)^T \left( \boldsymbol{\zeta} \left( \int_{(e_2,0)}^{(p_i,W_i)} \mathrm{d} \boldsymbol{u} + \boldsymbol{K}_\infty  \right) - 2( \boldsymbol{\eta}^{\prime}\boldsymbol{\varepsilon}^\prime + \boldsymbol{\eta}\boldsymbol{\varepsilon} )  - \frac12 \boldsymbol{\mathfrak{Z}}(p_i,W_i)  \right) \right.  \right.\\ 
& \left.\left. + \ln\frac{\sigma\left(  W^2(y)  \right)}{\sigma\left( W^1(y) \right)}
-  \ln \frac{\sigma\left(  W^2(y_{\rm in})  \right)}{\sigma\left( W^1(y_{\rm in}) \right)}  \right] + M_7(v - v_0) + M_8f_1(v - v_0) \right\rbrace
\end{split} \, ,
\end{equation}
where $W_i=\sqrt{Y(p_i)}$ and $W^{1,2}(y) = \int^{y}_{\infty}{d\boldsymbol{u}} \pm  \int_{(e_2,0)}^{(p_i,W_i)} \mathrm{d} \boldsymbol{u} - \boldsymbol{K}_\infty $.

\section{The orbits}
\label{sec:orbits}

One can think of ring coordinates as two pairs of polar coordinates
\begin{equation}
\begin{array}{l} 
x_1=r_1 \sin(\phi)\\
x_2=r_1 \cos(\phi)
\end{array}
\quad\text{and}\quad
\begin{array}{l} 
x_3=r_2 \sin(\psi)\\
x_4=r_2 \cos(\psi)
\end{array}
\end{equation}
where
\begin{equation}
 r_1=R\frac{\sqrt{1-x^2}}{x-y} \quad\text{and}\quad r_2=R\frac{\sqrt{y^2-1}}{x-y}
\end{equation}
(see \cite{Hoskisson:2007zk,Emparan:2006mm}). $x_1$, $x_2$, $x_3$, $x_4$ are four-dimensional Cartesian coordinates.\\
If $\psi$ is constant, the horizon of the black ring consists of two $S^2$ spheres. If we look at the rotational axis where $y=-1$, the coordinates $x_1$ and $x_2$ describe the plane between these two spheres, so the horizons and the singularity cannot be seen in this plane.\\
If $\phi$ is constant, the horizon has $S^1\times S^1$ topology. So if $x=\pm1$ the coordinates $x_3$ and $x_4$ describe the equatorial plane ``as seen from above'', so each horizon consists of two circles. The singularity is a circle of radius $1$.\\
If both angles are constant, the coordinates describe the $x$-$y$-plane. Here we can change from ring coordinates ($x$,$y$) to  polar coordinates ($\rho$, $\theta$) via the transformation
\begin{equation}
 \rho=\frac{R\sqrt{y^2-x^2}}{x-y}\, ,\quad \tan \theta =\sqrt{\frac{y^2-1}{1-x^2}} \, .
\end{equation}
Then conventional Cartesian coordinates take the form
\begin{equation}
 a = \rho\sin\theta \, ,\quad b = \rho\cos\theta
\end{equation}
(see \cite{Hoskisson:2007zk}). The singularity of the black ring is at $a=\pm 1$, $b=0$.\\
Note that $a=r_2$ and $b=r_1$ so that
\begin{equation}
\begin{array}{l} 
x_1=\rho\cos(\theta)\sin(\phi)\\
x_2=\rho\cos(\theta)\cos(\phi)\\
x_3=\rho\sin(\theta)\sin(\psi)\\
x_4=\rho\sin(\theta)\cos(\psi) \, .
\end{array}
\end{equation}

\begin{figure}
 \centering
 \subfigure[$x_1$-$x_3$-$x_4$-plot ($\phi=\frac{\pi}{2}$)]{
   \includegraphics[width=6cm]{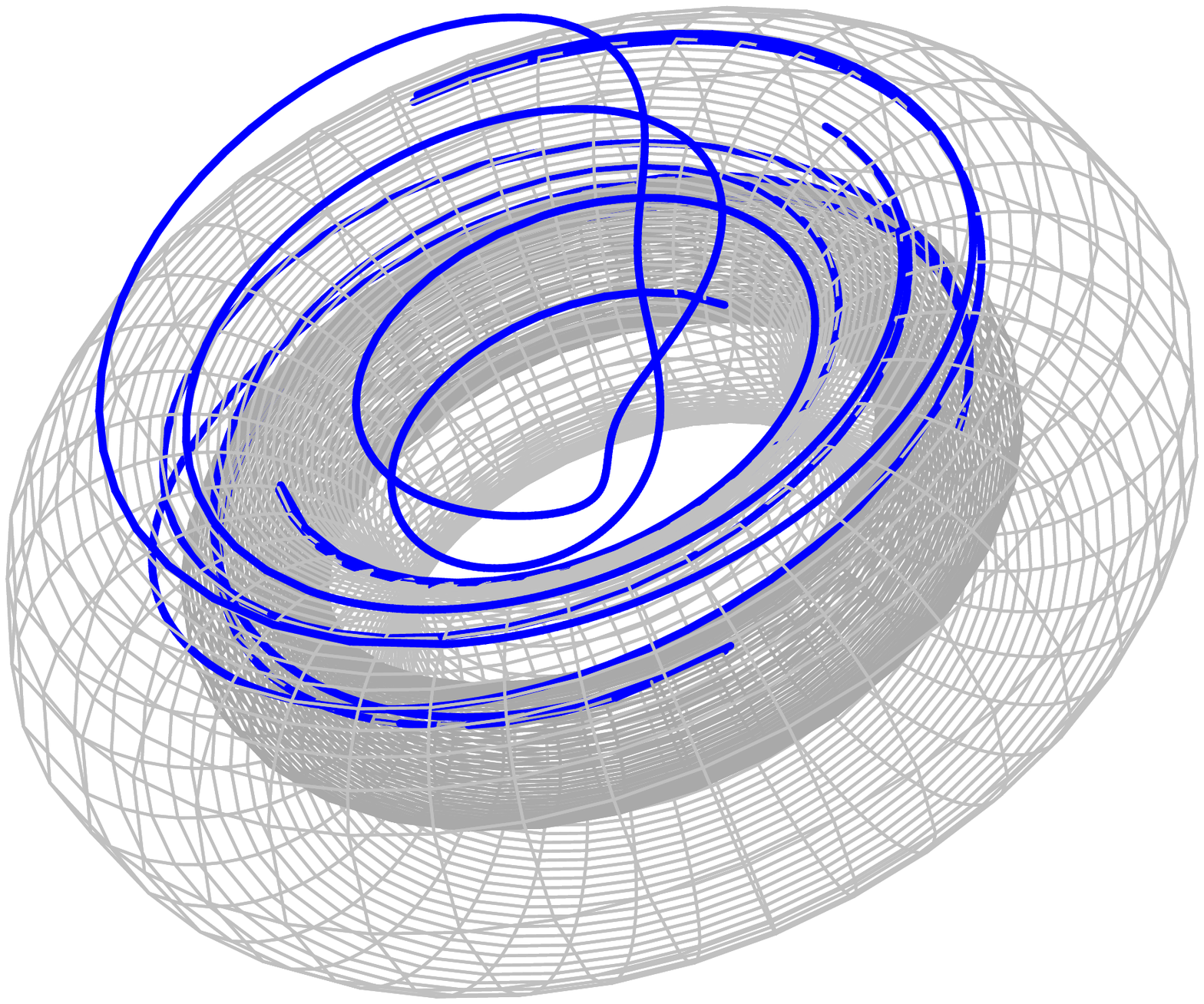}
 }
 \qquad\qquad
 \subfigure[$x_1$-$x_2$-$x_3$-plot ($\psi=\frac{\pi}{2}$)]{
   \includegraphics[width=6cm]{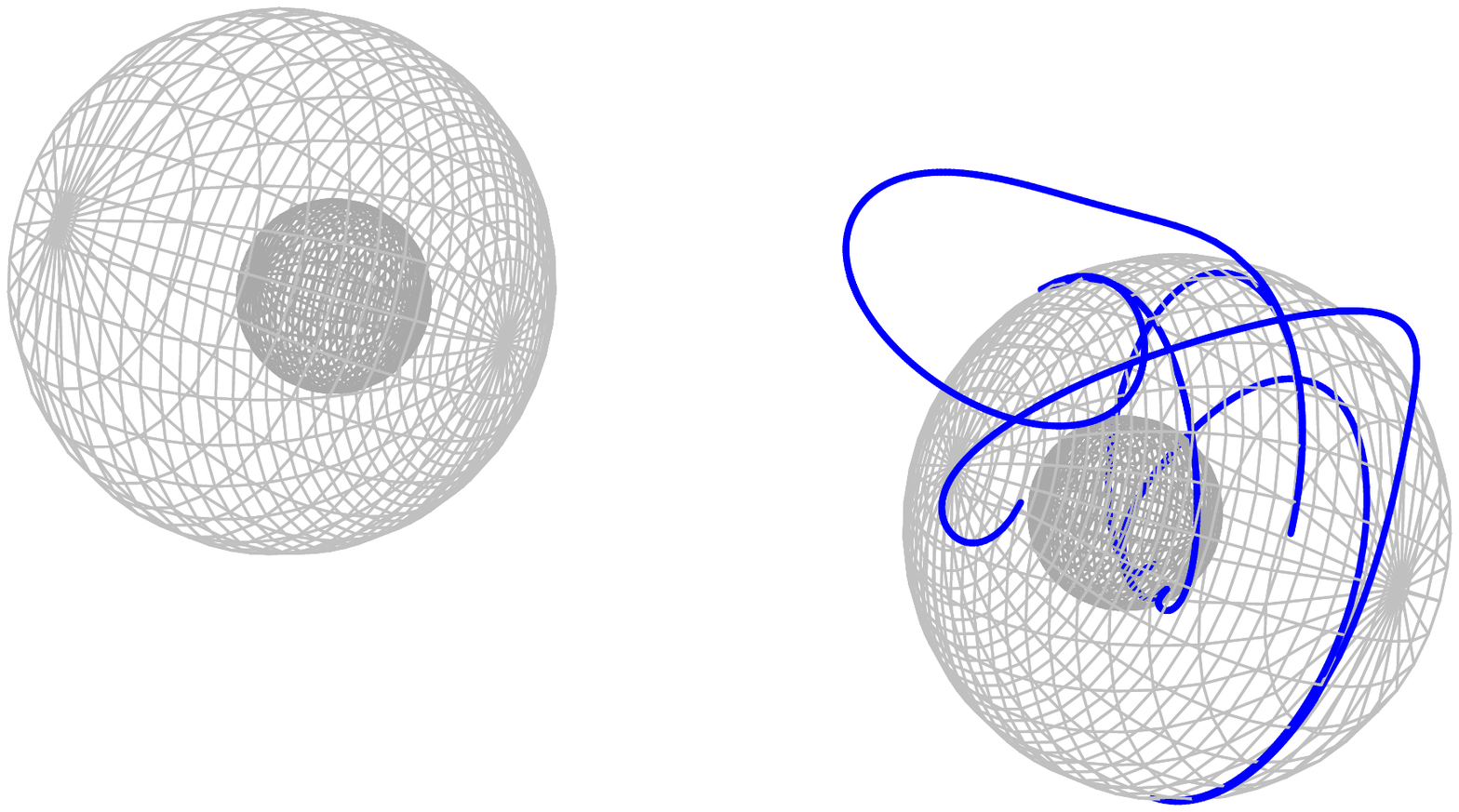}
 }
  \subfigure[$x_3$-$x_4$-plot]{
   \includegraphics[width=6cm]{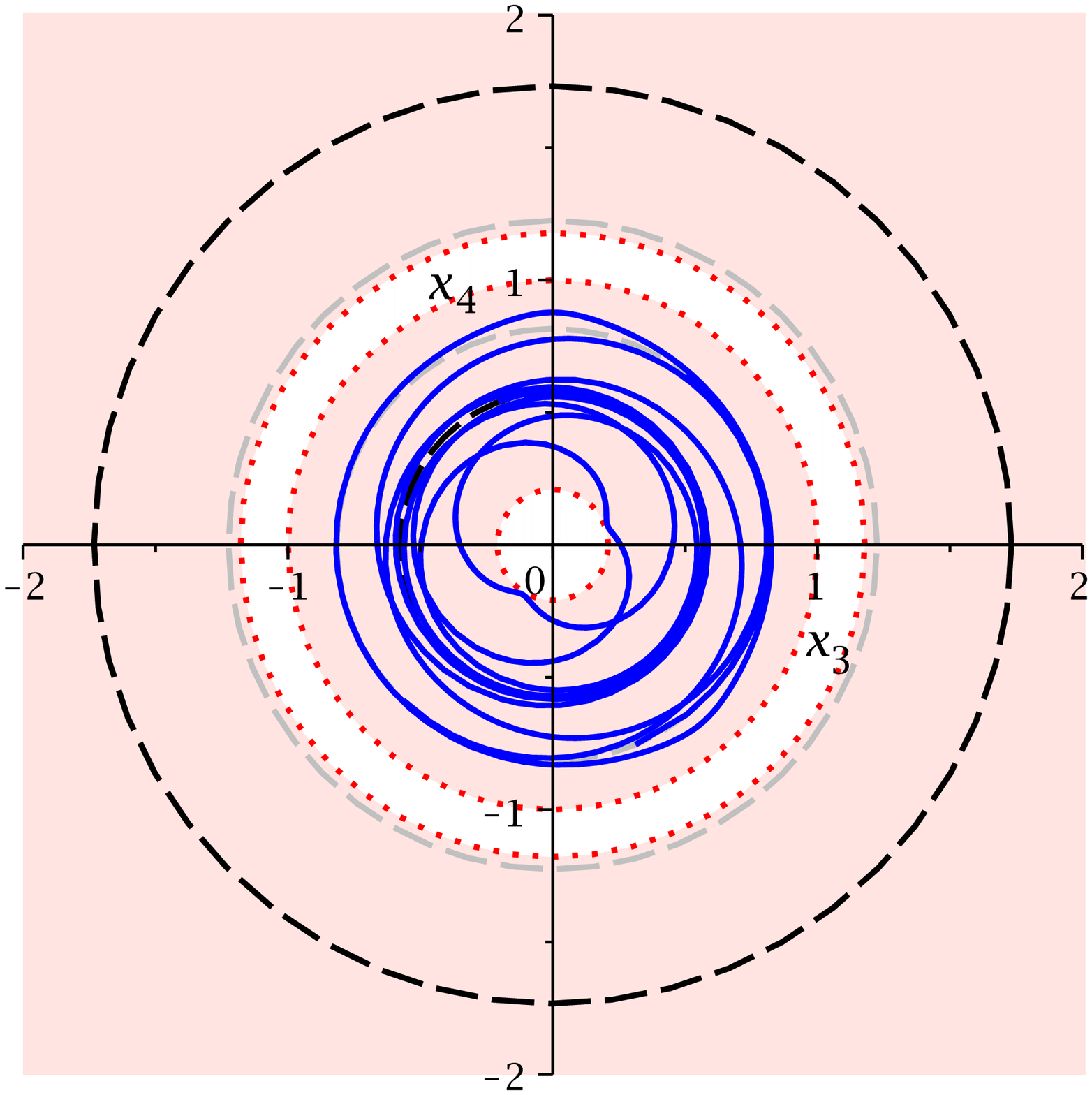}
 }
 \qquad
 \subfigure[$a$-$b$-plot ($\phi=\psi=\frac{\pi}{2}$)]{
   \includegraphics[width=7cm]{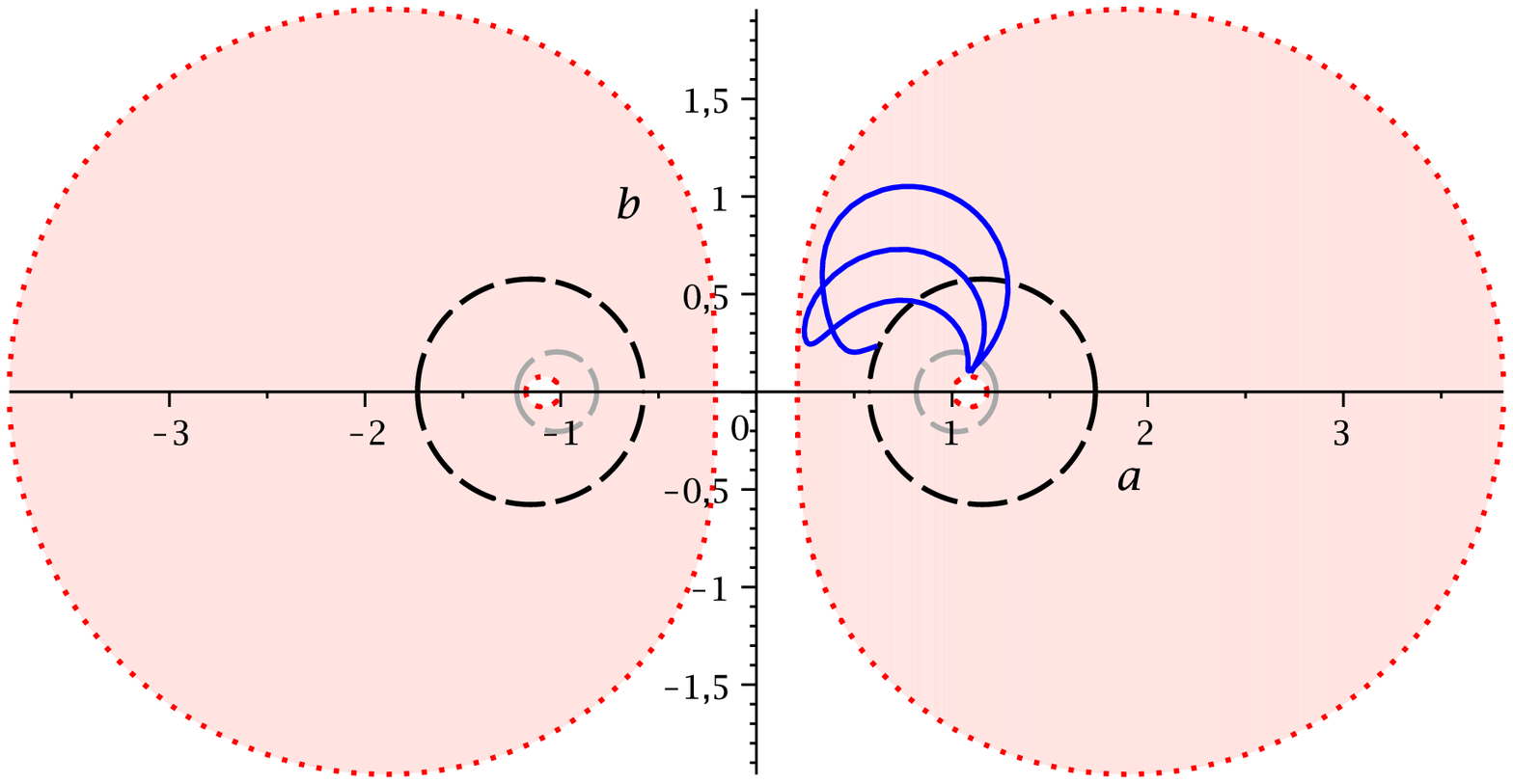}
 }
 \caption{$k=1$, $\nu=0.1$, $\lambda=0.7$, $\Phi=0.4$ and $\Psi=0.8$\newline
  Many-World Bound Orbit for light in the Ergosphere ($E=m=0$). The tori (a) or spheres (b) are the inner and outer horizons of the black ring. In (c) and (d) the ergosphere is depicted as a light red area with a red dotted border. The black and grey dashed circles in (c) and (d) are the inner and outer horizons.}
 \label{pic:em0-mbo}
\end{figure}

\begin{figure}
 \centering
 \subfigure[$x_1$-$x_3$-$x_4$-plot ($\phi=\frac{\pi}{2}$)]{
   \includegraphics[width=6cm]{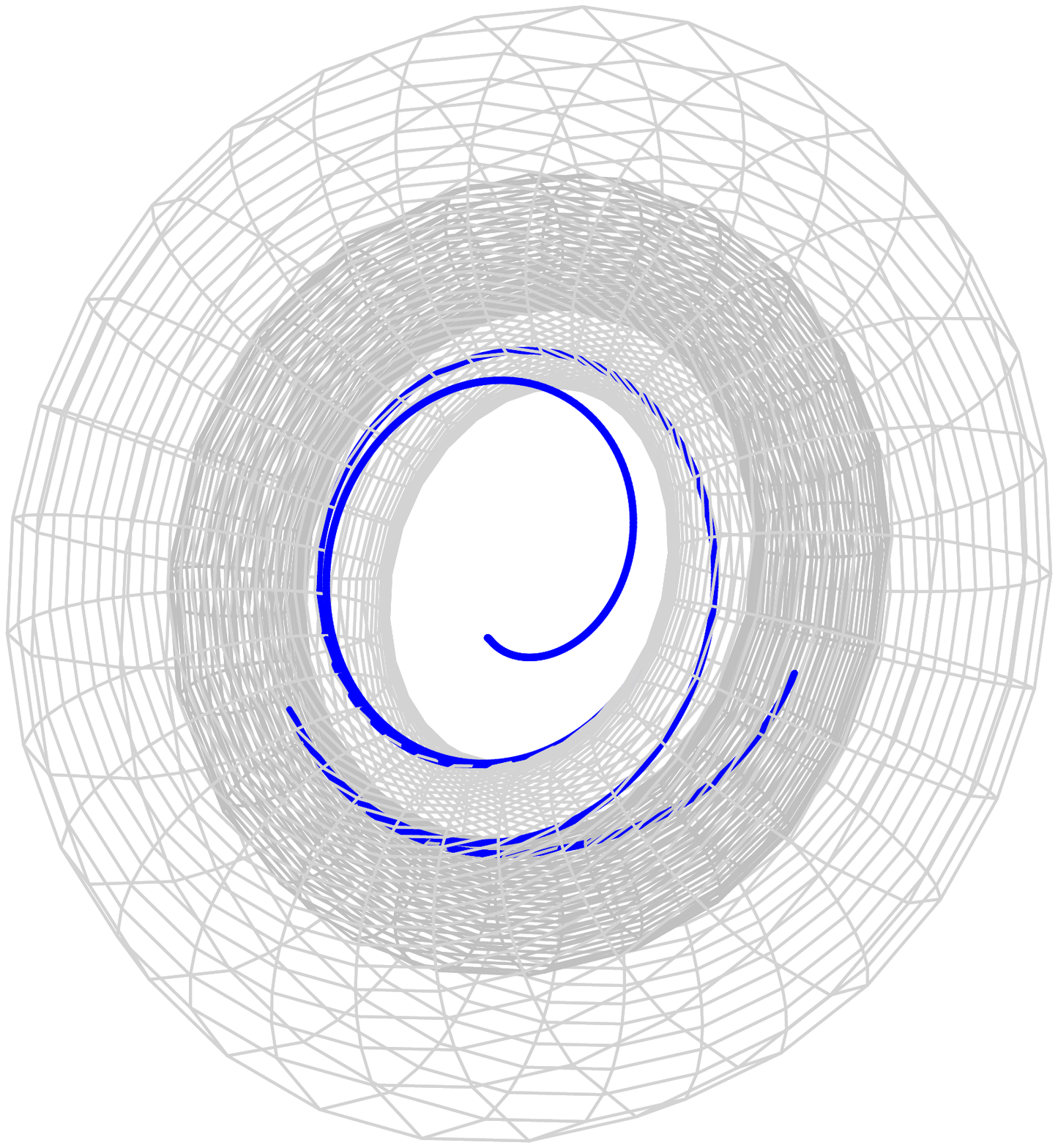}
 }
 \qquad\qquad
 \subfigure[$x_1$-$x_2$-$x_3$-plot ($\psi=\frac{\pi}{2}$)]{
   \includegraphics[width=6cm]{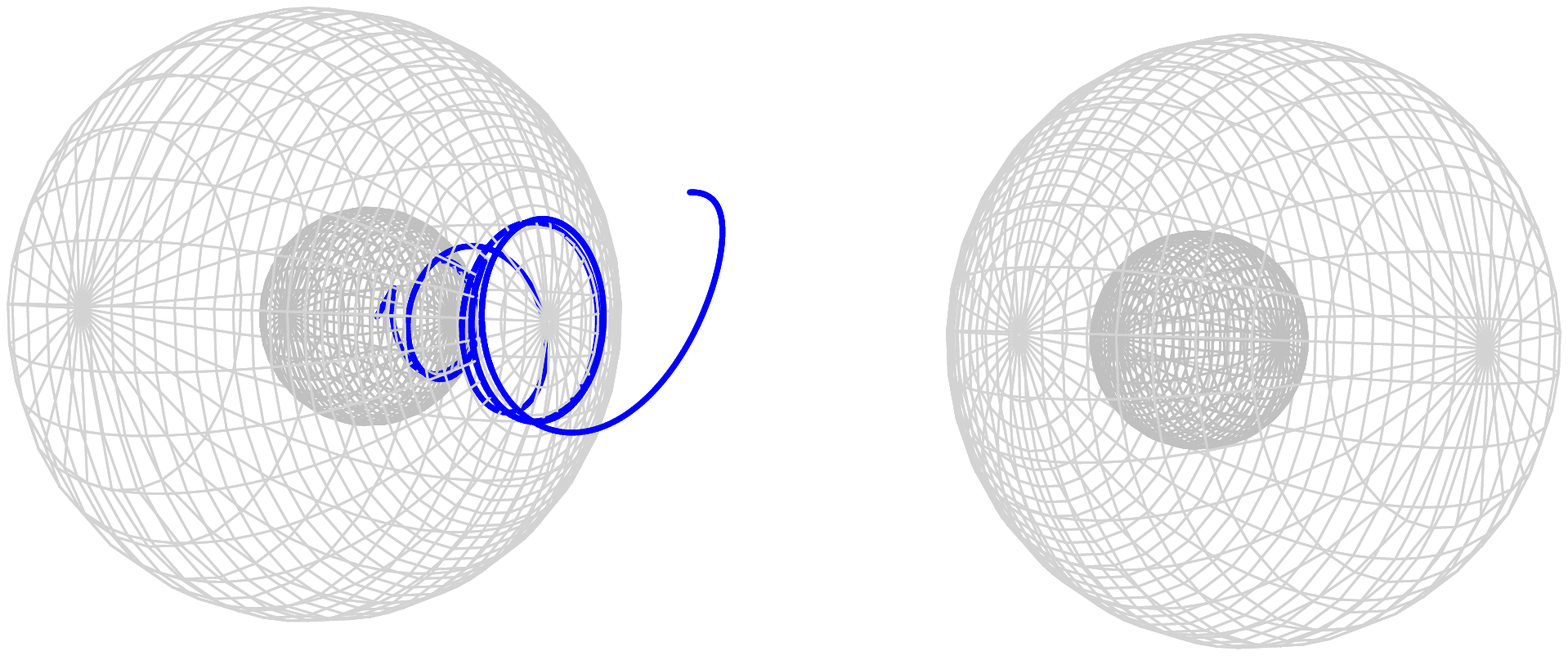}
 }
 \caption{$k=1$, $\nu=0.1$, $\lambda=0.7$, $\Phi=-0.2$ and $\Psi=3.6$\newline
  Termninating Orbit for light in the Ergosphere ($E=m=0$). The tori (a) or spheres (b) are the inner and outer horizons of the black ring.}
 \label{pic:em0-to}
\end{figure}

\begin{figure}
 \centering
 \subfigure[$x_1$-$x_3$-$x_4$-plot ($\phi=\frac{\pi}{2}$)]{
   \includegraphics[width=6cm]{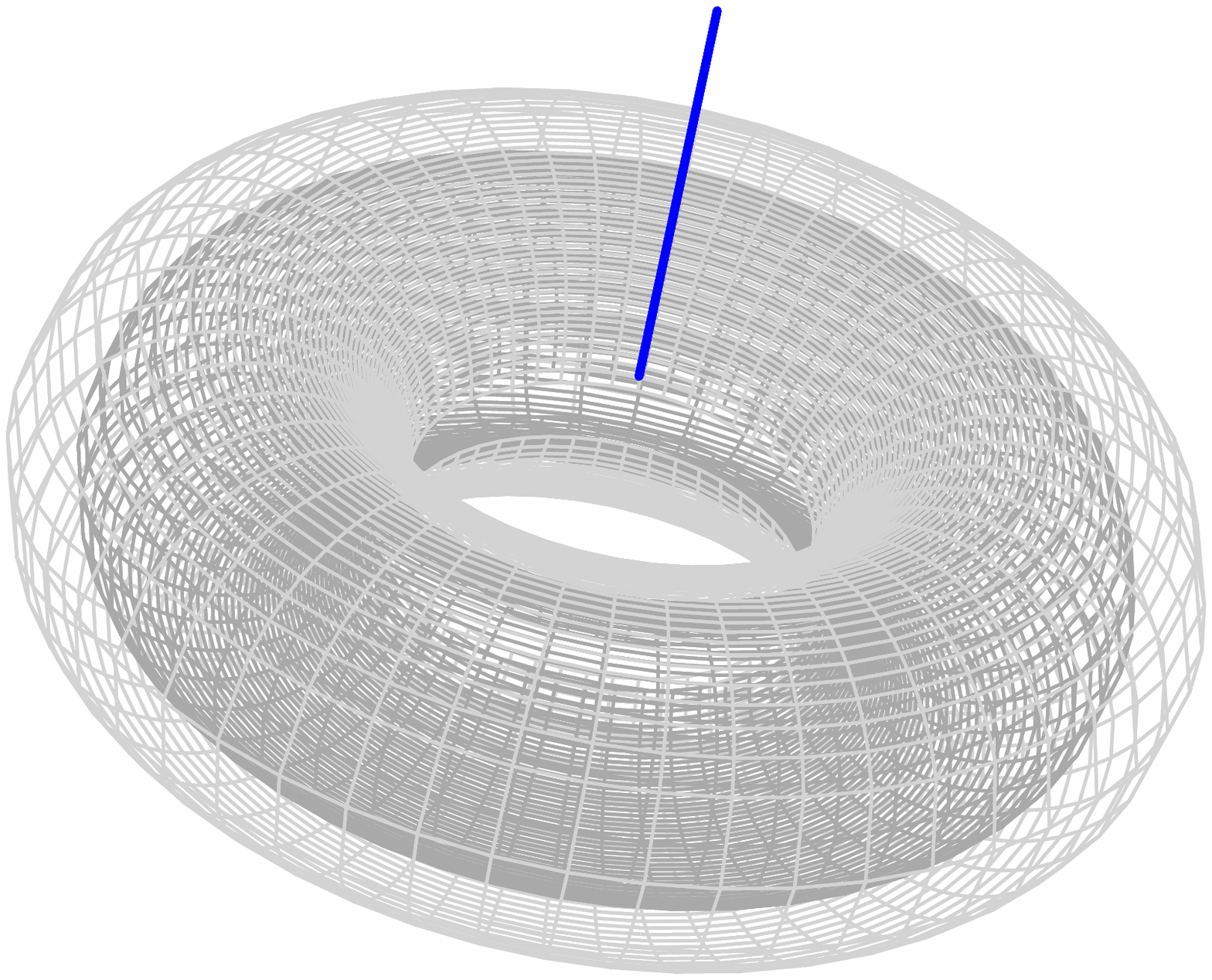}
 }
 \qquad\qquad
 \subfigure[$x_1$-$x_2$-$x_3$-plot ($\psi=\frac{\pi}{2}$)]{
   \includegraphics[width=6cm]{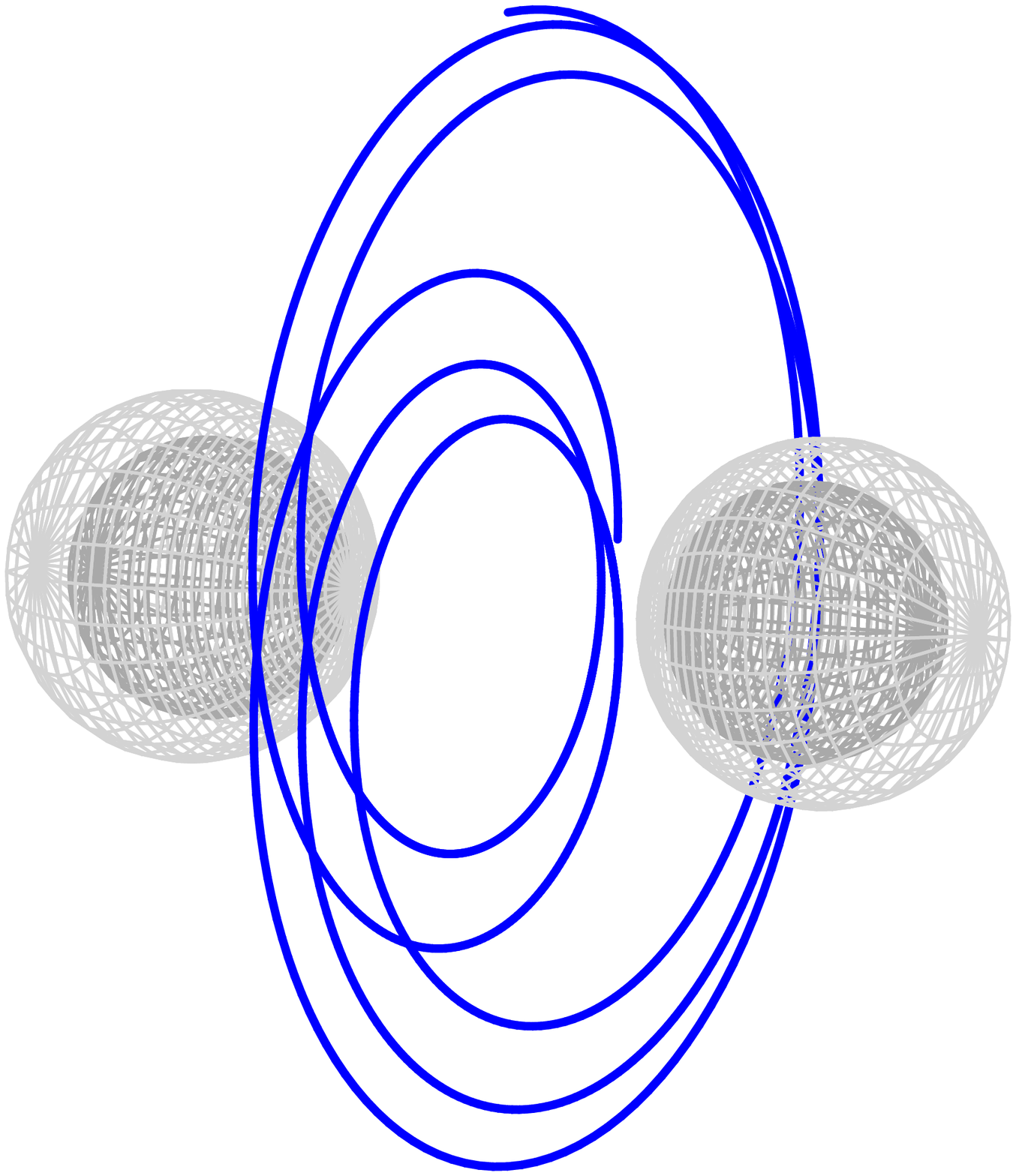}
 }
 \caption{$k=0.57$, $\nu=0.2$, $\lambda=0.9$, $\Phi=0.5$ and $\Psi=0$\newline
  Bound Orbit for light on the axis of $\psi$-rotation, i.e. $y=-1$, and also in the Ergosphere ($E=m=0$). The tori (a) or spheres (b) are the inner and outer horizons of the black ring.}
 \label{pic:em0-bo}
\end{figure}

\begin{figure}
 \centering
 \subfigure[$x_1$-$x_3$-$x_4$-plot ($\phi=\frac{\pi}{2}$)]{
   \includegraphics[width=6cm]{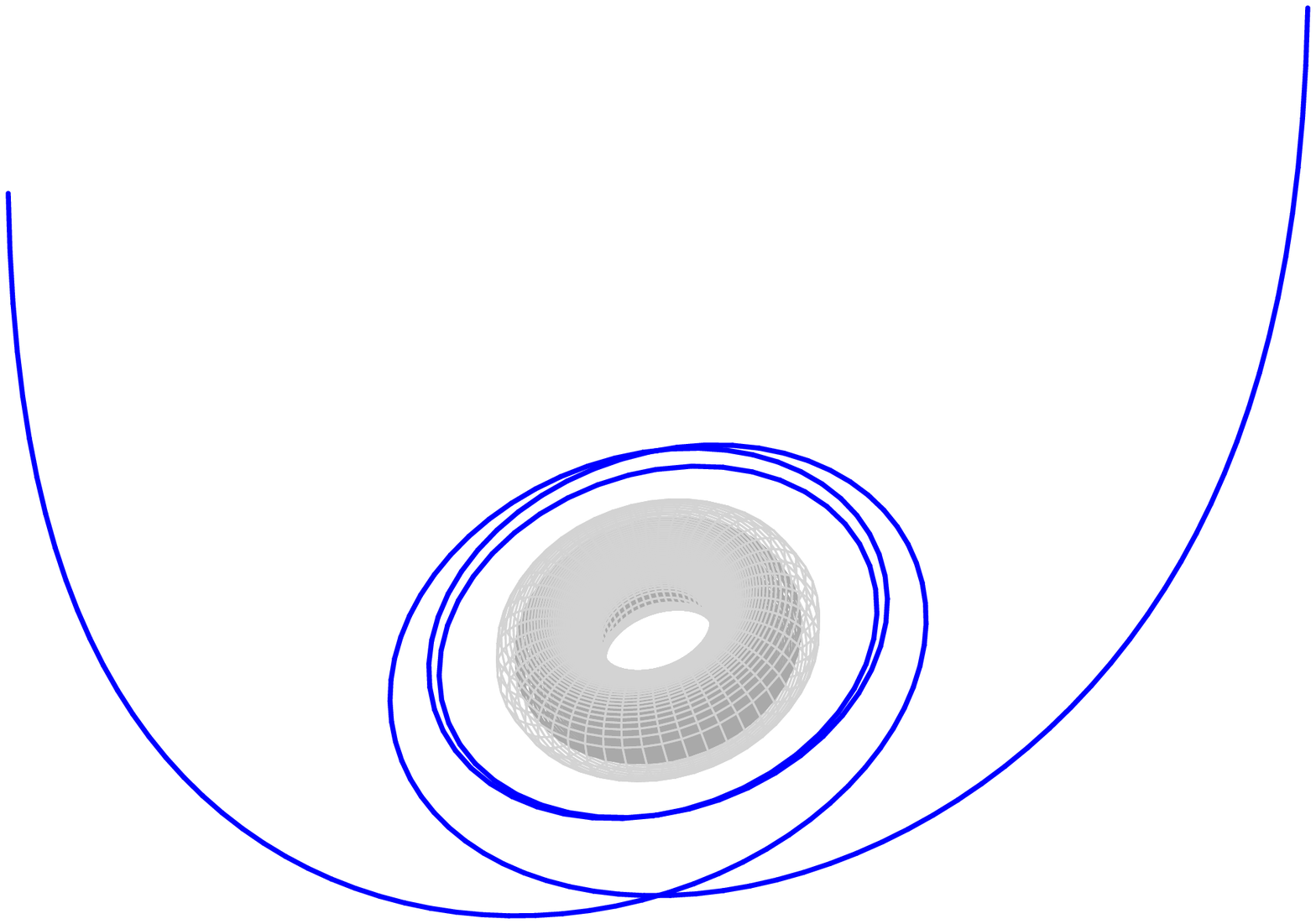}
 }
 \qquad\qquad
 \subfigure[$x_1$-$x_2$-$x_3$-plot ($\psi=\frac{\pi}{2}$)]{
   \includegraphics[width=6cm]{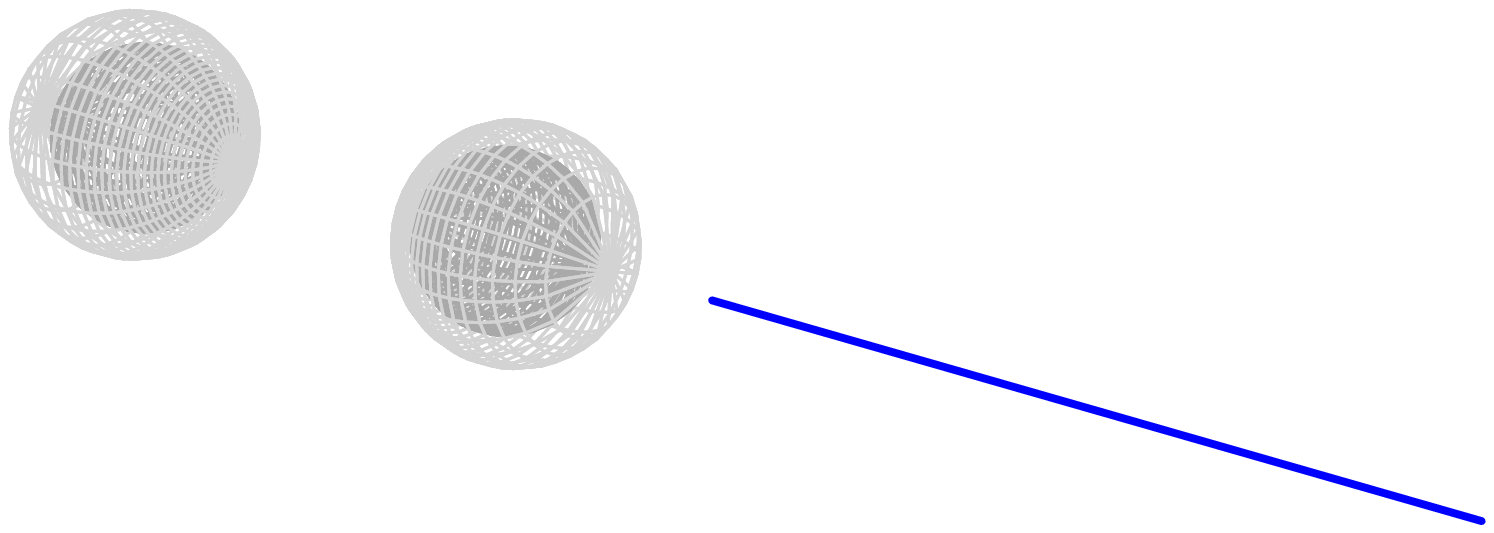}
 }
 \caption{$\nu=0.2$, $\lambda=0.9$,$\Psi=5$, $m=1$, $s=0$ and $E=1.38$\newline
  Escape Orbit for particles in the outer equatorial plane ($x=-1$ and $\Phi=0$) of an uncharged black ring. The tori (a) or spheres (b) are the inner and outer horizons of the black ring.}
 \label{pic:phiout-eo}
\end{figure}

\begin{figure}
 \centering
 \subfigure[$x_1$-$x_3$-$x_4$-plot ($\phi=\frac{\pi}{2}$)]{
   \includegraphics[width=6cm]{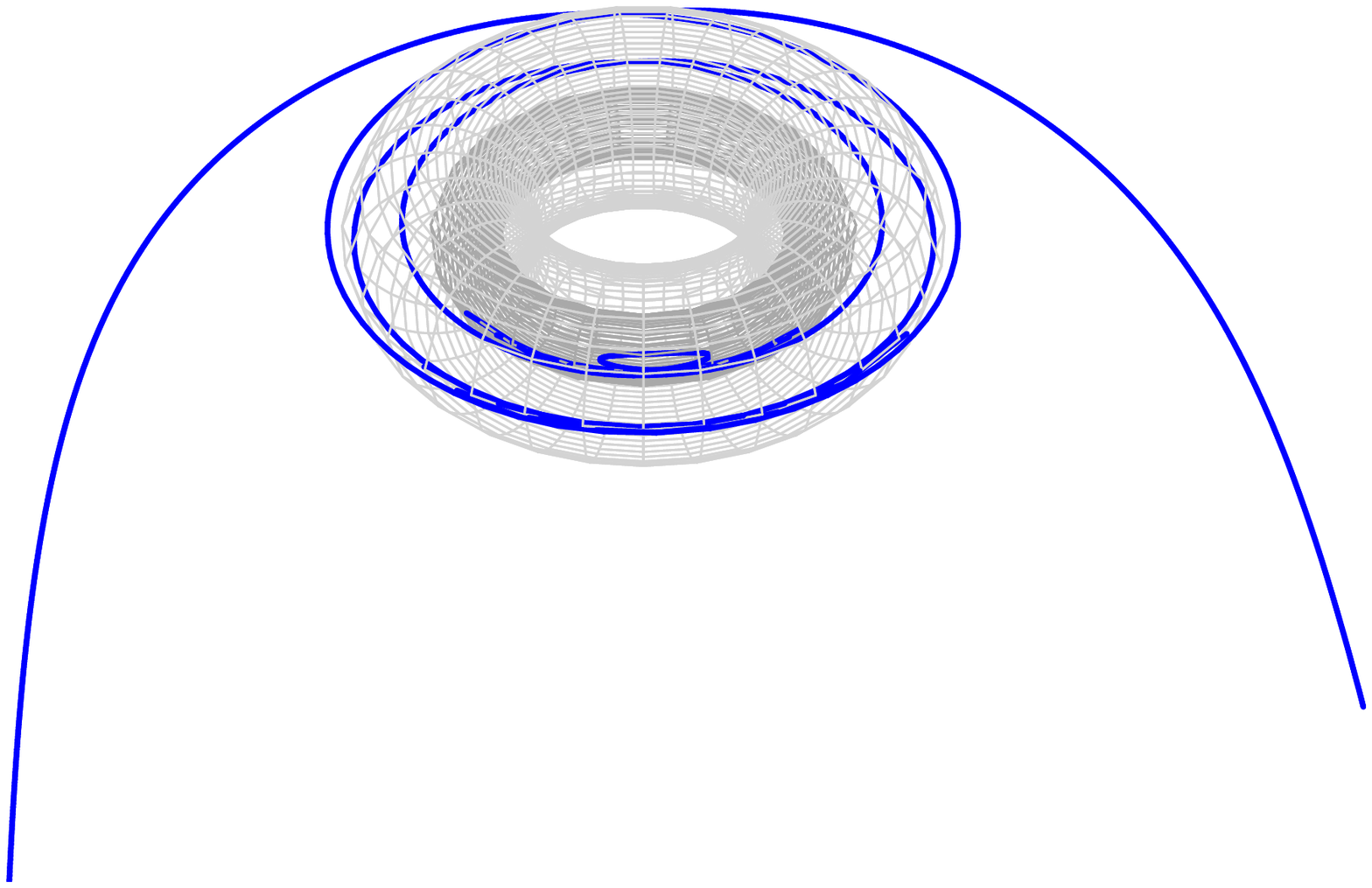}
 }
 \qquad\qquad
 \subfigure[$x_1$-$x_2$-$x_3$-plot ($\psi=\frac{\pi}{2}$)]{
   \includegraphics[width=6cm]{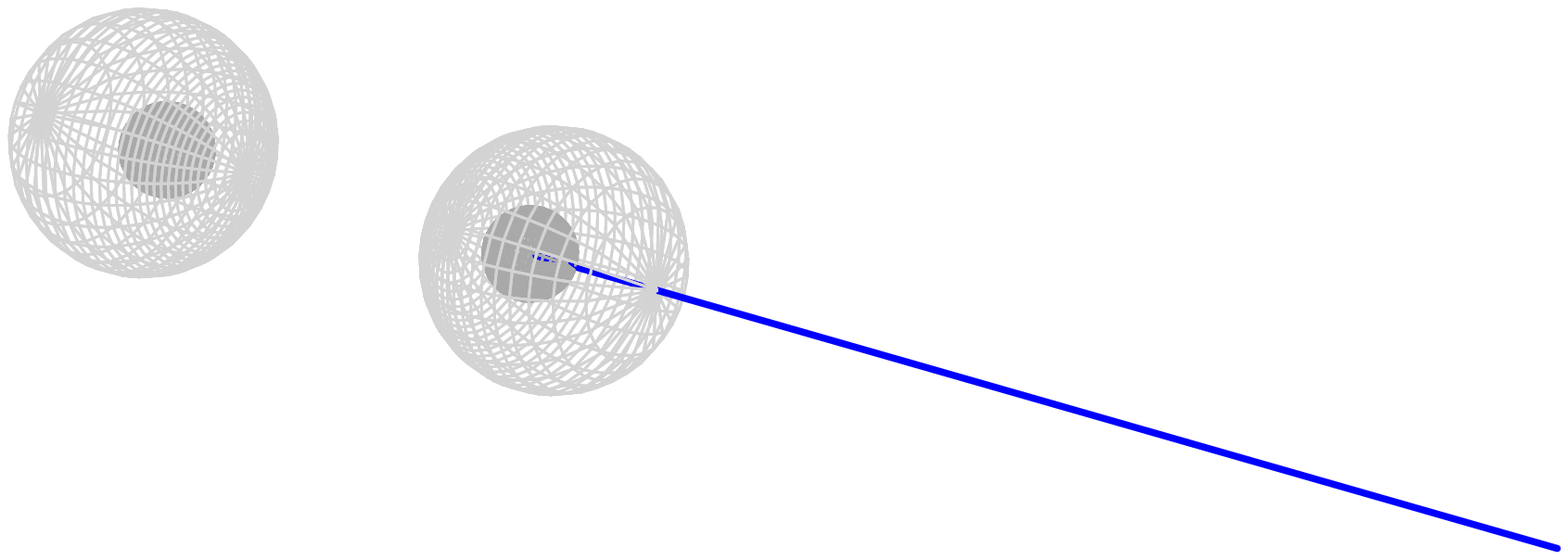}
 }
  \subfigure[$x_3$-$x_4$-plot]{
   \includegraphics[width=6cm]{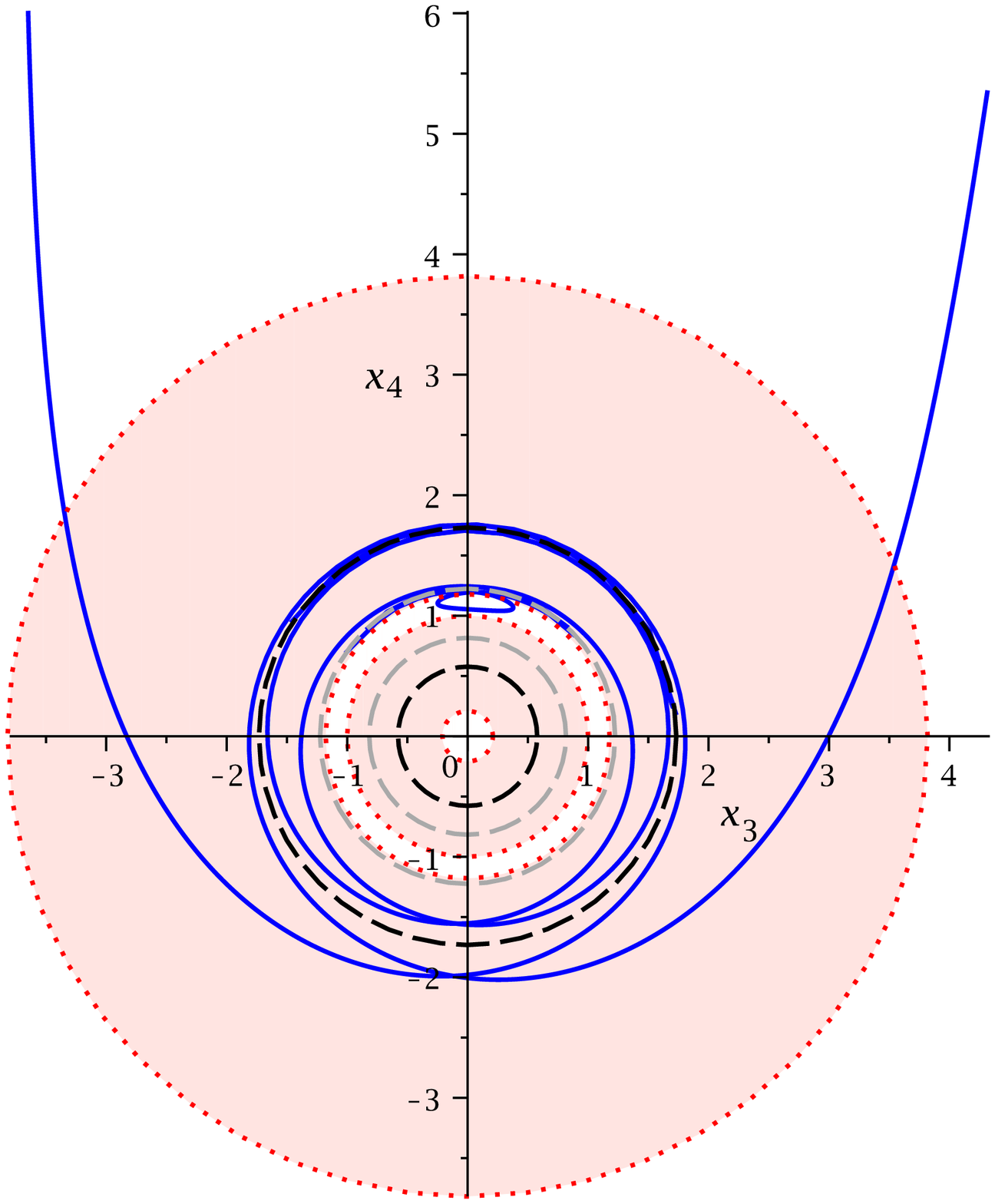}
 }
 \qquad
 \subfigure[$a$-$b$-plot ($\phi=\psi=\frac{\pi}{2}$)]{
   \includegraphics[width=7cm]{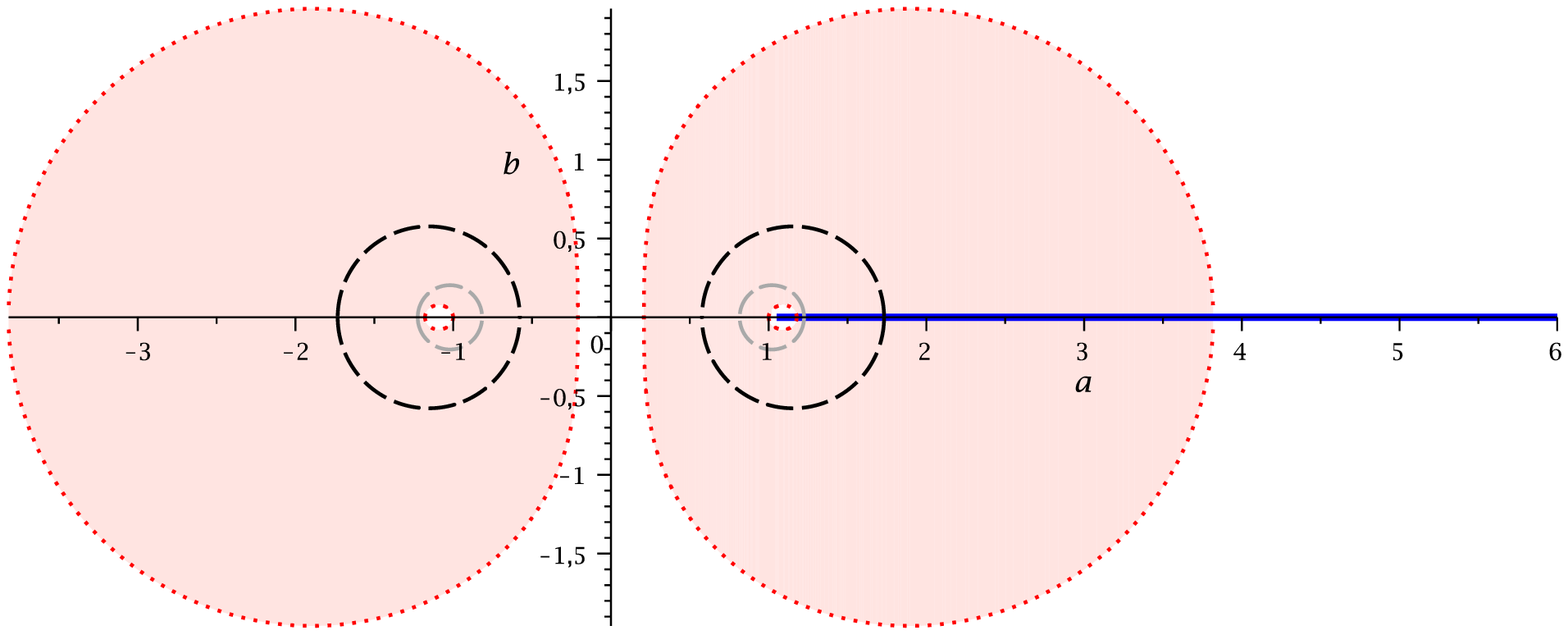}
 }
 \caption{$\nu=0.1$, $\lambda=0.7$,$\Psi=5$, $m=0$, $s=0.5$ and $E=2$\newline
  Two-World Escape Orbit for light in the outer equatorial plane ($x=-1$ and $\Phi=0$) of a charged black ring, i.e. $x=-1$ and $\Phi=0$. The tori (a) or spheres (b) are the inner and outer horizons of the black ring. In (c) and (d) the ergosphere is depicted as a light red area with a red dotted border. The black and grey dashed circles in (c) and (d) are the inner and outer horizons.}
 \label{pic:phiout-teo}
\end{figure}

\begin{figure}
 \centering
 \subfigure[$x_1$-$x_3$-$x_4$-plot ($\phi=\frac{\pi}{2}$)]{
   \includegraphics[width=6cm]{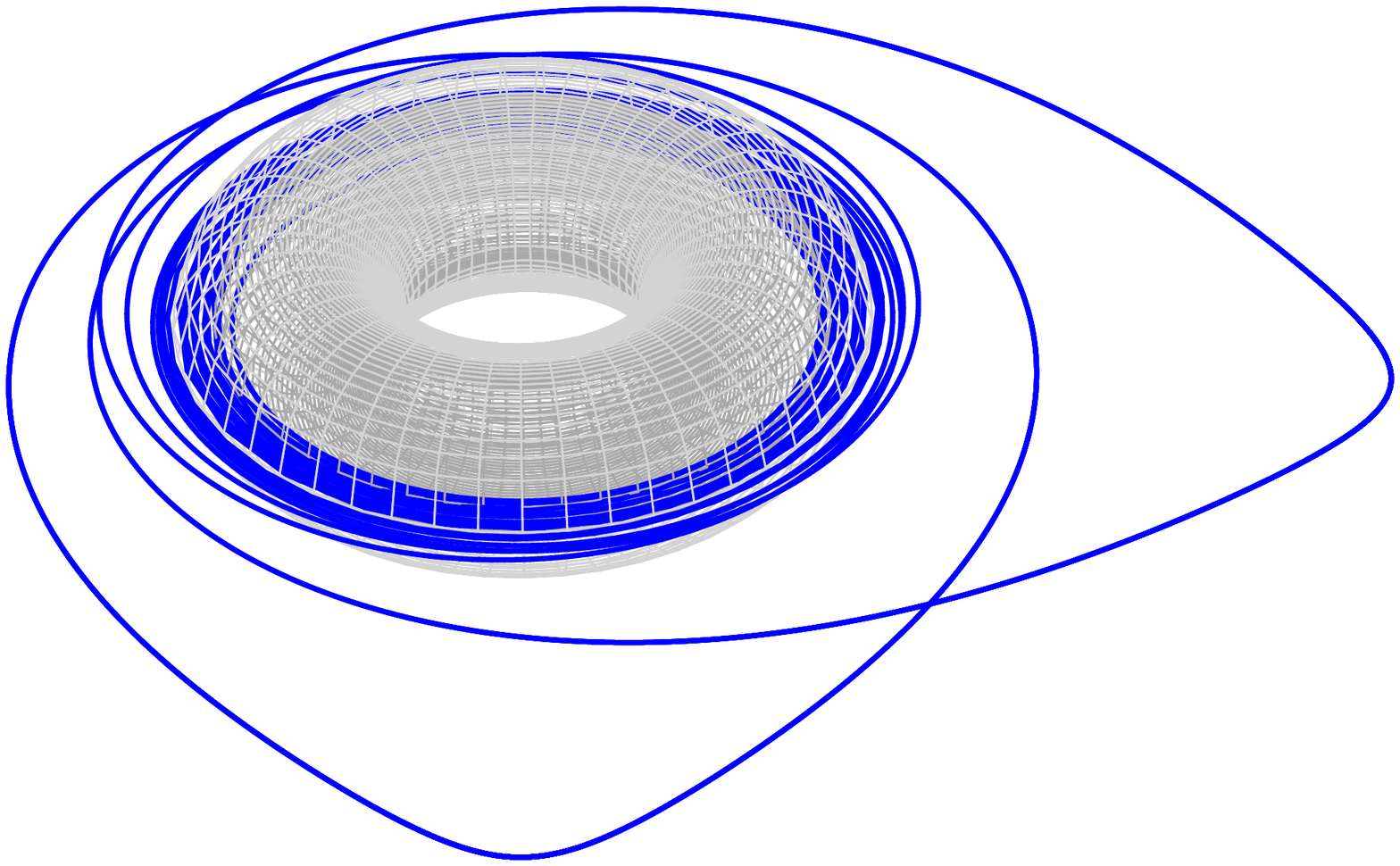}
 }
 \qquad\qquad
 \subfigure[$x_1$-$x_2$-$x_3$-plot ($\psi=\frac{\pi}{2}$)]{
   \includegraphics[width=6cm]{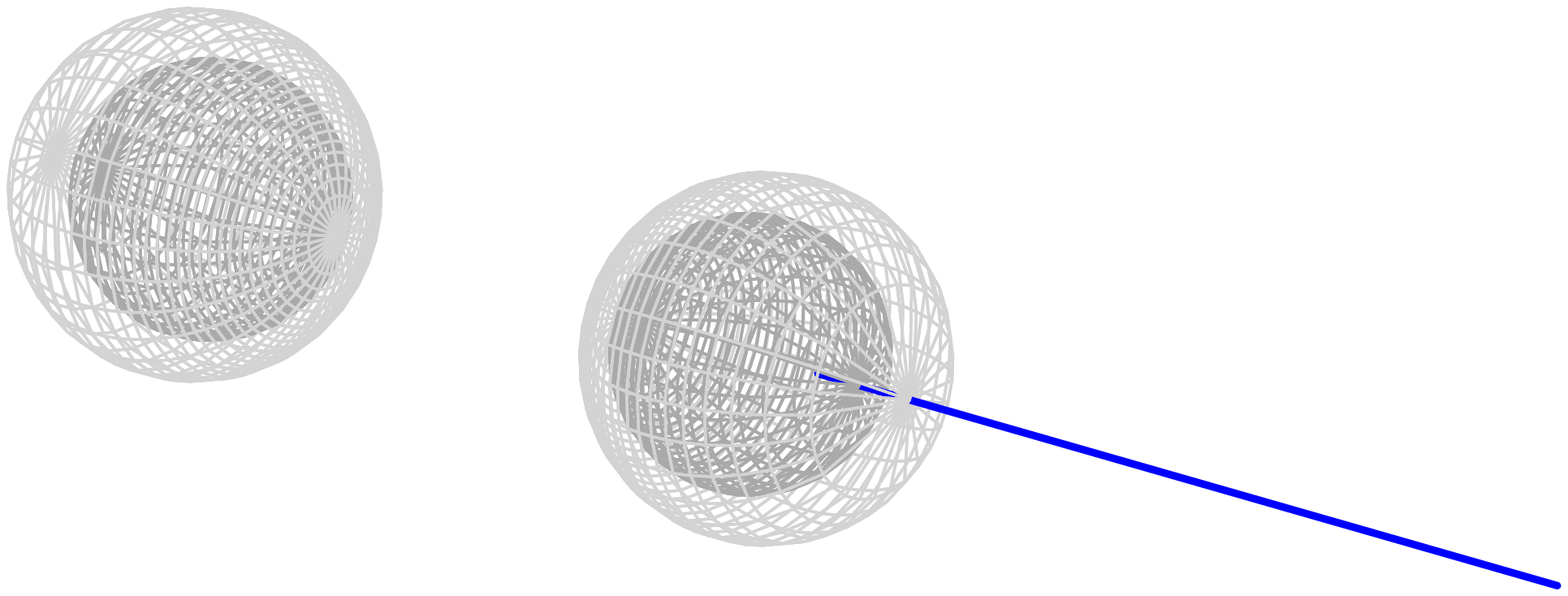}
 }
 \subfigure[$x_3$-$x_4$-plot]{
  \includegraphics[width=6cm]{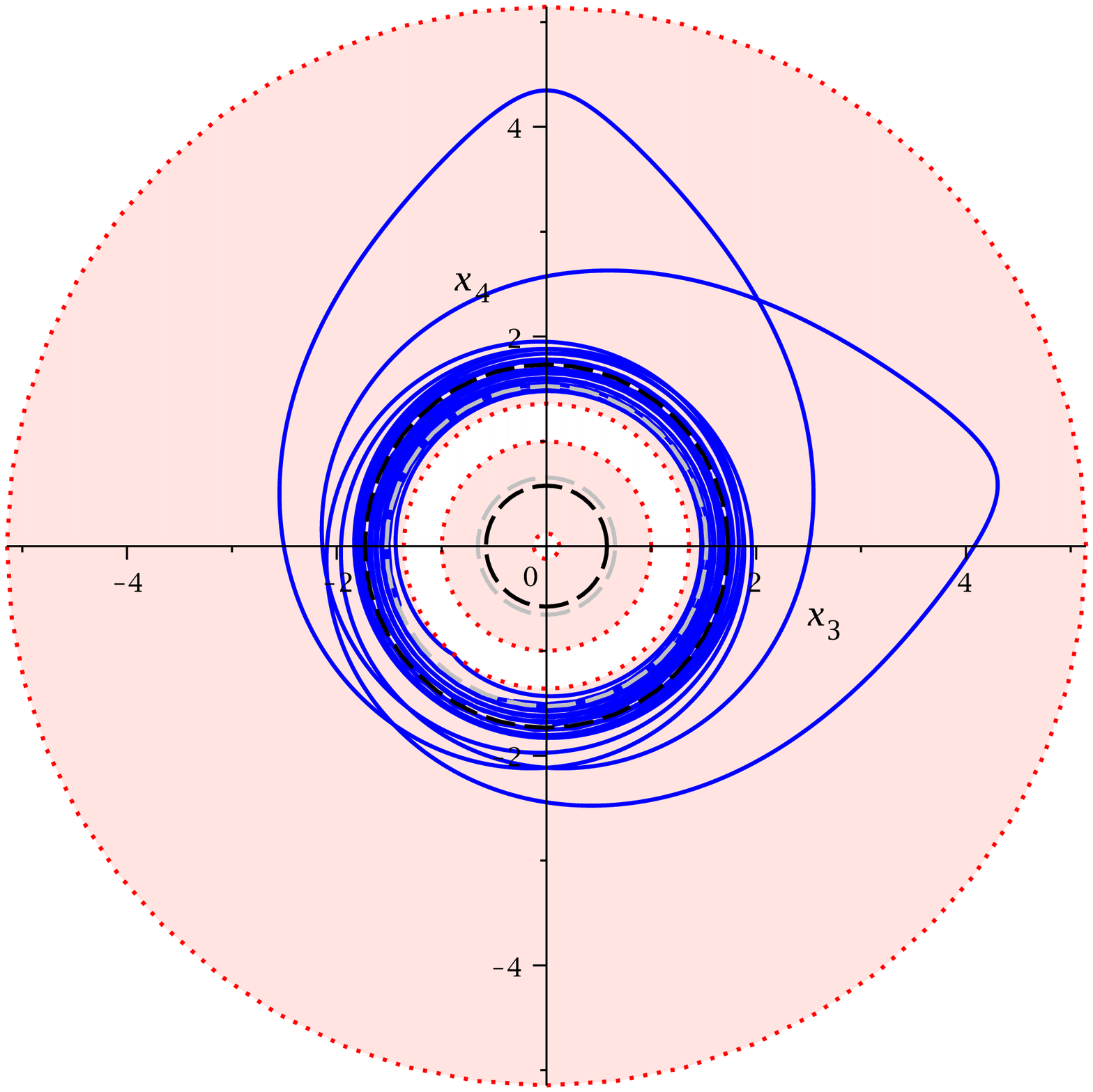}
 }
 \qquad
 \subfigure[$a$-$b$-plot ($\phi=\psi=\frac{\pi}{2}$)]{
   \includegraphics[width=7cm]{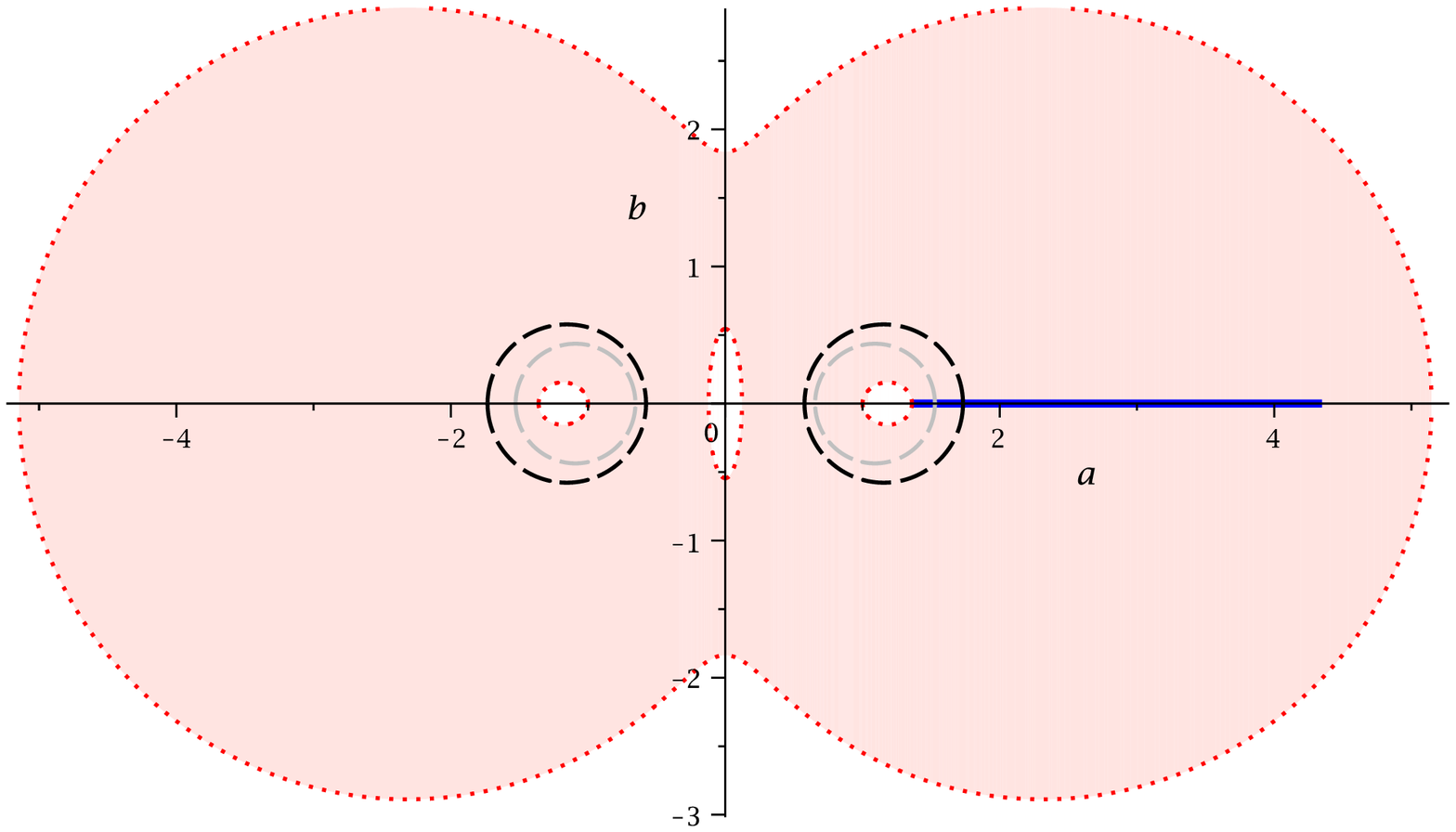}
 }
 \caption{$\nu=0.2$, $\lambda=0.9$,$\Psi=5$, $m=1$, $s=0$ and $E=0.1$\newline
  Many-World Bound Orbit for particles in the outer equatorial plane ($x=-1$ and $\Phi=0$) of an uncharged black ring. The tori (a) or spheres (b) are the inner and outer horizons of the black ring. In (c) and (d) the ergosphere is depicted as a light red area with a red dotted border. The black and grey dashed circles in (c) and (d) are the inner and outer horizons.}
 \label{pic:phiout-mbo}
\end{figure}

\begin{figure}
 \centering
 \subfigure[$x_1$-$x_3$-$x_4$-plot ($\phi=\frac{\pi}{2}$)]{
   \includegraphics[width=6cm]{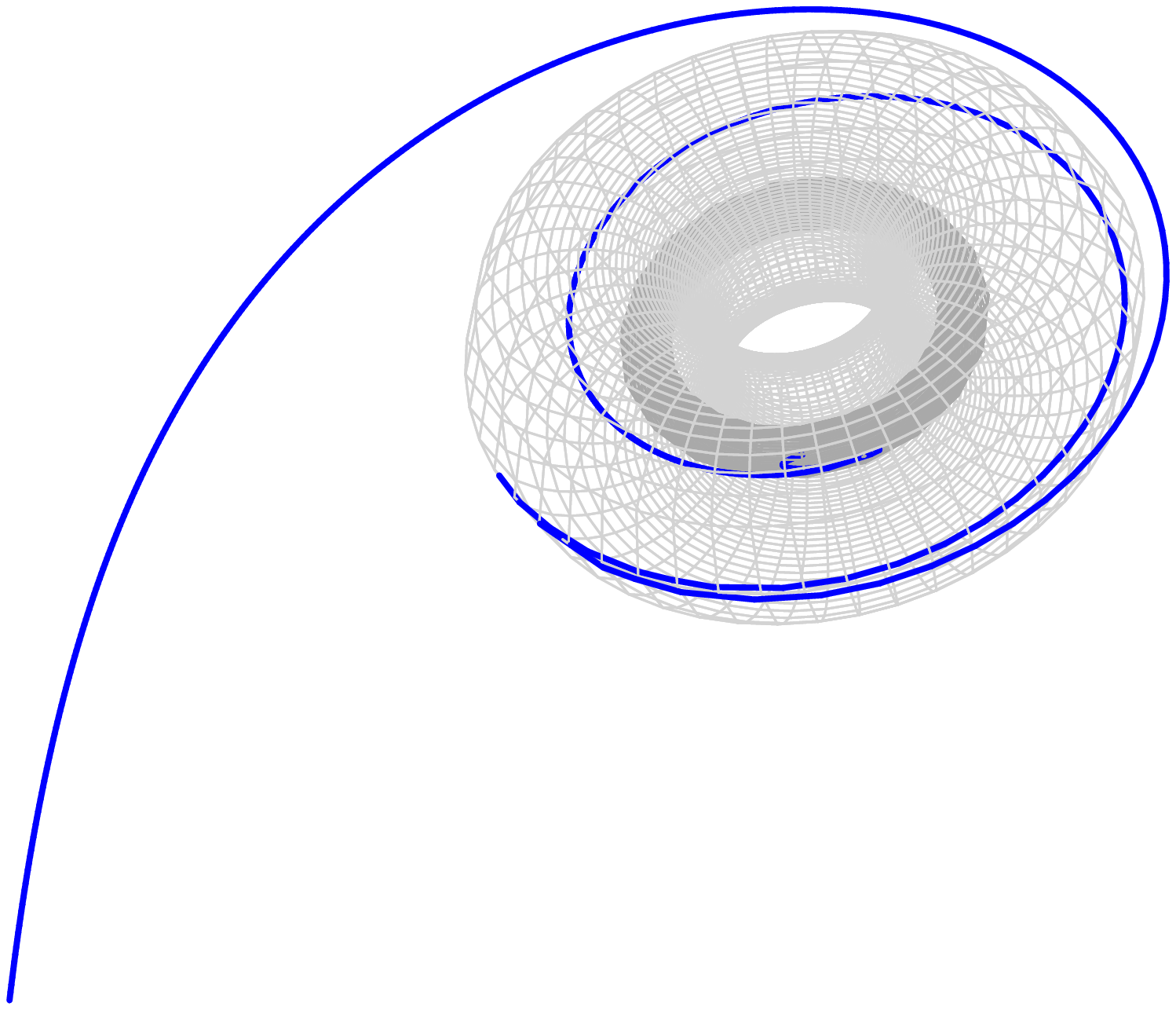}
 }
 \qquad\qquad
 \subfigure[$x_1$-$x_2$-$x_3$-plot ($\psi=\frac{\pi}{2}$)]{
   \includegraphics[width=6cm]{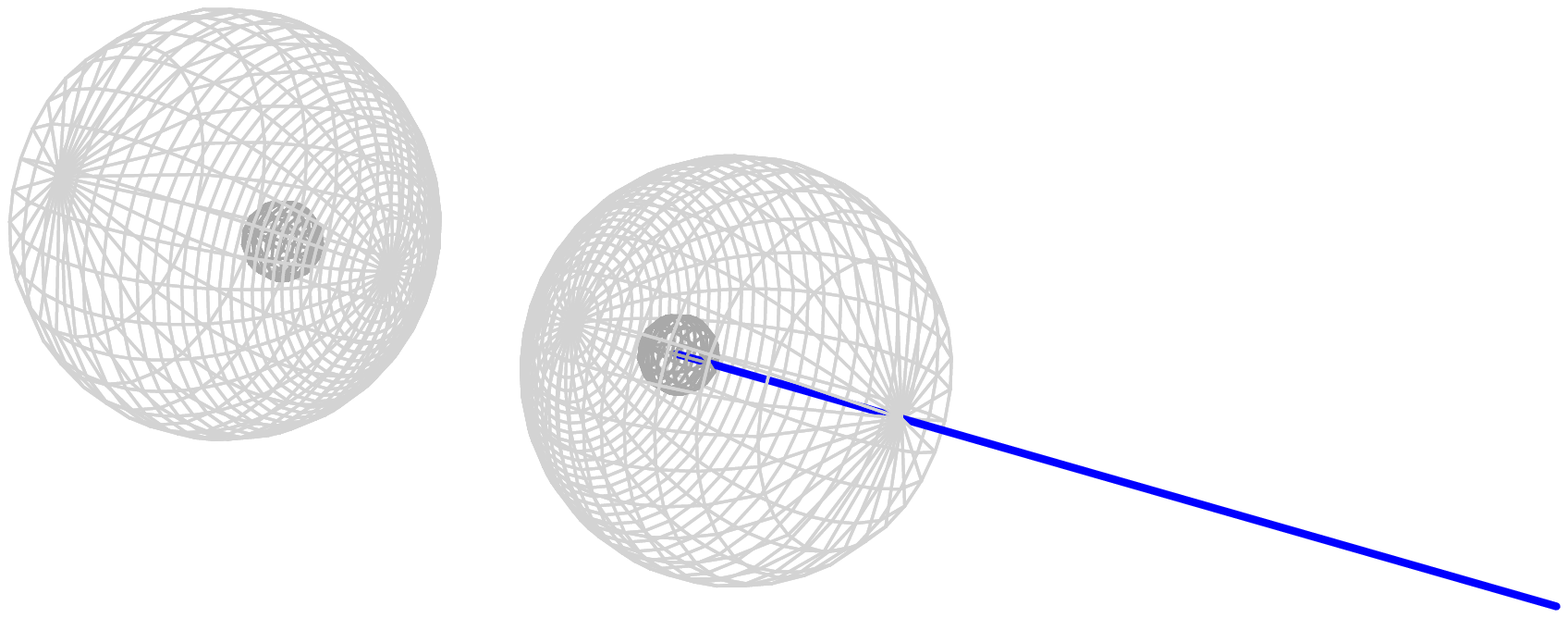}
 }
 \subfigure[$x_3$-$x_4$-plot]{
  \includegraphics[width=6cm]{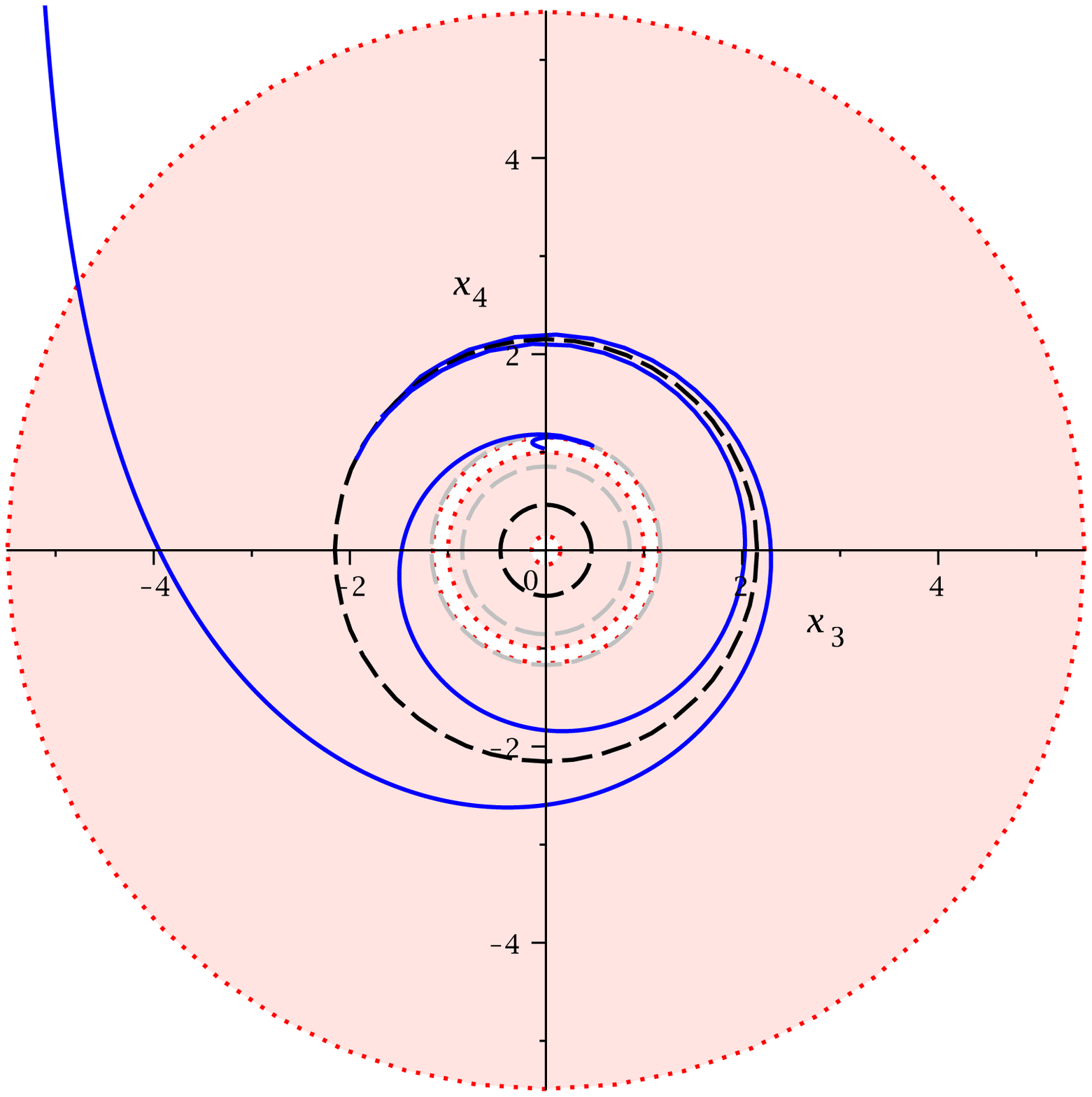}
 }
 \qquad
 \subfigure[$a$-$b$-plot ($\phi=\psi=\frac{\pi}{2}$)]{
   \includegraphics[width=7cm]{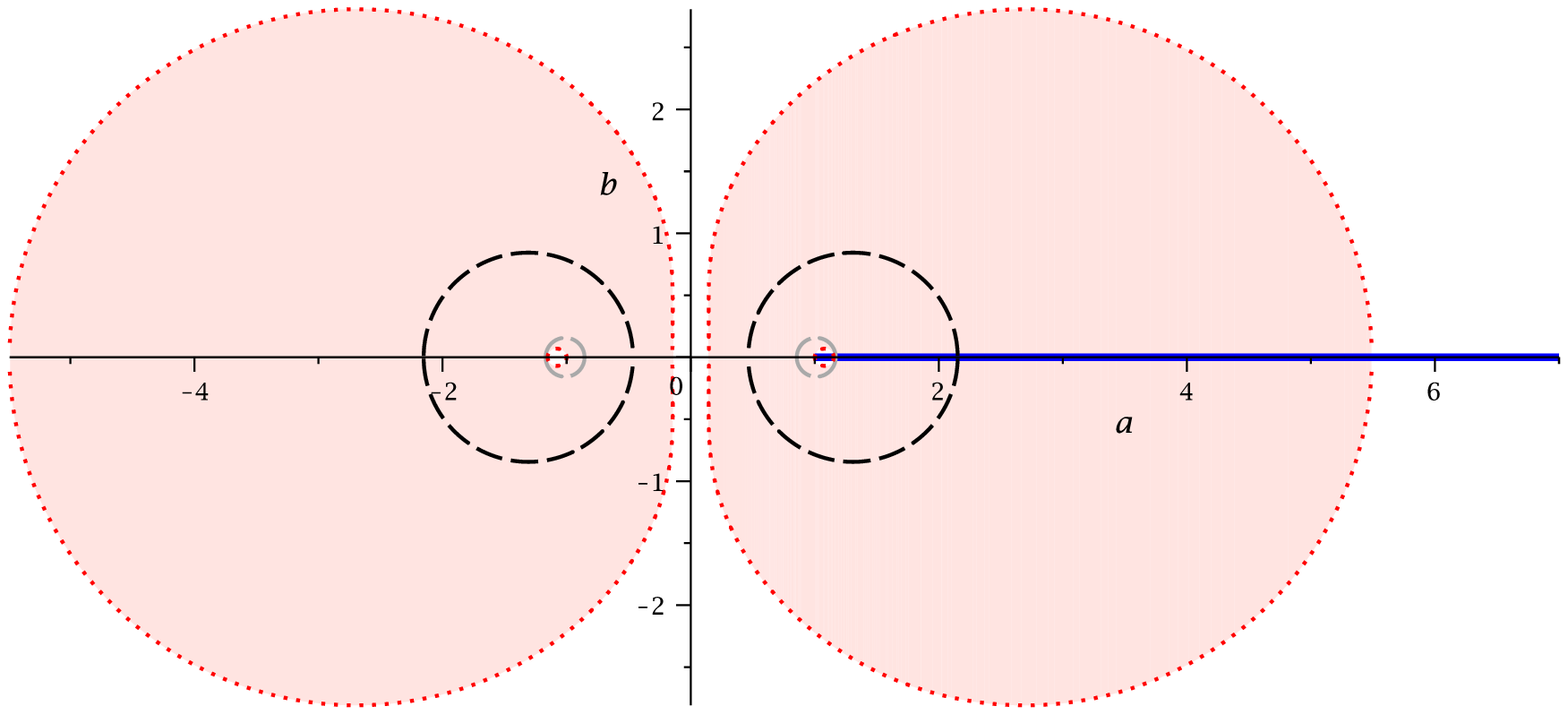}
 }
 \caption{$\nu=0.1$, $\lambda=0.8$,$\Psi=10$, $m=1$, $s=0$ and $E=3.34$\newline
  Terminating Orbit for particles in the outer equatorial plane ($x=-1$ and $\Phi=0$) of an uncharged black ring. The tori (a) or spheres (b) are the inner and outer horizons of the black ring. In (c) and (d) the ergosphere is depicted as a light red area with a red dotted border. The black and grey dashed circles in (c) and (d) are the inner and outer horizons.}
 \label{pic:phiout-to}
\end{figure}

\begin{figure}
 \centering
 \subfigure[$x_1$-$x_3$-$x_4$-plot ($\phi=\frac{\pi}{2}$)]{
   \includegraphics[width=6cm]{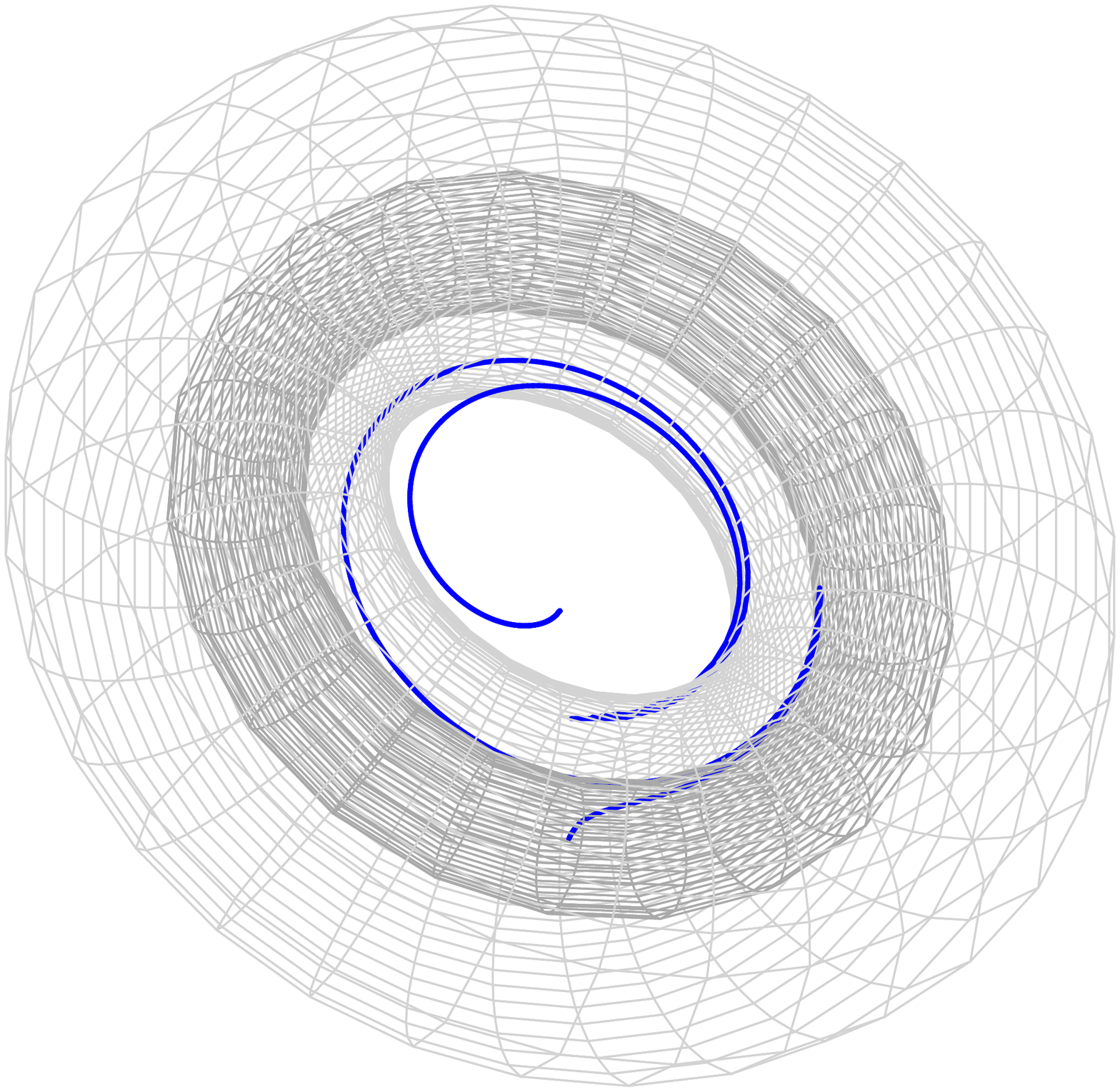}
 }
 \qquad\qquad
 \subfigure[$x_1$-$x_2$-$x_3$-plot ($\psi=\frac{\pi}{2}$)]{
   \includegraphics[width=6cm]{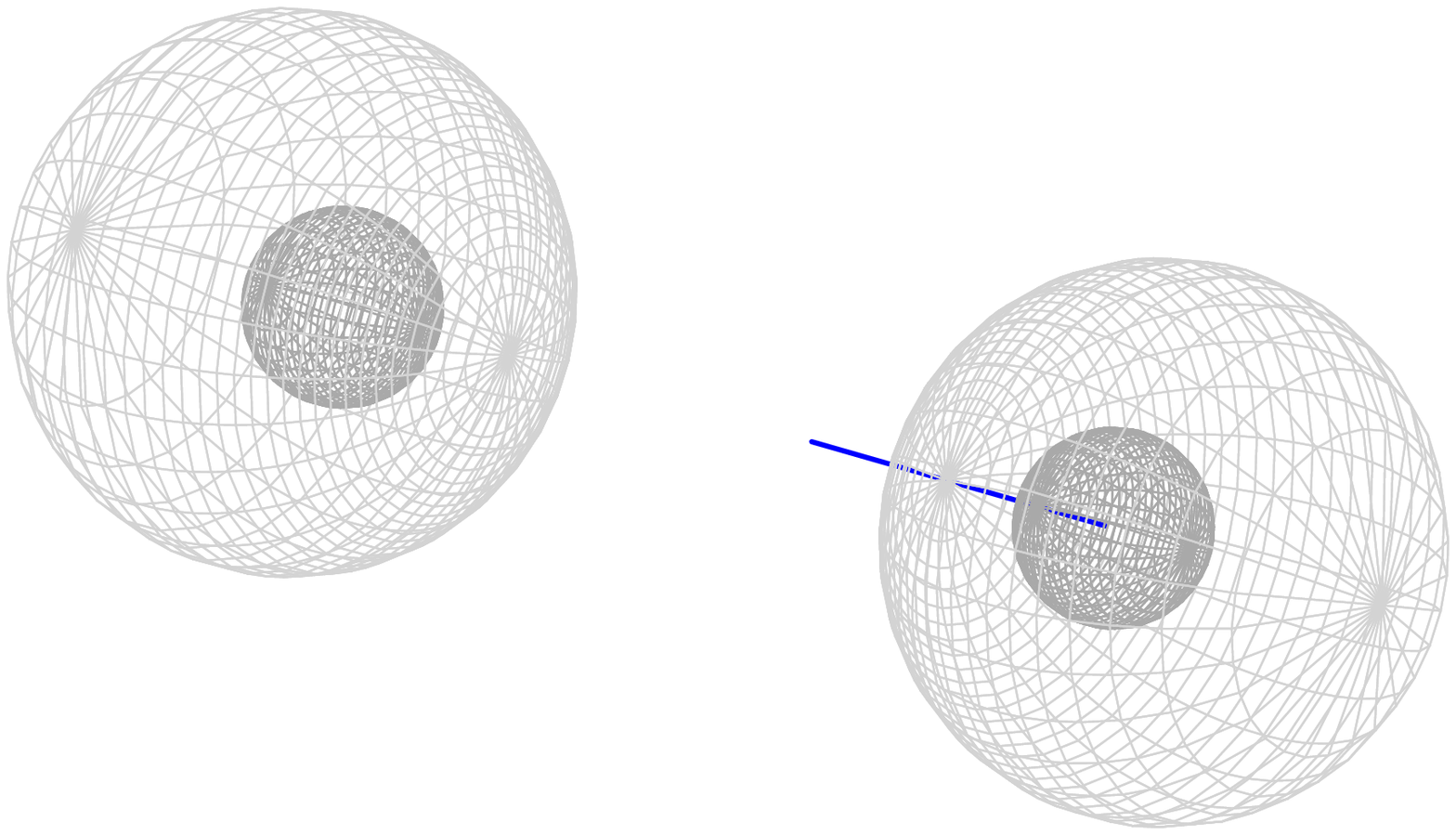}
 }
 \caption{$\nu=0.1$, $\lambda=0.7$,$\Psi=5$, $m=0$, $s=0.5$ and $E=0.1$\newline
  Terminating Orbit for light in the inner equatorial plane ($x=+1$ and $\Phi=0$) of a charged black ring. The tori (a) or spheres (b) are the inner and outer horizons of the black ring.}
 \label{pic:phiin-to}
\end{figure}

Figures \ref{pic:em0-mbo} - \ref{pic:phiin-to} show examples for possible orbits in the (charged) doubly spinning black ring spacetime. The orbits are shown for $\phi=\frac{\pi}{2}=\rm const.$ ($x_1$-$x_3$-$x_4$-plot) and $\psi=\frac{\pi}{2}=\rm const.$ ($x_1$-$x_2$-$x_3$-plot). Sometimes the orbit is also plotted in the equatorial plane ($x_3$-$x_4$-plot) and the plane of constant angles ($a$-$b$-plot), in these plots we have also marked the ergosphere.

From the plots where one or both angles were set constant one might get the impression that the motion takes place only on one ``side'' of the black ring. In the $x_1-x_3-x_4$-plots the orbit seems to be either above or below the equatorial plane, while in the $x_1-x_2-x_3$-plots the motion seems to take place only in the vicinity of one of the two spheres that show the horizon. However, this is an artifact of the chosen representation. When none of the angles is set constant, particles and light move all around the black ring with respect to their turning points. Nevertheless, we chose to exhibit the orbits with constant angle for reasons of clarity.\\

A MBO and a TO for light in the ergosphere are depicted in figure \ref{pic:em0-mbo} and \ref{pic:em0-to}. Figure \ref{pic:em0-bo} shows a BO for light on the axis of $\psi$-rotation, this orbit also lies in the ergosphere. An EO for particles in the equatorial plane of an uncharged black ring can be seen in figure \ref{pic:phiout-eo}. 

A TEO for light around a charged black ring is shown in figure \ref{pic:phiout-teo}. The orbit forms a little loop, which can be explained by the fact that there is an ergosphere-free region inside the black ring  (see section \ref{sec:spacetime}). After the light has left the ergosphere it is no longer dragged along by the rotation of the black ring and changes its direction. The light reaches its turning point and approaches again the ergosphere where it turns around a second time.

A MBO for particles in the equatorial plane of an uncharged black ring is depicted in figure \ref{pic:phiout-mbo}. TOs in the equatorial plane are shown in figure \ref{pic:phiout-to} (particles in the uncharged black ring spacetime) and figure \ref{pic:phiin-to} (light in the charged black ring spacetime). In figure \ref{pic:phiout-to} the effect of the ergosphere can be seen as the particle changes its direction after entering the ergosphere-free region inside the black ring 

\section{Conclusion}

In this paper we presented the analytical solutions of the geodesic equations of the charged doubly spinning black ring for special cases. Since the Hamilton-Jacobi equation seems not to be separable in general, we had to concentrate on the zero energy null geodesics in the ergosphere ($E=m=0$), geodesics on the  axis of $\psi$-rotation ($y=-1$) and geodesics on the equatorial plane ($x=\pm1$), which is the axis of $\phi$-rotation.\\
We discussed the general structure of the orbits and gave a complete classification of their types.

In the case $E=m=0$ terminating orbits, many-world bound orbits and bound orbits for $y={\rm const.}=-1$ are possible. The charge of the black ring has no effect on the orbits in the ergosphere. If the black ring is singly spinning ($\nu=0$) only terminating orbits exist (see \cite{Grunau:2012ai}).

On the axis of $\psi$-rotation $y$ is constant, so here the $x$-motion determines the type of orbit. We found escape orbits and bound orbits in the (charged) doubly spinning spacetime like in the singly spinning black ring spacetime.

The axis of $\phi$-rotation (the equatorial plane) is divided into two parts: the plane enclosed by the ring ($x=+1$) and the plane surrounding the ring ($x=-1$). On the plane enclosed by the ring  ($x=+1$) terminating orbits and many-world bound orbits are possible. If the doubly spinning black ring is charged, bound orbits behind the inner horizon exist. If the black ring is singly spinning only terminating orbits are possible.

On the plane surrounding the ring ($x=-1$) escape orbits, terminating orbits, two-world escape orbits and many-world bound orbits are possible. If the black ring is singly spinning, only escape orbits and terminating orbits exist.\\

The separability of the Hamilton-Jacobi equation is a coordinate related phenomenon, so one might think of a coordinate system in which it would be possible to separate the Hamilton-Jacobi equation in general. But recently Igata, Ishihara and Takamori found evidence of chaotic motion in the singly spinning black ring spacetime using the Poincar\'e map \cite{Igata:2010cd}. From that one could conclude that it is in general not possible to separate the Hamilton-Jacobi equation of singly spinning black rings in any coordinate system. It would be interesting to see if chaotic motion also appears in the (charged) doubly spinning black ring spacetime.\\

The methods shown in this paper can be applied to other black ring spacetimes like for example the supersymmetric black ring \cite{Elvang:2004rt}, \cite{Elvang:2004ds}. This will be done in future work.

\section{Acknowledgements}

We would like to thank Victor Enolski, Norman G\"urlebeck and Volker Perlick for helpful discussions. We gratefully acknowledge support by the DFG, in particular, also within the DFG Research Training Group 1620 ``Models of Gravity''.


\bibliographystyle{unsrt}

\end{document}